%% file: FourBosonRadius.tex
\documentclass[aps,prc,twocolumn,superscriptaddress,nofootinbib,tightenlines,
preprintnumbers,showkeys,floatfix]{revtex4-2}

\usepackage[utf8]{inputenc}
\usepackage[english]{babel}
\usepackage{amssymb,amsthm,amsmath,amstext,amsbsy,amsopn,mathrsfs}
\usepackage{bbm}
\usepackage{bm}
\usepackage{nicefrac}
\usepackage{suffix}
\usepackage{array}
\usepackage{slashed}
\usepackage{leftidx}
\usepackage{graphicx}
\usepackage{hyperref}
\usepackage{environ}
\usepackage{mathtools}
\usepackage{xspace}
\usepackage{xcolor}
\usepackage{scalerel,stackengine}
\usepackage{physics}
\usepackage{braket}
\usepackage{isotope}

\input{defs}

\begin{document}

\title{Three- and four-boson systems expanded around the unitarity limit:\\
Application to \isotope[4]{He} molecules}

\author{Feng Wu}
\email{feng.wu@ijclab.in2p3.fr}
\affiliation{Université Paris-Saclay, CNRS/IN2P3, IJCLab, 91405 Orsay, France}
\affiliation{Department of Physics,
University of Arizona, Tucson, Arizona 85721, USA}

\author{Xincheng Lin}
\email{xlin28@ncsu.edu}
\affiliation{Department of Physics and Astronomy,
North Carolina State University,
Raleigh, North Carolina 27695, USA}

\author{Ubirajara van Kolck}
\email{vankolck@ijclab.in2p3.fr}
\affiliation{European Centre for Theoretical Studies in Nuclear Physics and
Related Areas (ECT*), Fondazione Bruno Kessler, 38123 Villazzano (TN), Italy}
\affiliation{Université Paris-Saclay, CNRS/IN2P3, IJCLab, 91405 Orsay, France}
\affiliation{Department of Physics,
University of Arizona, Tucson, Arizona 85721, USA}

\author{Sebastian König}
\email{skoenig@ncsu.edu}
\affiliation{Department of Physics and Astronomy,
North Carolina State University,
Raleigh, North Carolina 27695, USA}

\begin{abstract}
 The three- and four-boson systems with a large scattering length and a short
 effective range in the two-body sector are studied in the framework of
 Short-Range Effective Field Theory (SREFT).
 The starting point (leading order) of the EFT is taken to
 be the universal unitarity limit, where the two-body sector is parameter-free
 and only one three-body parameter enters.
 In this limit, physical systems manifest discrete scale invariance.
 Deviations from universality arising from finite scattering-length and
 effective-range corrections, as well as a four-body force required by
 renormalization, are included perturbatively at next-to-leading order.
 The three-body ground state and its associated four-body ground and
 first-excited states are studied using the Faddeev-Yakubovsky (FY) formalism and a
 complementary diagrammatic approach.
 By employing techniques to remove contributions from deep trimers in tetramer
 calculations, we extend our analysis to larger cutoffs than previously
 accessible within the FY approach to SREFT.
 Our results for binding energies and radii of \isotope[4]{He}
 three- and four-atom systems converge well to results obtained with
 sophisticated phenomenological potentials.
 These successes suggest that the physics of \isotope[4]{He} atomic clusters is
 governed by only small deviations from discrete scale invariance.
\end{abstract}

\maketitle

\section{Introduction}
\label{sec:Introduction}

Systems with large scattering lengths and short effective ranges, including
nuclear clusters, halo nuclei, and cold atomic gases, provide a unique
playground for effective field theory (EFT) approaches to strongly interacting
quantum systems~\cite{Braaten:2004rn, Hammer:2019poc}.
Despite the differences in their underlying interactions and energy scales,
these systems exhibit universal properties due to their proximity to a
non-trivial fixed point of the renormalization group (RG) in the two-body
system~\cite{Weinberg:1991um}---the unitarity limit, where the two-body
scattering length diverges and the effective range vanishes.
The unitarity limit serves as a powerful starting point for understanding such
systems, as it strips the complex interactions at short distances down to their
bare essence.
However, any real physical system deviates from the exact
unitarity limit and a more refined description requires corrections, including
those from finite scattering lengths and effective ranges.

Short-Range (or Pionless, in nuclear physics) EFT, which only involves contact
interactions, provides a systematic way to perform an expansion around the
unitarity limit, as first suggested in Refs.~\cite{Konig:2015aka,
Konig:2016utl} for light nuclei.
At leading order (LO), when the two-body system is tuned to the unitarity limit,
scale invariance results.
In contrast to two-component fermions, for bosons or multicomponent fermions
renormalization beyond the two-body system requires the introduction at LO of a
three-body force with a single dimensionful parameter,
$\Lambda_\star$~\cite{Bedaque:1998kg, Bedaque:1998km}.
The resulting RG limit cycle manifests the (anomalous) breaking of continuous
scale invariance down to a discrete scale invariance
(DSI)~\cite{vanKolck:2017jon}.
This leads to the appearance of the Efimov effect~\cite{Efimov:1970zz},
where the three-body spectrum features an infinite geometric sequence of bound
states, known as Efimov trimers.
With any one regularization procedure, different physical three-body systems
near two-body unitarity differ essentially only in the value of $\Lambda_\star$,
which sets the scale for the position of the Efimov tower.

The fact that no additional interaction is needed for renormalization at LO
beyond the three-body system~\cite{Platter:2004he, Bazak:2016wxm} has profound
consequences.
Systems belonging to this universality class have all their properties
determined, in a first approximation, by the single parameter $\Lambda_\star$.
Each Efimov trimer is associated with two tetramer states: one lying just below
the threshold for breakup into an Efimov trimer and a free particle, and the
other approximately 4.6 times more deeply
bound~\cite{Hammer:2006ct,Deltuva:2010xd}.
For simplicity, we refer to them as the four-body excited and ground states,
respectively.
DSI then implies two geometric towers in the four-body spectrum.
Similar tower structures have also been observed in systems with more
particles~\cite{vonStecher:2011zz, Gattobigio:2011ey, Gattobigio:2012tk,
Kievsky:2014dua}.
Even more importantly, the ground-state energies of all systems with more than
three bodies are also determined by $\Lambda_\star$~\cite{vonStecher:2010XX,
Nicholson:2012zp, Kievsky:2014dua, Carlson:2017txq}, that is to say, correlated
with the three-body ground state~\cite{Platter:2004he, Hammer:2006ct,
Bazak:2016wxm}.

Deviations from unitarity start to appear at next-to-leading order (NLO).
At this order, the LO two-body interaction, a contact force with no derivatives
or free parameters, is corrected to account for the large scattering length,
while the finite effective range is introduced through a contact interaction
with two derivatives.
Since they are expected to generate small corrections in observables,
higher-order interactions must be amenable to (distorted-wave) perturbation
theory.
In the three-body system, these corrections induce only a
shift~\cite{Bedaque:1998km, Hammer:2001gh, Bedaque:2002yg, Ji:2011qg} in the
non-derivative three-body force that entered at LO.
A new, two-derivative three-body force enters only at N$^2$LO~\cite{Ji:2012nj}.
In contrast, a four-body force is required for renormalization of the four-body
system at NLO~\cite{Bazak:2018qnu}.
This four-body force needs to be fixed by a four-body datum, thereby introducing
a new, four-body scale.
Together with the two-body scattering length and effective range, this four-body
scale modifies the correlations among the properties of few-body systems.
The impact of these new scales on the correlations between the four-body
excited- and ground-state properties was investigated in Ref.~\cite{Wu:2023mhg}.

The idea of a perturbative expansion around the unitarity limit has been
successfully applied to nuclear few-body systems~\cite{Konig:2015aka,
Konig:2016utl, Konig:2016iny, Konig:2019xxk} within the framework of Pionless
EFT.
The expansion is found to compare well with the observed ground-state binding
energies and radii of the three- and four-nucleon systems, when
scattering-length corrections are included.
However, the NLO four-body force is most naturally determined from the four-body
ground-state energy, and other four-body observables are needed to gauge the
convergence of the theory.
Although the perturbative Faddeev (F)~\cite{Faddeev:1960su} and
Faddeev-Yakubovsky (FY)~\cite{Yakubovskii:1966} formalisms for calculating the
three- and four-body radii have already been developed~\cite{Konig:2019xxk}, a
systematic treatment of effective-range corrections and the four-body force at
NLO in calculations of four-body radii and excited-state energies has remained
underexplored.

In this work, we address this gap by applying Short-Range EFT (SREFT) to cold
\isotope[4]{He} atomic systems, which have been extensively studied with
potential models~\cite{Blume:2000monte, Lazauskas:2006He4, Hiyama:2011ge,
Hiyama:2012cj, Deltuva:2022zqd, Deltuva:2026dra} and EFT approaches~\cite{Bedaque:1998km, Platter:2004he,
Ji:2012nj, Bazak:2016wxm, Bazak:2018qnu, Lin:2023zqw, Contessi:2023yoz}.
The small energy of the \isotope[4]{He} dimer makes \isotope[4]{He} clusters an
ideal ground for investigating the many-body consequences of the proximity to
the two-body unitarity limit.
We determine the three-body ground-state radius, the four-body ground-state
binding energy and radius, and the four-body excited-state binding energy by
solving the F and FY equations as well as the integral equations obtained by a
diagrammatic method~\cite{Brodsky:2006d, Lin:2023zqw}.
At LO, the three-body force, with the only LO parameter $\Lambda_\star$, is
solved for exactly and determined from the \isotope[4]{He} trimer ground-state
energy.
The trimer radius is predicted to NLO without new parameters, providing an
alternative to the existing SREFT prediction away from the unitarity
limit~\cite{Qin:2020hsc}.
Using the tetramer ground-state binding energy as the input to determine the
four-body force, we perform detailed calculations for the four-body ground-state
radius and excited-state binding energy.
By incorporating all the corrections at NLO, we establish a more comprehensive
understanding of the interplay between the scattering length, effective range,
and four-body scale, and their collective impact on the properties of few-body
systems near unitarity.
Our results show that the expansion around the unitarity limit converges well
for both binding energies and radii in the three- and four-boson systems.

For comparison, we also present results where unitarity is not imposed at LO,
which then contains an additional scale given by the two-body scattering length.
In this ``standard'' approach, DSI is explicitly broken already at LO and its
role less transparent.
Preliminary results with and without imposing unitarity at LO have been
presented in meetings.\footnote{These include the talks ``Perturbation theory for four-body systems near unitarity" by S. K\"onig at the Institute for Nuclear Theory (October 21, 2024) and ``Bosonic few-body systems expanded around the unitarity limit" by F. Wu at the European Centre for Theoretical Studies in Nuclear Physics and Related Areas (June 11, 2025).} 
Recently, three- and four-nucleon radii have been calculated in Pionless EFT at
NLO without the unitarity expansion, using either the Stochastic Variational
Method~\cite{Mondal:2025wml} or the F/FY formalism~\cite{Lyu:2025yhz}.

The SREFT interactions are singular and require a regulator, typically implemented as a momentum cutoff of some form, 
for observables to acquire finite values; renormalization subsequently removes regulator dependence. Existing four-body calculations are restricted to relatively small cutoffs,
where deep trimers do not appear.
As pointed out in Ref.~\cite{Lin:2023zqw}, the cutoff dependence is more
complicated at larger cutoffs, necessitating relatively large cutoffs to achieve
converged results.
To extend our analysis to large cutoffs, we introduce techniques to remove
unphysical deep trimers from the three-body spectrum in both FY and diagrammatic
methods.
Without deep-trimer removal, the four-body states would become unstable and one
would have to consider decay channels, significantly complicating calculations.
With the larger values accessible after deep trimer removal, some quantities
show oscillations with the cutoff, which can be seen as a consequence of DSI.
The combined use of two different methods---the F/FY formalism and a
diagrammatic approach---allows us to benchmark calculations of observables
accessible to both methods as well as access observables more amenable to one
method than the other.

This paper is organized as follows.
We describe the F/FY formalism as it applies to SREFT in
Sec.~\ref{sec:MethodsFY} and the diagrammatic approach with auxiliary fields in
Sec.~\ref{sec:Diagrammatic}.
Additional details on solving the NLO bound-state equation are provided in
Appendices~\ref{appen:projection_NLO} and~\ref{app:diagrammatic}, respectively.
The renormalization of SREFT is tackled in Sec.~\ref{sec:renormalization}, with
details of regulator dependence of coupling constants and
observables relegated to Appendix~\ref{app:reg_dep}.
Section~\ref{sec:Results} presents our main results for the \isotope[4]{He}
trimer ground state and tetramer ground and excited states.
The dependence of observables on the unphysical parameter introduced to subtract
the deep trimers is analyzed in Appendix~\ref{app:eta_dep}, while
Appendix~\ref{app_sec:convergence} examines the convergence with the maximum
angular momentum imposed on the numerical calculations.
Conclusions are presented in Sec.~\ref{sec:Conclusion}.

\section{Short-Range EFT in the Faddeev-Yakubovsky Formalism}
\label{sec:MethodsFY}

We investigate bosonic systems near two-body unitarity within the framework of
SREFT, from which one can derive two- and few-body forces and currents.
These are contact interactions, requiring regularization with appropriate
regulators to ensure well-defined solutions.
With the regularized potential, one can solve the F/FY equations to obtain
few-body binding energies, wave functions, and other properties.
By employing a power-counting scheme that guarantees renormalizability order by
order, calculations are organized in a strict distorted-wave perturbative
manner.

We are interested in distance scales much larger than the size of a particle and
the range $R$ of its interactions.
Under these circumstances, we can describe the dynamics in terms of a particle
field $\psi$, here spinless, of mass $m$.
SREFT contains the most general short-range dynamics allowed by assumed
spacetime symmetries (rotation, translation, parity, time reversal, particle
number).
The relevant Lagrangian is
\begin{spliteq}
 \mathcal{L}_{\text{SREFT}} =
 &\psi^{\dagger}
 \left( i \partial_0 + \frac{\vec{\nabla}^2}{2m} \right)
 \psi - \frac{C_0}{2} \left( \psi^{\dagger} \psi\right)^2 \\
 & \null + \frac{C_2}{8}  \left(
  \psi^{\dagger} \psi^{\dagger} \psi \overleftrightarrow{\nabla}^2 \psi
  + {\rm H.c.}\right) \\
 & \null - \frac{D_0}{6} \left( \psi^{\dagger} \psi\right)^3
 - \frac{E_0}{24} \left( \psi^{\dagger} \psi\right)^4 + \cdots \,,
\label{L_SREFT}
\end{spliteq}
where $\overleftrightarrow{\nabla}\equiv
\overrightarrow{\nabla}-\overleftarrow{\nabla}$, $C_{0,2}$, $D_0$, and $E_0$ are
parameters or ``low-energy constants" (LECs), and ``$\cdots$'' stand for
interactions that contribute at orders beyond those considered in this work.

The LECs $C_{0,2}$ account for, respectively, the energy independence of the
two-particle interaction at LO and its correction linear in the energy at
NLO~\cite{vanKolck:1998bw}.
The non-derivative contact three-body operator with LEC $D_0$ is needed at LO to
renormalize the three-body system~\cite{Bedaque:1998kg, Bedaque:1998km}, while
the non-derivative four-body term with LEC $E_0$ is necessary to ensure
renormalizability once the two-derivative two-body term ($C_2$) is included at
NLO~\cite{Bazak:2018qnu}.
For bookkeeping, we note that all LECs are split into different orders,
even though only a single input is needed to overall fix each individual LEC.
For example, we write
\begin{equation}
 D_0 = D_0^{(0)} + D_0^{(1)} + \cdots \,.
\end{equation}
The LO $D_0^{(0)}$ is determined by one datum, and the higher-order terms
$D_0^{(\nu\geq 1)}$ are introduced to absorb the shifts of the input used to fix
$D_0^{(0)}$ due to other terms introduced at $\nu$-th order.
We decompose other LECs similarly.

Because the problem at hand is nonrelativistic, SREFT boils down to standard
quantum mechanics with contact potentials.
As these potentials are singular, they require regularization.
For the F/FY calculations, we consider separable regulators that in
momentum space have the (super-) Gaussian form
\begin{equation}
 g(k^2) = \exp({-}k^{2n}/\Lambda^{2n}) \,,
\end{equation}
where $\Lambda$ has the dimension of momentum and serves as the momentum cutoff,
and $n=1,2,3,\ldots$ correspond to, respectively, standard, quartic,
sextic~\etc (super-) Gaussians.
The exact choice of regulator is rendered arbitrary through
order-by-order renormalization.
The $\Lambda$ dependence of the LECs present at any given order ensures that the
effects of the regulator are no larger than higher-order contributions
originating in short-distance physics.

In the two-body sector, the LO potential in momentum space reads
\begin{equation}
 \braket{\vec{k} | V_2^{(0)} | \vec{k}' }
  = C_0^{(0)} \braket{ \vec{k} | g } \braket{ g | \vec{k}' } \,,
\label{V2_0}
\end{equation}
where $\vec{k}$ ($\vec{k}'$) denotes the initial (final) relative momentum in
the center-of-mass (c.m.) frame, and
\begin{equation}
 \braket{ \vec{k} | g } = g(k^2) \,.
\end{equation}
At NLO, there are corrections,
\begin{equation}
 \braket{ \vec{k}' | V_2^{(1)} | \vec{k} }
 = \left[C_0^{(1)} + \frac{C_2^{(1)}}{2} \left(k^2+{k}'^2\right) \right]
 g(k^2) g(k'^2) \,,
\label{V2_1}
\end{equation}
where $C_2^{(1)}$ is the first non-zero order of the $C_2$ term in
Eq.~\eqref{L_SREFT}.

With these interactions we solve the Lippmann-Schwinger equation for the
two-body $T$ matrix,
\begin{equation}
 t_2 = V_2 + V_2 G_{2} t_2 \,,
\end{equation}
where $G_{2}$ is the free two-body Green's function.
For separable regulators,
the calculation can be done analytically and involves the master integral
\begin{spliteq}
 I_{2n}(k) &\equiv \int\!\!\frac{\dd^3p}{(2\pi)^3}
 \frac{p^{2n} \, g^2(p^2) }{p^2-k^2-i\epsilon} \\
 &= \frac{k^{2n+1}}{4\pi} \left[ i g^2(k^2)
 + \sum_{l=0}^{\infty} \theta_{1+2(n-l)}
 \left(\frac{k}{\Lambda}\right)^{2(l-n)-1} \right] \,,
\label{Ig2n}
\end{spliteq}
where the $\theta_l$'s are regulator-dependent constants.
For (super-) Gaussian regulators,
\begin{equation}
 \theta_l = \frac{2^{-l/2n}}{n\pi} \, \Gamma\!\left(\frac{l}{2n}\right) \,.
\label{theta_nl}
\end{equation}
The two-body LECs $C_0$ and $C_2$ can be fixed by matching the $s$-wave
scattering amplitude derived from SREFT to the effective range expansion (ERE),
\begin{equation}
 k \cot\delta_s(k)= {-}\frac{1}{a_2} + \frac{r_2 }{2} k^2 + \cdots \,,
 \label{ERE}
\end{equation}
in which $k$ is the magnitude of the relative momentum in the c.m. frame,
$\delta_s$ is the $s$-wave phase shift, and $a_2$, $r_2$, $\ldots$, are the
scattering length, the effective range, \etc.
Near the unitarity limit, $a_2 \gg r_2/2 \sim R$ and the binding energy of the
single bound state is $B_2\simeq  1/2ma_2^2$.
For $a_2^{-1} \ll k\ll R^{-1}$, both scattering length and effective range can
be treated as NLO corrections, and the two-body system is tuned to the unitarity
limit at LO.

To go beyond the two-body system, we need to account for few-body interactions
which are tied to our low resolution scale.
In the three- and four-body sectors, we employ Jacobi coordinates defined as
\begin{subalign}
 \vec{u}_1 &= \frac{1}{2} (\vec{p}_1 - \vec{p}_2) \,, \\
 \vec{u}_2 &= \frac{2}{3}
 \left[ \vec{p}_3 - \frac{1}{2} (\vec{p}_1 + \vec{p}_2) \right] \,, \\
 \vec{u}_3 &= \frac{3}{4} \left[ \vec{p}_4 - \frac{1}{3}
 \left( \vec{p}_1 + \vec{p}_2 + \vec{p}_3 \right) \right] \,,
\end{subalign}
where $\vec{p}_i$ is the momentum of the particle $i$.
The regularized three-body potential to NLO is
\begin{equation}
 \braket{ \vec{u}_1 \vec{u}_2 | V_3^{(\nu)} | \vec{u}_1' \vec{u}_2' }
 = D_0^{(\nu)} \braket{ \vec{u}_1 \vec{u}_2 | \xi }
 \braket{ \xi | \vec{u}_1' \vec{u}_2' } \,,
\label{V3}
\end{equation}
in which $\nu = 0, 1$, and the regulator function is
\begin{equation}
 \braket{ \vec{u}_1 \vec{u}_2 | \xi } = g\Big(u_1^2 + \frac{3}{4} u_2^2\Big) \,.
\end{equation}
The four-body potential starts at NLO and has the form
\begin{equation}
 \braket{
  \vec{u}_1 \vec{u}_2 \vec{u}_3 | V_4^{(1)} | \vec{u}_1' \vec{u}_2' \vec{u}_3'
 }
 = E_0^{(1)}\braket{ \vec{u}_1 \vec{u}_2 \vec{u}_3 | \zeta }
 \braket{ \zeta | \vec{u}_1' \vec{u}_2' \vec{u}_3' } \,,
\label{V4}
\end{equation}
where
\begin{equation}
 \braket{ \vec{u}_1 \vec{u}_2 \vec{u}_3 | \zeta }
 = g\Big(u_1^2 + \frac{3}{4} u_2^2 + \frac{2}{3} u_3^2\Big) \,.
\end{equation}

Interchangeability between particles means that the total three-body wave
function can be written with the permutation operator $P\equiv P_{12} P_{23} +
P_{13} P_{23}$, where $P_{ij}$ exchanges particles $i$ and $j$, as
\begin{equation}
 \ket{\Psi_{3}} = (1+P) \ket{\psi_{3}} \,,
\label{Psi_3_tot}
\end{equation}
in terms of a single-component wave function.
The latter obeys the Faddeev equation with the three-body potential,
\begin{equation}
 \ket{\psi_3} = K_3 \ket{\psi_3} \,,
\label{Faddeev_eq}
\end{equation}
where
\begin{equation}
 K_3 = G_{3} t_2 P + \frac{1}{3}\left(1+G_{3} t_2\right) G_{3} V_3 (1+P) \,,
\end{equation}
in which $G_{3}$ denotes the free three-body Green's function.

The Faddeev equation is solved in a partial-wave basis
\begin{equation}
 \ket{u_1 u_2;(l_2 l_1)L} \,,
\end{equation}
where $L$ is the total angular momentum ($L=0$ for the states we are going to
study), and $l_i$ denotes the angular momentum corresponding to the Jacobi
momentum $\vec{u}_i$.
In practical calculations, we impose an angular momentum cutoff, $l_{i} \leq
l_{\text{max}}$, on the partial-wave basis.
After projecting onto this basis and discretizing the magnitudes and relative
angles of the various momenta, the kernel $K_3$ becomes a matrix that depends on the energy $E$.
The eigen-energies of three-body bound states can be found by the condition
\begin{equation}
 \det(1 - K_3({-}B_{3,i})) = 0 \,.
 \label{det3}
\end{equation}
The corresponding wave function can be obtained by singular value decomposition
(SVD) once the binding energy is found.

The extension to the four-body system stems from the realization that two types
of cluster configurations are possible, (3+1) and (2+2), with the latter
implemented through the permutation operator $\tilde{P} \equiv P_{13} P_{24}$.
The total wave function of the four-boson system can then be written as
\begin{equation}
 \ket{\Psi_4} = (1+P) \left[
  (1+P_{34}(1+P)) \ket{\psi_A} + (1+\tilde{P}) \ket{\psi_B}
 \right] \,,
 \label{Psi_4}
\end{equation}
where $\ket{\psi_{A, B}}$ are the two FY components that correspond to,
respectively, the (3+1) and (2+2) cluster
configurations~\cite{Kamada:1992odh}
(see also Ref.~\cite{Gloeckle:1983} for a pedagogical discussion of the FY
components).
With the three- and four-body forces we use, the FY equations are
\begin{align}
 \ket{\psi_A}
 &= G_{4} t_2 P
 \left[ (1+P_{34}) \ket{\psi_A} + \ket{\psi_B} \right] \nonumber \\
  & \null \quad + \frac{1}{3}\left(1+G_{4} t_2\right) G_{4}
  \left( V_3 + \frac{V_4}{4} \right) \ket{\Psi_4} \,, \\
 \ket{\psi_B}
 &= G_{4} t_2 \tilde{P}
 \left[ (1+P_{34}) \ket{\psi_A} + \ket{\psi_B} \right] \,,
\end{align}
where $G_{4}$ denotes the free four-body Green's function.
Alternatively, we can write the FY equations in a matrix form,
\begin{equation}
 \ket{\hat{\psi}} = \hat{K}_4 \ket{\hat{\psi}}\,,
 \label{FY_eq_Mat}
\end{equation}
where $\ket{\hat{\psi}} = ( \ket{\psi_A}, \ket{\psi_B} )^T$ and the kernel
\begin{equation}
 \hat{K}_4 = G_{4} t_2
 \hat{P} + \frac{1}{3}(1+G_{4} t_2) G_{4}
 \left( V_3 + \frac{1}{4} V_4 \right) \hat{P}_3 \,,
\label{kernel4}
\end{equation}
in which
\begin{align}
 \hat{P} &= \begin{pmatrix}
  P & 0\\0 & \tilde{P}
 \end{pmatrix}
 \begin{pmatrix}
  1+P_{34} & 1\\1+P_{34} &1
 \end{pmatrix}\,,\\
  \hat{P}_3 &= (1+P) \begin{pmatrix}
   1+ P_{34}(1+P) & 1+\tilde{P}\\0&0
  \end{pmatrix} \,.
\end{align}

Similar to Refs.~\cite{Kamada:1992odh, Gloeckle:1993vr, Platter:2004he,
Hadizadeh:2006bs, Hadizadeh:2011ke, Konig:2019xxk}, we solve the FY equations,
too, in a partial-wave momentum basis.
Corresponding to the (3+1) and (2+2) configurations, denoted by $\alpha$ and
$\beta$ respectively, there are two distinct bases,
\begin{align}
 \ket{u_1 u_2 u_3; \alpha }
 &= \ket{ u_1 u_2 u_3; ((l_2 l_1)l_{12} l_3)L } \,,
 \label{bas_a} \\
 \ket{v_1 v_2 v_3; \beta }
 &= \ket{ v_1 v_2 v_3; ((\lambda_2 \lambda_1) \lambda_{12}\lambda_3)L} \,,
 \label{bas_b}
\end{align}
where $l_{12}$ ($\lambda_{12}$) is the coupled angular momentum of $l_1$
($\lambda_1$) and $l_2$ ($\lambda_2$), and $\lambda_i$ represents the angular
momentum associated with another set of Jacobi momenta $\vec{v}_i$, defined as
\begin{subalign}
 \vec{v}_1 &= \frac{1}{2}(\vec{p}_1 -\vec{p}_2) \,, \\
 \vec{v}_2 &= \frac{1}{2}(\vec{p}_1 +\vec{p}_2)
 - \frac{1}{2}(\vec{p}_3 +\vec{p}_4)\,,\\
 \vec{v}_3 &= \frac{1}{2}(\vec{p}_3 -\vec{p}_4) \,.
\end{subalign}
After projecting onto these bases truncated with $l_{i}, \lambda_i \leq l_{\text{max}}$
and discretizing the magnitudes and relative angles of the various momenta, the
kernel $\hat{K}_4$ also becomes a matrix that depends
on the energy $E$.
The binding energies of the four-body system satisfy
\begin{equation}
 \det(1 - \hat{K}_4({-}B_{4,i})) = 0 \,,
 \label{det4}
\end{equation}
and the corresponding wave function is obtained by SVD.

To avoid the Wigner bound~\cite{Wigner:1955zz, Phillips:1996ae} and ensure
renormalizability, the two-body Lippmann-Schwinger equation, the three-body
Faddeev equation, and the four-body FY equations should be all solved in
distorted-wave perturbation theory, as implemented in Ref.~\cite{Konig:2019xxk}.

\subsection{LO}

By solving the Lippmann-Schwinger equation at LO and imposing it satisfies the unitarity limit, one obtains the LO two-body $T$ matrix
\begin{equation}
 t_2^{(0)}(E;\vec{k},\vec{k}') = g(k^2) \, \tau_2^{(0)}(E) \, g(k'^2) \,,
\label{t0z}
\end{equation}
where
\begin{align}
 \tau_2^{(0)}(E)
 &= \left(
  1/C_0^{(0)} - \braket{ g | G_{2}(E) | g}
 \right)^{{-}1} \nonumber \\
 &= \left( 1/C_0^{(0)} + m I_{0}(\sqrt{m E}) \right)^{{-}1} \label{eq:tau0}\\
 &= \frac{4\pi}{m \sqrt{mE}}
 \left(
  \ii g^2(mE) + \theta_{{-}1} \frac{\sqrt{mE}}{\Lambda} + \cdots
 \right)^{{-}1} \nonumber \,,
\end{align}
for
\begin{equation}
 C_0^{(0)} = -\frac{4\pi}{m \theta_1 \Lambda} \,.
\label{C0_0}
\end{equation}
The $\Lambda^{{-}1}$ term in Eq.~\eqref{eq:tau0} is no larger than NLO
corrections for $\Lambda\simge R^{{-}1}$ and ``$\cdots$'' stand for terms that
are smaller still.
There is no dimensionful parameter at LO, reflecting the universality associated
with (continuous) scale invariance.

To obtain a renormalized solution of the three-body system, the LO three-body
potential~\eqref{V3} must be included, and the kernel of the Faddeev equation is
\begin{equation}
 K_3^{(0)} = G_{3} t_2^{(0)} P + \frac{1}{3}(1+G_{3} t_2^{(0)}) G_{3} V_3^{(0)} (1+P) \,.
\end{equation}
Once the three-body LEC $D_0^{(0)}(\Lambda)$ is fixed by a three-body datum,
which we choose to be the three-body ground-state binding energy $B_{3,0}$,
other trimers can be obtained by finding the roots of Eq.~\eqref{det3}.
The binding energies of these trimers form an approximate, semi-geometric series
with ratio $\simeq 515$, corresponding to a truncated Efimov tower.
As the cutoff $\Lambda$ increases, deeper trimers with larger binding energies
appear.
The LO trimer wave functions $\ket{\Psi_{3,i}^{(0)}}$, where $i$ labels the different
trimer states, are obtained by solving Eq.~\eqref{Faddeev_eq} with the LO kernel
$K_3^{(0)}$ and substituting the resulting solution $\ket{\psi_{3,i}^{(0)}}$ into
Eq.~\eqref{Psi_3_tot}.

In the four-body problem, since the four-body force does not enter at LO, the kernel of the FY equations simplifies to
\begin{equation}
 \hat{K}_4^{(0)} = G_{4} t_2^{(0)} \hat{P}
 + \frac{1}{3}(1+G_{4} t_2^{(0)}) G_{4} V_3^{(0)} \hat{P}_3 \,.
\label{kernel4}
\end{equation}
Once $\ket{\hat{\psi}^{(0)}} = ( \ket{\psi_A^{(0)}}, \ket{\psi_B^{(0)}})^T$ is
obtained by solving the FY equations in the partial-wave basis,
Eqs.~\eqref{bas_a} and~\eqref{bas_b}, the LO four-body wave function can be
calculated from Eq.~\eqref{Psi_4}.

The LO calculations in the F/FY formalism were first carried out in
Refs.~\cite{Platter:2004he,Hammer:2006ct}, but were limited to cutoffs below the
first deep-trimer threshold.
To remove the deep trimers from the spectrum, we extend the pseudopotential
projection method, which is often used to exclude spurious two-body bound
states~\cite{Lehman:1982zz, Nogga:2005hy, Song:2016ale}, to three-body systems.
With this technique, the LO three-body potential entering $\hat{K}_4^{(0)}$ is modified according to $V_3^{(0)} \rightarrow V_3^{(0)} + V_{P,3}$,
with
\begin{equation}
 V_{P,3}=\eta \sum_{i=1}^{N_3} B_{3,{-}i}^{(0)}
 \ket{\Psi_{3,-i}^{(0)}} \bra{\Psi_{3,{-}i}^{(0)}} \,,
\label{pseudopotential}
\end{equation}
where $\eta$ is a large, positive number and $i$ runs through the $N_3$ deep
trimers to be removed, counted from the ground state ($i=0$).
As $\eta$ goes to infinity, the deep trimers are
subtracted. In our calculation, we use a finite but sufficiently large $\eta$.
A similar method based on the three-body scattering matrix, where one
can take the limit $\eta \rightarrow \infty$, was sketched in
Ref.~\cite{Platter:thesis}.

\subsection{NLO}

In the two-body sector, we introduce $C_0^{(1)}$ and $C_2^{(1)}$ at NLO to
reproduce the scattering length and effective range. The corresponding LECs are
\begin{align}
 \frac{C_2^{(1)}}{C_0^{(0)}} &= \frac{mC_0^{(0)}}{4\pi} \left(
  \frac{r_2}{2} + \frac{\theta_{-1}}{\Lambda}
 \right) \,,
\label{C2_1} \\
 \frac{C_0^{(1)}}{C_0^{(0)}} &= \frac{m C_0^{(0)}}{4\pi} \left(
  \theta_3 \Lambda^3 \frac{C_2^{(1)}}{C_0^{(0)}} - \frac{1}{a_2}
 \right) \,.
\label{C0_1}
\end{align}
The correction to the two-body $T$ matrix is
\begin{equation}
 t_2^{(1)}(E;\vec{k},\vec{k}') = g(k^2) \, \tau_2^{(1)}(E; k,k') \, g(k'^2) \,,
\label{t1z}
\end{equation}
where
\begin{align}
  \tau_2^{(1)}(E;k,k') = & \left[
   \frac{C_0^{(1)}}{C_0^{(0)2}}
   - \frac{C_2^{(1)}}{C_0^{(0)}} \, m I_{2}(\sqrt{m E})
  \right]\tau_2^{(0)2}(E) \nonumber\\
  & + \frac{C_2^{(1)}}{C_0^{(0)}} \frac{k^2+k'^2}{2}\tau_2^{(0)}(E) \nonumber\\
  =& \frac{m}{4\pi} \left\{ {-}\frac{1}{a_2}
  + \left(\frac{r_2}{2} + \frac{\theta_{{-}1}}{\Lambda} \right) \right. \nonumber\\ 
  &\times \left[
    \frac{k^2+k'^2}{2} + \frac{4\pi}{\theta_1 \Lambda} I_{0}(\sqrt{mE}) \right. \nonumber\\
    & \times \left. \left.\left( mE - \frac{k^2+k'^2}{2} \right)
  \right] \right\} \tau_2^{(0)2}(E) \,.
\end{align}
When on-shell ($k=k'=\sqrt{mE}$),
\begin{equation}
 \tau_2^{(1)}(E;k,k) = \frac{m}{4\pi} \left[
  {-}\frac{1}{a_2} + \left(\frac{r_2}{2}
  + \frac{\theta_{{-}1}}{\Lambda} \right)k^2
 \right] \tau_2^{(0)2}(E) \,.
\end{equation}
The $\Lambda^{{-}1}$ term here cancels, up to terms of N$^2$LO, the
$\Lambda^{{-}1}$ term in Eq.~\eqref{eq:tau0}.
We are left with corrections representing short-range physics in the form of
$r_2$ and long-range physics in the form of $a_2$.
In the standard approach to SREFT, the latter contributions are resummed and
included among the LO terms.
While this can be done with no harm to renormalization, it adds the scale
$a_2^{{-}1}$ at LO that obscures the role of DSI.

These corrections in the two-body sector would modify the three-body
ground-state binding energy that is used to fix $D_0^{(0)}$.
To retain the three-body ground state at the right position, the NLO energy
shift of this state, denoted by $B_{3,0}^{(1)}$, should vanish.
As mentioned already, the NLO three-body LEC $D_0^{(1)}(\Lambda)$ is
therefore fixed by the condition
\begin{equation}
 B_{3,0}^{(1)} = {-}\braket{
  \Psi_{3,0}^{(0)} | 3 V_2^{(1)} + V_3^{(1)} | \Psi_{3,0}^{(0)}
 } = 0 \,,
 \label{B30_1}
\end{equation}
in which $V_2^{(1)}$ represents the two-body interaction between any two
particles, and the factor 3 arises from the symmetry of the three-body wave
function under the exchange of particles.

Likewise, there is a shift $B_{4,0}^{(1)}$ in the four-body ground-state binding
energy.
This shift depends sensitively on the cutoff $\Lambda$, unless the four-body
potential~\eqref{V4} is included, in which case
\begin{equation}
 B_{4,0}^{(1)} = {-}\braket{
  \Psi_{4,0}^{(0)} | 6V_2^{(1)} + 4V_3^{(1)}+ V_4^{(1)}| \Psi_{4,0}^{(0)}
 } \,,
 \label{B40_1}
\end{equation}
where $V_2^{(1)}$ is the same as that in Eq.~\eqref{B30_1}, $V_3^{(1)}$
represents the three-body interaction among any three particles, and the factors
6 and 4 arise from the symmetry of the four-body wave function under exchange of
particles.
The corresponding four-body LEC $E_0^{(1)}(\Lambda)$ is determined by the
four-body ground-state binding energy
\begin{equation}
 B_{4,0} = B_{4,0}^{(0)} + B_{4,0}^{(1)} \,.
\end{equation}
It should be noted that the relevant matrix element
$\braket{\zeta|\Psi_4^{(0)}}$ is a function of $\Lambda$ and can vanish at
certain cutoff values.
Consequently, the four-body LEC can develop poles at these cutoffs.

To obtain other observables at NLO, such as radii, we need the NLO wave
functions.
The F/FY framework for a perturbative treatment of subleading corrections was
developed in Ref.~\cite{Konig:2019xxk} and applied to two-body scattering-length
corrections.
We generalize this approach to the complete NLO.

The NLO Faddeev equation for a state associated with an index $i$
reads
\begin{equation}
 (1-K_{3,i}^{(0)}) \ket{\psi_3^{(1)}} = K_{3,i}^{(1)} \ket{\psi_3^{(0)}} \,,
 \label{F_NLO}
\end{equation}
in which
\begin{equation}
 K_{3,i}^{(1)} = \left(1 + G_{3} t_2^{(0)}\right) G_{3}
  \left[
  V_2^{(1)} (1+P) +V_3^{(1)} + B_{3,i}^{(1)}
 \right] \,.
\label{K3_1}
\end{equation}
The above equation should be evaluated at $E = {-}B_{3,i}^{(0)}$, where
$1-K_{3,i}^{(0)}$ is singular.
$B_{3,i}^{(1)}$ is the NLO energy shift, and can be calculated in
the same manner as $B_{3,0}^{(1)}$ in Eq.~\eqref{B30_1}, but using the LO wave
function of the corresponding state, once the NLO LECs have been fixed.
Similarly, at NLO the FY equation becomes
\begin{equation}
 (1-\hat{K}_{4,i}^{(0)}) \ket{\hat{\psi}^{(1)}} = \hat{K}_{4,i}^{(1)} \ket{\hat{\psi}^{(0)}} \,,
 \label{FY_NLO}
\end{equation}
where
 \begin{align}
  \hat{K}_{4,i}^{(1)} = & (1+G_{4} t_2^{(0)})G_{4} \left[
   B_{4,i}^{(1)}  +\frac{1}{3} \left(V_3^{(1)}
   + \frac{1}{4}V_4^{(1)}\right) \hat{P}_3
 \right] \nonumber\\
 &+ G_{4} t_2^{(1)} \left[
  \hat{P}+\frac{1}{3} G_{4} V_3^{(0)} \hat{P}_3
 \right] \,.
\label{K1}
\end{align}
In the above equation, $G_{4}$, $t_2^{(0)}$, and $t_2^{(1)}$ should be
evaluated at $E= {-}B_{4,i}^{(0)}$, where $1-\hat{K}_{4,i}^{(0)}$ is singular.
$B_{4,i}^{(1)}$ denotes the NLO energy shift and can be calculated
in the same manner as $B_{4,0}^{(1)}$ in Eq.~\eqref{B40_1}.
The NLO three- and four-body wave functions can be obtained by the projection
method described in Appendix~\ref{appen:projection_NLO}.

\section{Diagrammatic Approach with Auxiliary Fields}
\label{sec:Diagrammatic}

Besides the tetramer ground state, the four-body excited state is also
interesting and relevant to study.
This can be done, in principle, within the FY formalism discussed in the
previous section.
However, as this state involves vastly different distance scales, such
calculations become increasingly resource-intensive due to the need for momentum
interpolations after discretization, which significantly increase the
dimensionality of the numerical problem.
Therefore, we consider in addition a diagrammatic
approach to solving the four-body problem~\cite{Brodsky:2006d, Lin:2023zqw},
which avoids momentum interpolations.
This not only makes it possible to study the excited tetramer state (albeit at
present without access to the radius), but also to
cross-check results for the ground state and to extend the analysis to even
larger cutoffs.

For this approach, we use the SREFT Lagrangian in the auxiliary-field formalism
(see, \eg, Refs.~\cite{Kaplan:1996nv,Bedaque:1998kg,Bedaque:2002yg}), and
employ a sharp momentum cutoff $\Lambda$ as our regulator.
Up to and including
NLO, the Lagrangian in this formulation can be written as
\begin{spliteq}
\label{L-SREFT-AF}
 \mathcal{L}^{\text{AF}}_{\text{SREFT}}
 &= \psi^{\dagger}\left(
  \ii \partial_0 + \frac{\vec{\nabla}^2}{2m}
 \right) \psi \\
 &\quad\null + d^\dagger\left[
  \Delta - c_d\left( \ii\partial_0 + \frac{\nabla^2}{4m} \right)
 \right] d + t^\dagger \Omega t \\
 &\quad\null - \sqrt{\frac{2\pi}{m}}\left(d^\dagger\psi\psi + \text{H.c.}\right) \\
 &\quad\null - \sqrt{\frac{4\pi}{m}}\left(t^\dagger d\psi + \text{H.c.}\right)
 + g_4\psi^\dagger t^\dagger t\psi +\ldots \,,
\end{spliteq}
where $d$ and $t$ are dimer and trimer auxiliary fields, respectively, and
``\ldots'' stand for terms that only appear beyond NLO.
The LECs in Eq.~\eqref{L-SREFT-AF} can be matched order by order to those in
Eq.~\eqref{L_SREFT} by integrating out the auxiliary fields and redefining the
field $\psi$~\cite{Kaplan:1996nv,Bedaque:1998kg,Bedaque:2002yg}.\footnote{
Compared to Refs.~\cite{Kaplan:1996nv,Bedaque:1998kg,Bedaque:2002yg}, the dimer
and trimer auxiliary fields in Eq.~\eqref{L-SREFT-AF} are rescaled to eliminate
redundant terms.
}

\begin{figure}[t]
\centering
\includegraphics[width=0.95\linewidth]{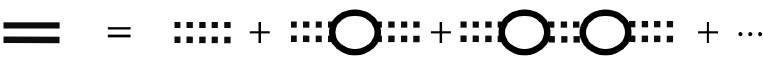}
\caption{Dressed dimer propagator (double solid line) as a sum of bare dimer
 propagators (double dotted lines) with insertions of loops with boson
 propagators (single solid lines).
\label{fig:dressed-dimer}
}
\end{figure}

The matching of two-body LECs up to NLO yields
\begin{align}
 \Delta^{(0)} &= \frac{4\pi}{mC_0^{(0)}} \,,
\label{Delta_0_match} \\
 \Delta^{(1)} &= -\Delta^{(0)}\frac{C_0^{(1)}}{C_0^{(0)}}\,,
\label{Delta_1_match} \\
 c^{(1)}_d &= 2m\Delta^{(0)}\frac{C_2^{(1)}}{C_0^{(0)}} \,.
\label{cd_1_match}
\end{align}
These LECs can be used to obtain the dressed dimer propagator, shown in
Fig.~\ref{fig:dressed-dimer}, as a sum of bare dimer propagators with insertions
of two-boson loop diagrams.
The boson and bare dimer propagators are the inverse of the corresponding
kinetic terms in Eq.~\eqref{L-SREFT-AF}.
Summing over the diagrams on the right-hand side of Fig.~\ref{fig:dressed-dimer}
yields the dressed dimer propagator.
Around unitarity, in the limit $\Lambda\to\infty$, it is
\begin{spliteq}
 \label{dressed-dimer-prop}
 \ii D_d(p_0, \vec{p})
 &= \frac{\ii}{ {-}\sqrt{{\vec p}^{2}/4 - mp_0 - \ii\epsilon}} \\
 &\quad\null \times \Bigg(
  1 + \frac{1}{a_2 \sqrt{{\vec p}^{2}/4 - mp_0 -\ii\epsilon}} \\
 &\quad\hspace{2.5em}\null + \frac{r_2}{2}\sqrt{{\vec p}^{2}/4 - mp_0} + \cdots
 \Bigg) \,,
\end{spliteq}
in which $p_0$ and $\vec{p}$ are the energy and momentum of the dimer.

%%%%%%%%%%%%%%%%%%%%%%%%%%%%%%%%%%%%%%%%%%%%%%%%%%%%%%%%%%%%%%%%%%%%%%%%%%%%%%%%
%%%%%%%%%%%%%%%%%%%%%%%%%%%%%%%%%%%%%%%%%%%%%%%%%%%%%%%%%%%%%%%%%%%%%%%%%%%%%%%%

The matching of three- and four-body LECs up to NLO  yields
\begin{align}
 \Omega^{(0)}
 &= \frac{3C_0^{(0)2}}{D_0^{(0)}} \,,
\label{Omega_0_match} \\
 \Omega^{(1)}
 &= \Omega^{(0)}\left(
  \frac{2C_0^{(1)}}{C_0^{(0)}} - \frac{D_0^{(1)}}{D_0^{(0)}}
 \right) \,,
\label{Omega_1_match} \\
 g^{(1)}_4 &= {-}\frac{\Omega^{(0)2}}{12C_0^{(0)2}} \, E^{(1)}_0 \,.
\label{g_1_match}
\end{align}
In practice, $\Omega$ and $g_4$ can be determined by fitting to trimer and
tetramer binding energies, respectively.
A detailed discussion of the three-body system and three-body force using the
diagrammatic approach can be found in Ref.~\cite{Bedaque:1998km}.
In this work, we will study both the three- and four-body forces, with
particular emphasis on the running of the LECs.

%%%%%%%%%%%%%%%%%%%%%%%%%%%%%%%%%%%%%%%%%%%%%%%%%%%%%%%%%%%%%%%%%%%%%%%%%%%%%%%%
\begin{figure*}[tb!]
\centering \includegraphics[width=0.7\linewidth]{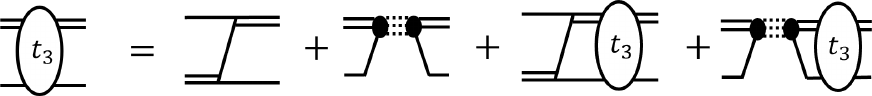}
\caption{Diagrammatic representation of the integral equation for the
 boson-dimer amplitude $t_3$.
 The bare trimer propagator is represented by a triple dotted line. Other lines
 as in Fig.~\ref{fig:dressed-dimer}.
 \label{fig:t3body}
}
\end{figure*}
%%%%%%%%%%%%%%%%%%%%%%%%%%%%%%%%%%%%%%%%%%%%%%%%%%%%%%%%%%%%%%%%%%%%%%%%%%%%%%%%

%%%%%%%%%%%%%%%%%%%%%%%%%%%%%%%%%%%%%%%%%%%%%%%%%%%%%%%%%%%%%%%%%%%%%%%%%%%%%%%%
\begin{figure*}[tb!]
\centering
\includegraphics[width=0.8\linewidth]{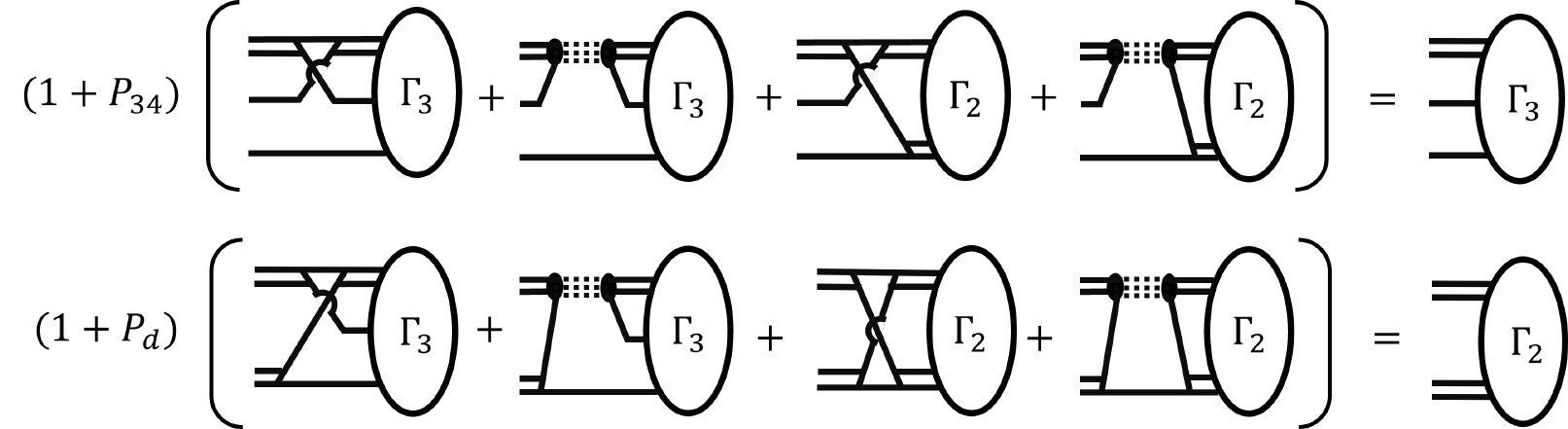}
\caption{ Diagrammatic representation of the four-body integral equations.
 Lines as in Figs.~\ref{fig:dressed-dimer} and~\ref{fig:t3body}.
 \label{fig:4Bdiagram1}
}
\end{figure*}
%%%%%%%%%%%%%%%%%%%%%%%%%%%%%%%%%%%%%%%%%%%%%%%%%%%%%%%%%%%%%%%%%%%%%%%%%%%%%%%%

%%%%%%%%%%%%%%%%%%%%%%%%%%%%%%%%%%%%%%%%%%%%%%%%%%%%%%%%%%%%%%%%%%%%%%%%%%%%%%%%
\begin{figure}[tb!]
\centering
\hspace*{-0.05\linewidth}
\includegraphics[width=1.0\linewidth]{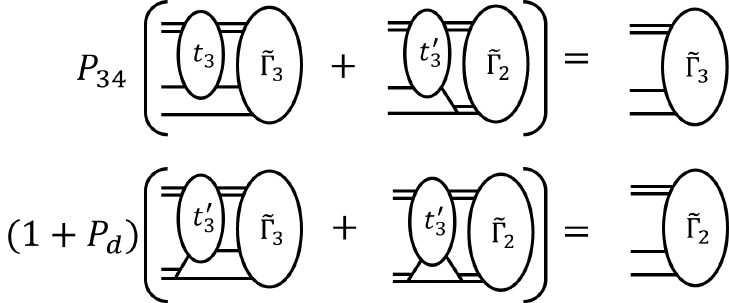}
\caption{ Diagrammatic representation of the four-body integral equations in
 terms of three-body amplitudes $t_3$ and $t_3'$.
 Both single-boson legs attached to $t_3$ are on-shell, whereas for $t_3'$ only
 the single-boson leg on the  side of the (3+1) fragmentation is
 on-shell~\cite{Lin:2023zqw}.
 Lines as in Figs.~\ref{fig:dressed-dimer} and~\ref{fig:t3body}.
 \label{fig:4Bdiagram2}
}
\end{figure}
%%%%%%%%%%%%%%%%%%%%%%%%%%%%%%%%%%%%%%%%%%%%%%%%%%%%%%%%%%%%%%%%%%%%%%%%%%%%%%%%

\subsection{LO}

At LO we must resum all diagrams involving not only the unitarity term in the
dimer propagator, Eq.~\eqref{dressed-dimer-prop}, but also the boson-dimer
coupling to the trimer field and the trimer propagator (given by the LEC
$\Omega^{(0)}$)~\cite{Bedaque:2002yg}.
The diagrammatic representation of the boson-dimer amplitude $t_3$ is shown in
Fig.~\ref{fig:t3body}.
It results~\cite{Bedaque:1998kg, Bedaque:1998km} in the
Skorniakov--Ter-Martirosian (STM) equation~\cite{Skorniakov:1957kgi}.
The only dimensionful parameter in this amplitude is $\Omega^{(0)}$, which is
determined by a three-body energy, such as the ground state's $B_{3,0}$.

The absence of a four-body force at LO means the four-body system can be
described in terms of two- and three-body auxiliary fields.
The four-body equations in the diagrammatic approach~\cite{Lin:2023zqw} can be
written in terms of two four-body vertex functions: $\Gamma_2$ with two outgoing
dimers, and $\Gamma_3$ with one outgoing dimer and two single particles.
The diagrammatic representation of the homogeneous four-body integral equation
is shown in Fig.~\ref{fig:4Bdiagram1}, where $P_{34}$ permutes the bottom two
single-boson lines and $P_{d}$ permutes the two dressed dimers.
The operator form of this equation can be written as
\begin{equation}
 \label{4bodydiageqn-opform}
 \begin{pmatrix}
  1+P_{34} & 0\\
  0&  1+P_{d}
 \end{pmatrix}
 \begin{pmatrix}
  K_{33} & K_{32}\\
  K_{23}&  K_{22}
 \end{pmatrix}
 \begin{pmatrix}
  \ket{\Gamma_3} \\
  \ket{\Gamma_2}
 \end{pmatrix}
 =
 \begin{pmatrix}
  \ket{\Gamma_3} \\
  \ket{\Gamma_2}
 \end{pmatrix}\,,
\end{equation}
where the subscripts ``3'' or ``2'' of the kernel $K_{ab}$ indicate (3+1) or
(2+2) fragmentation, respectively.

In order to explicitly subtract the deep trimer states that turn the tetramers
into resonances at large cutoffs, the homogeneous four-body integral equation
can be brought into a form where the kernel explicitly involves
three-body amplitudes, as shown in Fig.~\ref{fig:4Bdiagram2}.
In operator form~\cite{Lin:2023zqw},
\begin{equation}
\label{4bodydiageqn2-opform}
 \begin{pmatrix}
  P_{34} & 0 \\
       0 & 1+P_{d}
 \end{pmatrix}
 \begin{pmatrix}
  \widetilde{K}_{33} & \widetilde{K}_{32} \\
  \widetilde{K}_{23} & \widetilde{K}_{22}
 \end{pmatrix}
 \begin{pmatrix}
  \ket{\widetilde{\Gamma}_3} \\
  \ket{\widetilde{\Gamma}_2}
 \end{pmatrix}
 = \begin{pmatrix}
  \ket{\widetilde{\Gamma}_3} \\
  \ket{\widetilde{\Gamma}_2}
 \end{pmatrix}\,,
\end{equation}
which is related to Eq.~\eqref{4bodydiageqn-opform} through (see
Appendix~\ref{app:diagrammatic})
\begin{equation}
 \begin{pmatrix}
  \ket{\widetilde{\Gamma}_3} \\
  \ket{\widetilde{\Gamma}_2}
 \end{pmatrix}
 = \begin{pmatrix}
  1-K_{33} & -K_{32} \\
         0 &  1
 \end{pmatrix}
 \begin{pmatrix}
  \ket{\Gamma_3} \\
  \ket{\Gamma_2}
 \end{pmatrix} \,,
\end{equation}
and
\begin{subalign}
 \widetilde{K}_{33} &= {K}_{33}(1-{K}_{33})^{-1} \,, \label{4bodydiageqn2-kerneldef-1}
\\
 \widetilde{K}_{23} &= {K}_{23}(1-{K}_{33})^{-1} \,, \\
 \widetilde{K}_{32} &= (1-{K}_{33})^{-1}{K}_{32} \,, \\
 \widetilde{K}_{22} &= {K}_{22} + {K}_{23}(1-{K}_{33})^{-1}{K}_{32}  \label{4bodydiageqn2-kerneldef-4}\,.
\end{subalign}

Using Eq.~\eqref{4bodydiageqn2-opform}, all the deep trimer states enter as
singularities of $(1-{K}_{33})^{{-}1}$, which can be isolated and addressed
using the Cauchy principal-value prescription (see
Appendix~\ref{app:diagrammatic}).
In addition, Eq.~\eqref{4bodydiageqn2-opform} can be simplified by substituting
$\ket{\widetilde{\Gamma}_3}$ in the equation for $\ket{\widetilde{\Gamma}_2}$,
yielding
\begin{spliteq}
\label{4bodydiageqn3-opform}
\ket{\widetilde{\Gamma}_2}&=(1+P_d) \\
&\quad\times\left[\widetilde{K}_{23}\left(1 - P_{34}\widetilde{K}_{33}\right)^{{-}1}
  P_{34}\widetilde{K}_{32}
  + \widetilde{K}_{22}\right]\ket{\widetilde{\Gamma}_2}
  \,,
\end{spliteq}
which is used to find the tetramer binding energies as long as
$(1 - P_{34}\widetilde{K}_{33})$ is invertible.
The diagrammatic results in this paper are obtained using
Eq.~\eqref{4bodydiageqn3-opform} for numerical convenience.

Tetramer binding energies can be obtained by solving Eq.~\eqref{det4} with the
FY kernel replaced by the diagrammatic one in Eq.~\eqref{4bodydiageqn2-opform}.
To do so, one needs to choose a set of bases to evaluate the matrix elements of
the operators therein.
In the diagrammatic approach with the dimer auxiliary field, the (2+2)
fragmentation is effectively a two-body system made of two dimers and, for total
angular momentum equal to zero, can be described with one scalar momentum and
one energy variable.
In practice, the energy is integrated from ${-}\ii\infty$ to $\ii\infty$ to
avoid the two dimer branch cuts~\cite{Brodsky:2006d}.
Analogously, the (3+1) fragmentation is effectively a three-body system made of
one dimer and two single bosons and, for total angular momentum equal to zero,
can be described with one angular momentum and two scalar
momenta~\cite{Lin:2023zqw,Brodsky:2006d}.
This angular momentum corresponds to that between the dimer and the single-boson
state in the three-body subsystem as well as that  between the three-body
subsystem and the last boson.
The explicit expressions for Eq.~\eqref{4bodydiageqn-opform} in momentum space
with single-particle coordinates are given in
Refs.~\cite{Brodsky:2006d,Lin:2023zqw}.
Those for Eq.~\eqref{4bodydiageqn2-opform} are given in Ref.~\cite{Lin:2023zqw},
where the matrix element of $(1-{K}_{33})^{{-}1}$ is evaluated using Jacobi
momentum of the corresponding three-body subsystem.
This ensures that the locations of the trimer poles transform properly when
boosting the three-body subsystem.
Following Ref.~\cite{Lin:2023zqw}, a sharp momentum cutoff $\Lambda$ and an
angular momentum cutoff $l_{\text{max}}$ are used.

\subsection{NLO}
\label{sec:diagrammatic-NLO}

At NLO, the finite two-body scattering length and effective range lead to
corrections to the dressed dimer propagator in Eq.~\eqref{dressed-dimer-prop}.
These corrections require the shift $\Omega^{(1)}$ in the three-body parameter
and a four-body force corresponding to the $g_4^{(1)}$ term in
Eq.~\eqref{L-SREFT-AF}

Determining the corrections to trimer or tetramer binding energies in the
diagrammatic approach is slightly more complicated than in the F/FY equations,
as the solution to the diagrammatic equation does not directly yield the
bound-state wave functions.
Instead, we obtain the corrections to trimer and tetramer binding
energies by using directly the integral equations.

We first symmetrize the matrix form of the integral equation.
In the four-body case, we multiply each side of Eq.~\eqref{4bodydiageqn3-opform}
by two dimer propagators for the (2+2) configuration and the weights (\ie, the
discretized volume element) for the momentum or energy integration variables.
Denoting this matrix by $M$, the homogeneous four-body integral equation has the
matrix form of a generalized eigenvalue problem,
\begin{equation}
 M\widetilde{K}\ket{\widetilde{\Gamma}_2^{i}}
 = \lambda_i M\ket{\widetilde{\Gamma}_2^{i}} \,,
\end{equation}
where $\lambda_i$'s are the eigenvalues and $\ket{\widetilde{\Gamma}_2}$ with
$\lambda_0=1$ corresponds to a bound state.
$\widetilde{K}$ is the shorthand for the kernel in
Eq.~\eqref{4bodydiageqn3-opform}, which is real (complex) if the deep trimers
are subtracted from (included in) the four-body system.
$M$ and $M\widetilde{K}$ are both real and symmetric matrices.
The first-order perturbative correction to $\lambda_i$ is
\begin{equation}
 \label{eigenvalue_variation}
 \delta\lambda_i
 = \frac{\braket{
 \widetilde{\Gamma}_2^{i}|M(\delta\widetilde{K})
  + (\delta M)\widetilde{K}|\widetilde{\Gamma}_2^{i}
 } - \lambda_i\braket{
  \widetilde{\Gamma}_2^{i}|\delta M|\widetilde{\Gamma}_2^{i}
 }}{\braket{\widetilde{\Gamma}_2^{i}|M|\widetilde{\Gamma}_2^{i}}} \,,
\end{equation}
where $\delta\widetilde{K}$ and $\delta M$ include variations from various
sources.
To obtain the NLO correction to the tetramer ground-state binding energy, one
can fix $\delta\lambda_0 = 0$ at NLO,
\begin{spliteq}
\label{4body-lambda-pert}
 0 &= \delta\lambda_0 \\
 &=  \left[
  \sum_i \delta \xi_i\left(\frac{\delta \lambda_0}{\delta \xi_i}\right)
 - B^{(1)}_{4,0}\left(\frac{\delta \lambda_0}{\delta E}\right)
 \right]_{ E = {-}B^{(0)}_{4,0}} \,,
\end{spliteq}
where $E$ is the four-body energy and $\xi_i$ in this case consists of four
variables: the finite two-body scattering length, the two-body effective range,
the NLO correction to the LO three-body force, and the four-body force.
The first three of these do not require new diagrams.
In particular, corrections from the scattering length and effective range enter
$M$ and $\widetilde{K}$ through corrections to the dressed dimer propagator.
The NLO correction to the three-body force enters $\widetilde{K}$ through the
same diagrams that entered Eq.~\eqref{4bodydiageqn-opform}.
When the LEC $\Omega^{(1)}$ is adjusted to keep the trimer ground-state energy
fixed, other trimer energies change.
Contributions from these three sources to $\delta M$ and $\delta\widetilde{K}$
are obtained using the chain rule, and their impact on the subtraction of the
deep trimer(s) is discussed in Appendix~\ref{app:diagrammatic}.\footnote{ The
explicit NLO expressions are not provided here because, in practice, the
corrections to the diagrams are obtained by programmatically automating
the chain rule, as opposed to keeping track of the perturbations to every
diagram manually.}
Corrections from the four-body force require new diagrams and will be discussed
shortly.
Finally, ${\delta M}/{\delta E}$ and ${\delta\widetilde{K}}/{\delta E}$ can be
obtained by taking their derivatives with respect to $E$ at $E =
{-}B_{4,0}^{(0)}$.

\begin{figure}[tb!]
\centering
\includegraphics[width=0.8\linewidth]{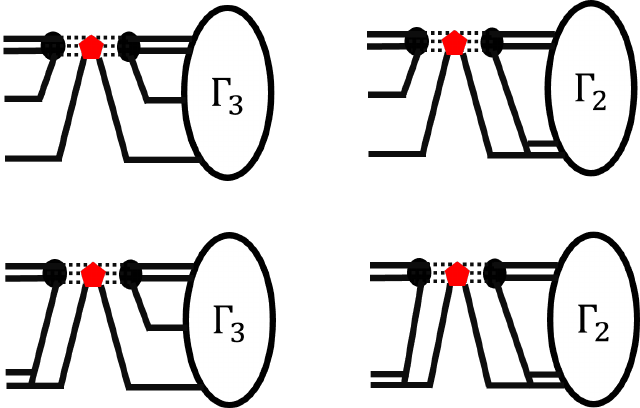}
\caption{Contributions of the four-body force $g_4$, represented by a red
 pentagon, in the diagrammatic four-body integral equation  depicted in
 Fig.~\ref{fig:4Bdiagram1}.
\label{fig:4bodydiagrams-w4BF}
}
\end{figure}

To find the contribution of the four-body force to the integral equation, one
can replace all the appearances of a bare trimer propagator plus a spectator in
Fig.~\ref{fig:4Bdiagram1} with this four-body force.
This results in the diagrams shown in Fig.~\ref{fig:4bodydiagrams-w4BF}.
Note that three of the four diagrams contain divergent loops with two
single-boson propagators and a bare trimer one.
While they have similar structures to the two-body loops in the dressed dimer
propagator, here they are regularized using a finite sharp cutoff.
The choice of the regulator should have no impact on observables in the
large-cutoff limit.

%%%%%%%%%%%%%%%%%%%%%%%%%%%%%%%%%%%%%%%%%%%%%%%%%%%%%%%%%%%%%%%%%%%%%%%%%%%%%%%%
%%%%%%%%%%%%%%%%%%%%%%%%%%%%%%%%%%%%%%%%%%%%%%%%%%%%%%%%%%%%%%%%%%%%%%%%%%%%%%%%

Similar to other NLO corrections, the perturbative correction from the
four-body force can be obtained by first including it non-perturbatively in the
integral equation and then expanding the relevant terms using the chain rule.
The diagrams in Fig.~\ref{fig:4bodydiagrams-w4BF} can be written as
\begin{equation}
 \begin{pmatrix}
  K^{g_4}_{33}\ket{\Gamma_3} & K^{g_4}_{32}\ket{\Gamma_2}\\
  K^{g_4}_{23}\ket{\Gamma_3}&  K^{g_4}_{22}\ket{\Gamma_2}
 \end{pmatrix} \,.
\label{eq:4BFinDA}
\end{equation}
These terms can be included in the four-body integral equation,
Eq.~\eqref{4bodydiageqn-opform}, by modifying the kernel
\begin{multline}
 \begin{pmatrix}
  K_{33} & K_{32} \\
  K_{23} & K_{22}
  \end{pmatrix} \to \begin{pmatrix}
  K^{\text{w4BF}}_{33} & K^{\text{w4BF}}_{32} \\
  K^{\text{w4BF}}_{23} & K^{\text{w4BF}}_{22}
 \end{pmatrix} \\
 \equiv
 \begin{pmatrix}
  K_{33} + K^{g_4}_{33} & K_{32} + K^{g_4}_{32} \\
  K_{23} + K^{g_4}_{23} & K_{22} + K^{g_4}_{22}
 \end{pmatrix} \,.
\label{eq:4BFinDAeffect}
\end{multline}

Similar to Eq.~\eqref{4bodydiageqn3-opform}, it is possible to transform the four-body
integral equation into a form that only involves $\ket{\widetilde{\Gamma}_2}$,
\begin{equation}
\label{4bodydiageqn3-opform-w4BF-main}
 \ket{\widetilde{\Gamma}_2} =
 \widetilde{K}^\text{w4BF}\ket{\widetilde{\Gamma}_2} \,,
\end{equation}
where (see Appendix~\ref{app:diagrammatic})
\begin{multline}
\label{K_tilde_w4BF-main}
 \widetilde{K}^\text{w4BF} = (1+P_d)\Big[
 \widetilde{K}^\text{w4BF}_{22}
  + \widetilde{K}^\text{w4BF}_{23} \\
  \null \times \left(1 -  P_{34}
   \widetilde{K}^\text{w4BF}_{33}\right)^{{-}1}
   P_{34}\widetilde{K}^\text{w4BF}_{32}
 \Big] \,,
\end{multline}
and
\begin{subalign}
 \widetilde{K}^\text{w4BF}_{33}
 &= {K}^{\text{w4BF}}_{33} (1-{K}^{\text{w4BF}}_{33})^{-1} \,, \label{4bodydiageqn2-kerneldef-w4BF-1}\\
 \widetilde{K}^\text{w4BF}_{23}
 &= {K}^\text{w4BF}_{23}(1-{K}^{\text{w4BF}}_{33})^{-1} \,, \\
 \widetilde{K}^\text{w4BF}_{32}
 &= (1-{K}^{\text{w4BF}}_{33})^{-1}{K}^\text{w4BF}_{32} \,, \\
 \widetilde{K}^\text{w4BF}_{22}
 &= {K}^\text{w4BF}_{22}
 + {K}^\text{w4BF}_{23}(1-{K}^{\text{w4BF}}_{33})^{-1}{K}^\text{w4BF}_{32} \,.
\label{4bodydiageqn2-kerneldef-w4BF-4}
\end{subalign}
Finally, the contribution from the four-body force to the eigenvalue
perturbation in Eq.~\eqref{eigenvalue_variation} can be written as
\begin{equation}
 \label{eigenvalue_variation_from_4BF}
 \delta\lambda_i^\text{4BF} = \frac{\braket{
   \widetilde{\Gamma}_2^{i}|M(\delta\widetilde{K}^\text{w4BF})|
   \widetilde{\Gamma}_2^{i}
 }}{\braket{\widetilde{\Gamma}_2^{i}|M|\widetilde{\Gamma}_2^{i}}} \,,
\end{equation}
where $\delta\widetilde{K}^\text{w4BF}$ is the perturbative expansion of
$\widetilde{K}^\text{w4BF}$ by keeping only the NLO piece of the four-body
force.
By construction, $\delta\lambda_i^\text{4BF}$ must be linear in $g_4^{(1)}$.
One can determine $g_4^{(1)}$, for example, by fixing $B_{4,0}^{(1)}$ in
Eq.~\eqref{4body-lambda-pert} and then use it to predict the NLO correction to
other four-body observables, such as the energy shift $B_{4,1}^{(1)}$ of the
four-body excited state.
In particular, the numerator of Eq.~\eqref{eigenvalue_variation_from_4BF} and
thus $\delta\lambda_i^\text{4BF}$ are not guaranteed to be non-zero.
This means that $g_4^{(1)}$ could diverge when $\delta\lambda_i^\text{4BF}$
vanishes.

\section{Renormalization}
\label{sec:renormalization}

At LO in the expansion around the two-body unitarity limit, there is a single
dimensionful parameter $\Lambda_\star$, which determines, through
renormalization, the trimer ground state
\begin{equation}
 \kappa_{3,0}  \equiv \sqrt{m B_{3,0}} = \Lambda_\star/b_0 \,,
\label{b_def}
\end{equation}
where $b_0$ is a real number that depends on the regularization procedure.
It also fixes the binding energies $B_A$ of larger
systems~\cite{vonStecher:2010XX, Nicholson:2012zp, Kievsky:2014dua,
Carlson:2017txq}, or alternatively the binding momenta $\kappa_A\equiv
\sqrt{mB_A}$.
For example, the tetramer ground- and excited-state binding momenta at LO,
\begin{equation}
 \kappa_{4,i}^{(0)} \equiv \sqrt{m B_{4,i}^{(0)}}
 \equiv \xi_i^{(0)} \kappa_{3,0} \,,
\label{eq:4intermsof3}
\end{equation}
are given by the universal numbers $\xi_0^{(0)}\simeq 2.15$ and
$\xi_1^{(0)}\simeq 1.00114$~\cite{Deltuva:2010xd}.

In contrast, at NLO, three additional dimensionful parameters appear:
$a_2^{-1}$, $r_2/2$, and $\kappa_{4,0} -\kappa_{4,0}^{(0)}$.
While LO is universal, higher-order corrections depend on the details of the
system under consideration.

\subsection{$^{\mathbf{4}}$He input}
\label{subsec:Input}

We present our results for the particular case of cold \isotope[4]{He} systems,
so $\psi$ represents the \isotope[4]{He} atom field and $m \simeq 0.0825$
K$^{{-}1}$$\mathrm{\AA}^{{-}2}$ (we have set $\hbar = 1$, as we do throughout the paper).
The range of the atom-atom interaction, $R\sim l_{\text{vdW}}\equiv (mC_6)^{1/4}
\simeq 5.4~$\AA, is set by the coefficient $C_6$~\cite{Zhang:2006XX} of the
$-r^{{-}6}$ tail of the van der Waals potential.

Experimentally, the \isotope[4]{He} diatomic system features a single, loosely
bound state with a binding energy $B_2= 1.76(15)$ mK measured in Coulomb
explosion~\cite{Zeller:2016mwo}.
From diffraction by a transmission grating, the \isotope[4]{He}-\isotope[4]{He}
scattering length is found to be $a_2= 104^{+8}_{-18}$ \AA
~\cite{Grisenti:2000zz}, which is more than an order of magnitude larger than
the interaction range.
Therefore, the two-\isotope[4]{He} system is close to the unitarity limit and
should be amenable to a systematic expansion around it.
A shallow, first-excited state of the trimer has also been observed at
$B_{3,1}-B_2 = 0.98(2)$ mK~\cite{Kunitski:2015qth}.

To gauge the convergence of the unitarity expansion, we use data from the LM2M2
potential~\cite{Aziz:1991LM2M2}, one of the most widely used realistic
\isotope[4]{He}-\isotope[4]{He} interactions.
With this potential, the two-body system is characterized by $a_2 \simeq 100$
\AA, $r_2 \simeq 7.326$ \AA ~\cite{Janzen1995Hepot}, and $B_2 \simeq 1.30348$
mK~\cite{Hiyama:2011ge}.
Theoretical calculations~\cite{Hiyama:2011ge} predict two trimer states with
binding energies $B_{3,0} \simeq  126.40$ mK and $B_{3,1}\simeq 2.2706$ mK, and
two tetramer states at $B_{4,0} \simeq 558.98$ mK and $B_{4,1} \simeq 127.33$
mK.
These values translate to $\kappa_{3,0}\simeq 0.1021$ $\mathrm{\AA}^{-1}$,
$\kappa_{4,0} \simeq 0.2147$ $\mathrm{\AA}^{-1}$, and $\kappa_{4,1}\simeq 0.1025$
$\mathrm{\AA}^{-1}$.
With some of these quantities used as renormalization conditions, the running of
the LECs can be determined and other observables predicted.

%%%%%%%%%%%%%%%%%%%%%%%%%%%%%%%%%%%%%%%%%%%%%%%%%%%%%%%%%%%%%%%%%%%%%%%%%%%%%%%%
\begin{figure*}[tb!]
\centering
\includegraphics[width=0.45\linewidth]{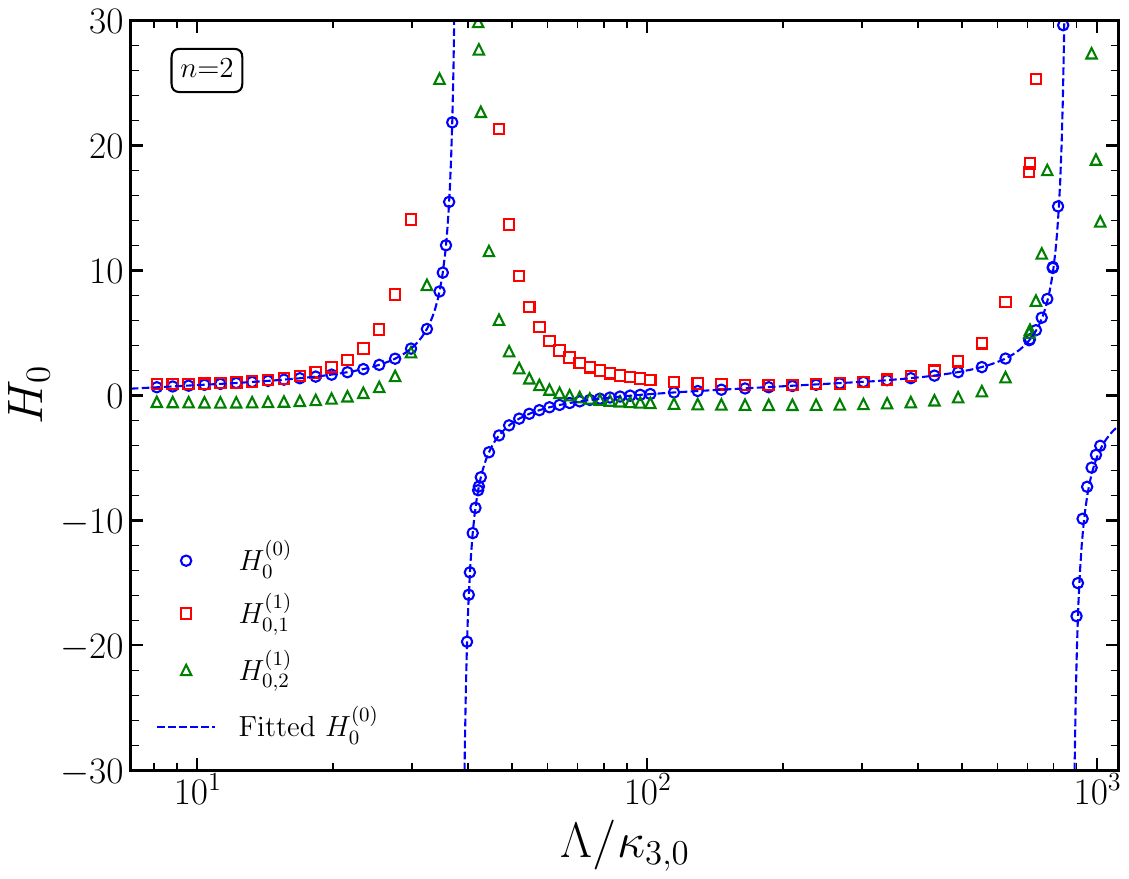} \hfil
\includegraphics[width=0.45\linewidth]{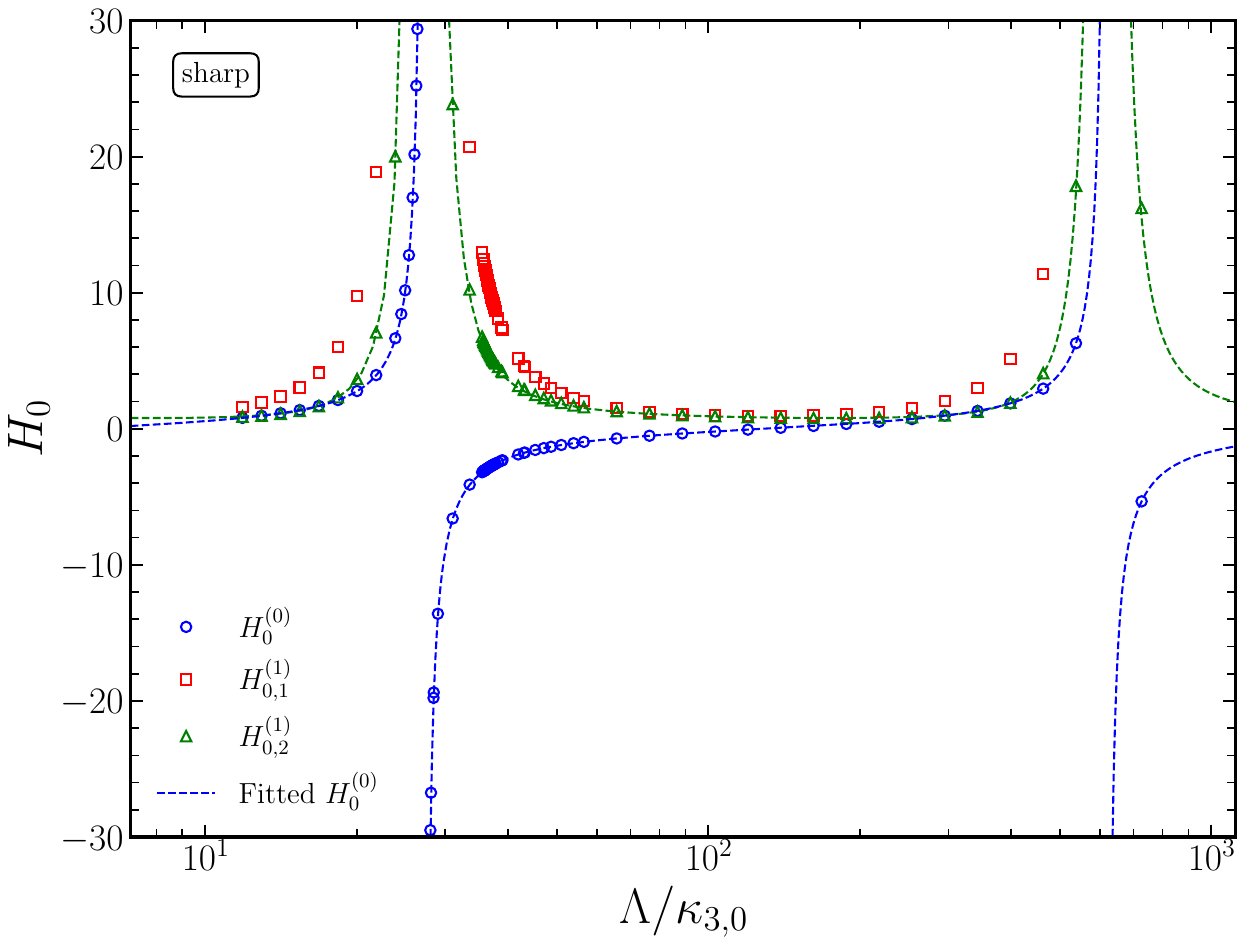}
\caption{Running of the dimensionless three-body LEC $H_0^{(0)}$ (blue circles)
 and the dimensionless components $H_{0,1}^{(1)}$ (red squares) and
 $H_{0,2}^{(1)}$ (green triangles) of its correction $H_0^{(1)}$, as functions
 of the momentum cutoff $\Lambda$ (in units of the trimer three-body binding
 momentum $\kappa_{3,0}$) for the quartic super-Gaussian ($n=2$) regulator in
 the Faddeev calculation (left panel) and a sharp-cutoff regulator in the
 diagrammatic approach (right panel).
 The blue dashed line on the left panel is determined by fitting the blue
 circles with Eq.~\eqref{H0_0_expr}.
 On the right panel, the blue dashed line uses parameters from the
 literature~\cite{Bedaque:1998kg,Bedaque:1998km,
 Braaten:2011sz,Braaten:2004rn,Ji:2015hha} and the green dashed line represents
 Eq.~\eqref{H02_1}.
\label{fig:H0_reg4}
}
\end{figure*}
%%%%%%%%%%%%%%%%%%%%%%%%%%%%%%%%%%%%%%%%%%%%%%%%%%%%%%%%%%%%%%%%%%%%%%%%%%%%%%%%

\subsection{Low-energy constants}
\label{subsec:LECs}

It is well-known that the LO three-body LEC displays a limit-cycle
behavior~\cite{Bedaque:1998kg, Bedaque:1998km}, which is apparent when we write
the three-body LEC in dimensionless form,
\begin{equation}
 H_0 \equiv \frac{\Lambda^2 D_0}{6 m C_0^{2}}
 = \frac{\Lambda^2}{2m\Omega} \,.
\label{H0_def}
\end{equation}
Expanding the right-hand side order by order,
\begin{subalign}
 H_0^{(0)} &= \frac{\Lambda^2 D_0^{(0)}}{6 m C_0^{(0)2}}
 = \frac{\Lambda^2}{2m\Omega^{(0)}} \,, \\
 H_0^{(1)} &= H_0^{(0)} \left(
  \frac{D_0^{(1)}}{D_0^{(0)}} - \frac{2 C_0^{(1)}}{C_0^{(0)}}
 \right) = {-}H_0^{(0)} \frac{\Omega^{(1)}}{\Omega^{(0)}} \,.
\end{subalign}

It has been shown that $H_0^{(0)}$ is a function of $\Lambda/\Lambda_\star$ on a
limit cycle of the RG with the analytical form~\cite{Bedaque:1998kg,
Bedaque:1998km, Braaten:2011sz, Chen:2025rti, Chen:2025iqp}
\begin{equation}
 H_0^{(0)}(\Lambda/\Lambda_\star)
 = h_0 \frac{ \sin(s_0 \ln (\Lambda/\Lambda_\star) - \delta_0 ) }
 { \sin(s_0 \ln (\Lambda/\Lambda_\star) + \delta_0 ) } \,,
 \label{H0_0_expr}
\end{equation}
where $s_0 \simeq$ 1.00624 is a constant characterizing the period of the limit
cycle, $h_0$ is an overall factor, and $\delta_0$ is a phase.
In the STM equation with a sharp-cutoff regulator, $\delta_0=\arctan(1/s_0)
\simeq 0.7823$~\cite{Bedaque:1998kg, Bedaque:1998km, Chen:2025rti}\footnote{Note that $\Lambda_\star$ is defined here as a three-body-force
parameter, not as a parameter in the asymptotic wave function as done in
Ref.~\cite{Chen:2025rti}.}
and $h_0\simeq 0.879$~\cite{Braaten:2011sz}.
For general separable regulators in the STM and Faddeev equations,
it was recently shown that the functional form in Eq.~\eqref{H0_0_expr} remains
valid, with the parameters 
$\delta_0$ and $h_0$, as well as $b_0$ defined in Eq.~\eqref{b_def}, depending on the
specific form of the employed regulator~\cite{Chen:2025iqp}.
An analysis of the RG equation satisfied by $H_0^{(0)}$, which takes a quadratic
form, further reveals a richer structure of the limit cycle: $h_0$ and
$\delta_0$ characterize the complex-conjugate pair of fixed points, and
$\Lambda_\star$ (or, equivalently, $b_0$) sets the boundary condition that
anchors the limit cycle to a concrete Efimov tower.

Figure~\ref{fig:H0_reg4} shows the running of the three-body LEC $H_0^{(0)}$ as
a function of the momentum cutoff $\Lambda$ for the quartic super-Gaussian in
the Faddeev equation and for a sharp cutoff in the diagrammatic approach.
Results for the standard and sextic (super-)Gaussians in the Faddeev formalism
are given in Appendix~\ref{app:reg_dep}.
While the various regulators differ in detail, their results are qualitatively
similar.
The LO three-body LEC $H_0^{(0)}$ displays a clear log-periodic behavior for all
the regulators we have considered. $H_0^{(0)}$ initially increases to keep the
reference state at the right energy, then at some point switches sign to replace
the state with a new one at the same energy.
When the reference state is the ground state, the old state becomes a deep
trimer outside the regime of validity of the EFT.
The process repeats with the accretion of further deep trimers.
Equation~\eqref{H0_0_expr} provides an excellent fit with parameters $\{b_0,
\delta_0, h_0\} \simeq $ $\{2.6236, 0.4455, 1.0189\}$ for the quartic
super-Gaussian regulator.
For the sharp-cutoff regulator in the diagrammatic approach, our numerical
results agree remarkably well with parameters $\{ 2.61, 0.7823, 0.879\}$ from the 
literature~\cite{Bedaque:1998kg,Bedaque:1998km,Braaten:2011sz,Braaten:2004rn,Ji:2015hha}.

The log-periodic behavior of $H_0^{(0)}(\Lambda/\Lambda_\star)$ is a reflection
of a discrete scale invariance where all momenta and $\Lambda$ are scaled by
powers of the same factor $\exp (\pi/s_0)\simeq 22.7$, but the parameter
$\Lambda_\star$ remains fixed (``active transformation'').
Alternatively, all momenta and $\Lambda$ remain fixed, but $\Lambda_\star$
changes by the (inverse of the) same factor (``passive
transformation'').
DSI is broken explicitly at subleading orders, with the appearance of
dimensionful parameters.
Because these contributions are taken into account in perturbation theory,
DSI-breaking parameters appear in a power series.
Once they are made dimensionless using $\kappa_{3,0}$ or a related momentum such
as $\kappa_{4,0}^{(0)}$, their coefficients are DSI-invariant functions of
$\Lambda/\Lambda_\star$.
In the running of subleading LECs, these functions are log-periodic, up to
overall factors of $\kappa_{3,0}/\Lambda$.

At NLO, the dimensionless three-body LEC receives two corrections, one from the
two-body scattering length, the other from the effective range:
\begin{spliteq}
 H_0^{(1)}(\Lambda/\Lambda_\star) &= \left(a_2 \kappa_{3,0}\right)^{{-}1}
 \, H_{0,1}^{(1)}(\Lambda/\Lambda_\star) \\
 & \quad + \frac{r_2 \Lambda}{2} \, H_{0,2}^{(1)}(\Lambda/\Lambda_\star)
 +\cdots \,,
 \label{H0_1}
\end{spliteq}
where the dimensionless functions $H_{0,1}^{(1)}$ and $H_{0,2}^{(1)}$ are
determined by the LO dynamics and can only depend on $\Lambda/\Lambda_\star$.
Residual cutoff dependence, represented by ``$\cdots$'', is comparable to other
higher-order contributions.

The runnings of the $H_{0,1}^{(1)}$ and $H_{0,2}^{(1)}$ that leave $B_{3,0}$
unchanged are also shown in Fig.~\ref{fig:H0_reg4} for the same regulators
employed at LO, with other regulators considered in Appendix~\ref{app:reg_dep}.
Although $H_{0,2}^{(1)}$ is somewhat smaller than $H_{0,1}^{(1)}$, the two
components of the NLO three-body LEC have the same order of magnitude and share the
log-periodicity of $H_0^{(0)}(\Lambda/\Lambda_\star)$.
$H_{0,1}^{(1)}(\Lambda/\Lambda_\star)$ is everywhere positive: this part of the
three-body force contributes repulsively (attractively) to counteract the
attractive (repulsive) part of the two-body NLO interaction stemming from
$a_2>0$ ($a_2<0$).
On the other hand, $H_{0,2}^{(1)}(\Lambda/\Lambda_\star)$ is everywhere positive
in the diagrammatic approach with a sharp cutoff but changes sign in the Faddeev
formalism with (super-) Gaussian regulators.
The effects of the range correction are more complicated than those of the
scattering length, as they can be attractive or repulsive, depending on the
regulator.

%%%%%%%%%%%%%%%%%%%%%%%%%%%%%%%%%%%%%%%%%%%%%%%%%%%%%%%%%%%%%%%%%%%%%%%%%%%%%%%%
\begin{figure*}[tb!]
\centering
\includegraphics[width=0.45\linewidth]{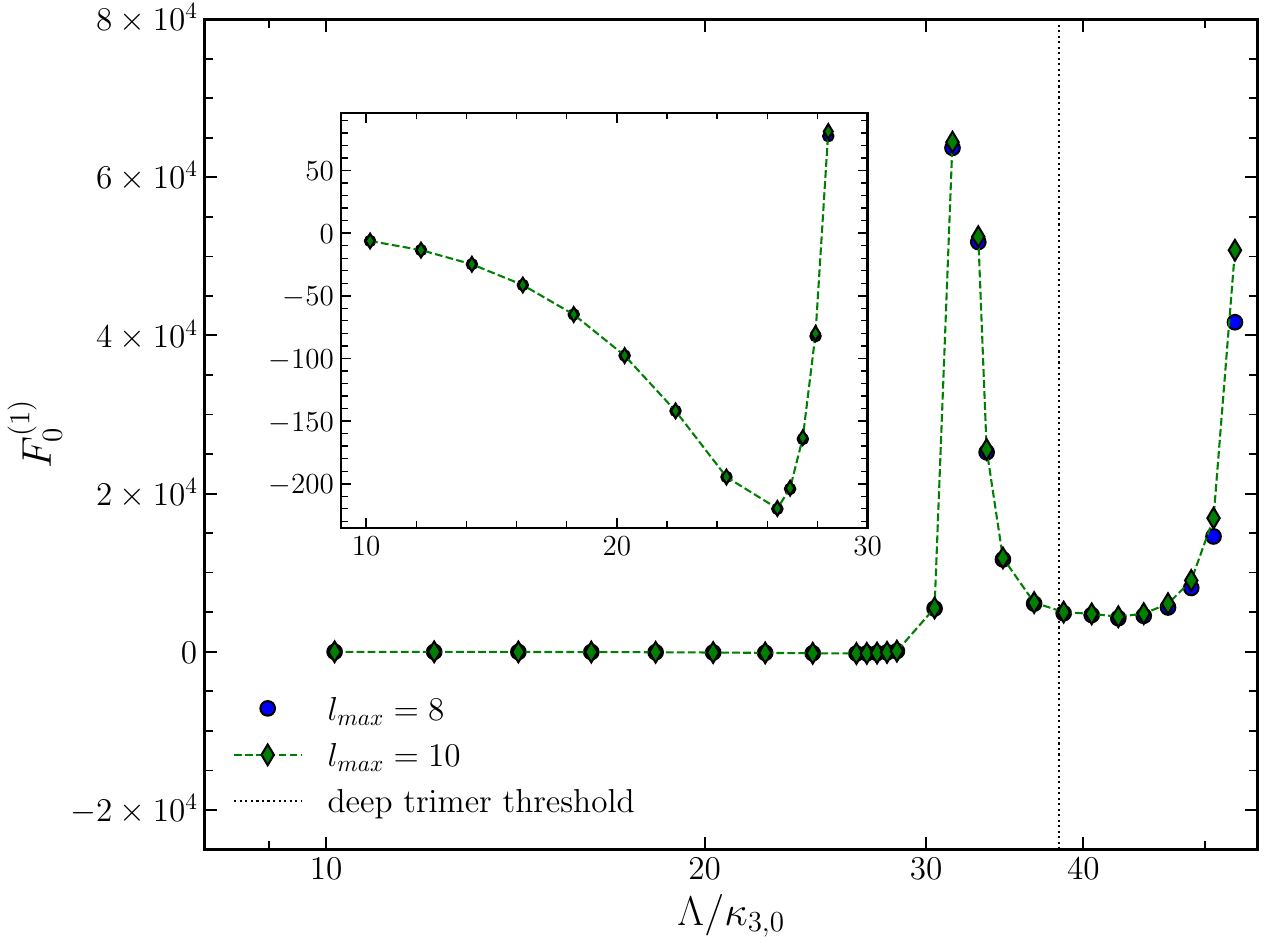}\hfil
\includegraphics[width=0.45\linewidth]{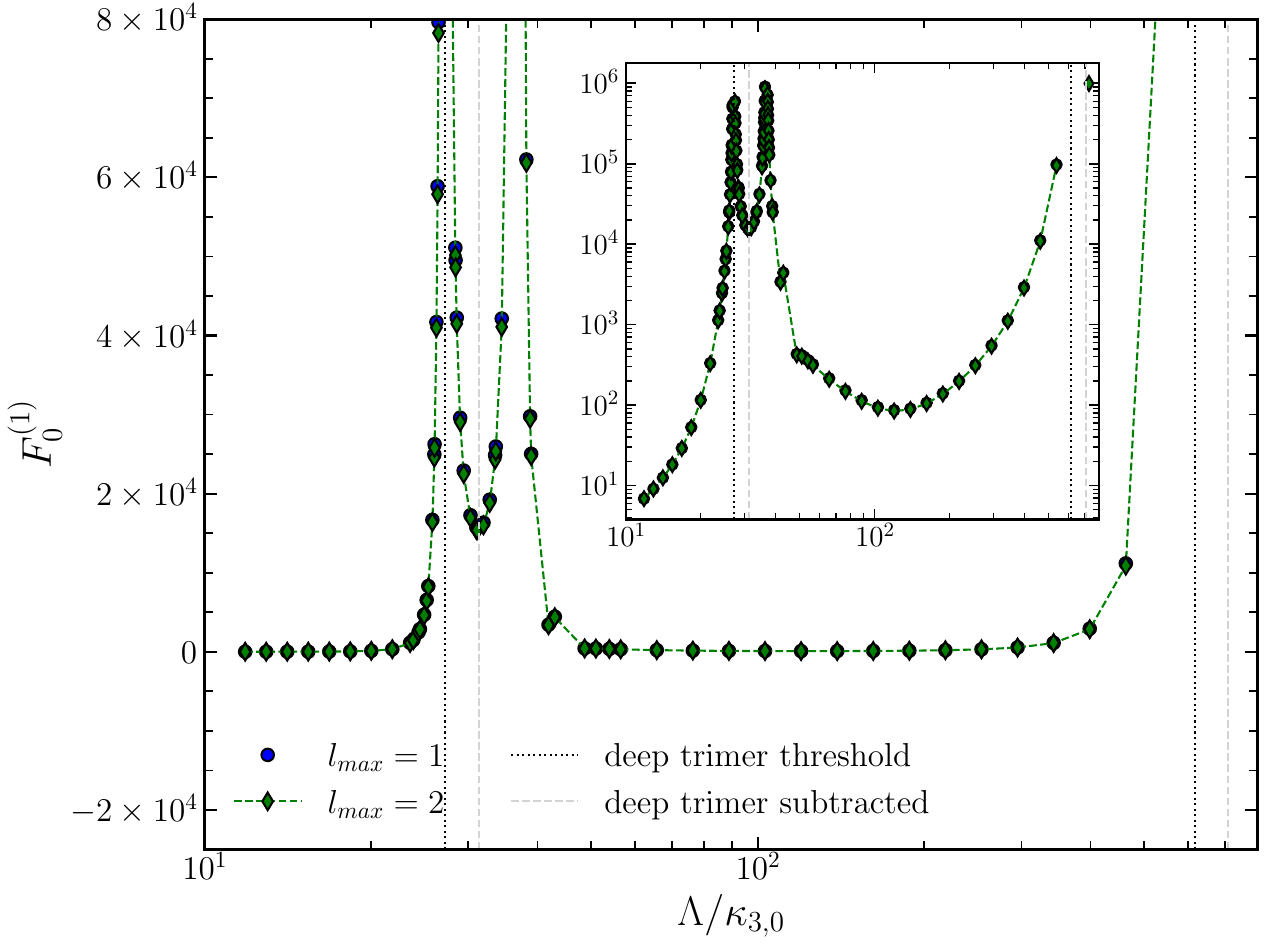}
\caption{
 Running of the dimensionless four-body LEC $F_0^{(1)}$ as a function of the
 momentum cutoff $\Lambda$ (in units of $\kappa_{3,0}$) for the quartic
 super-Gaussian ($n=2$) regulator in the Faddeev-Yakubovsky formalism (left
 panel) and a sharp-cutoff regulator in the diagrammatic approach (right panel).
 On the left panel, blue circles and green diamonds represent the results for
 $l_{\text{max}} =$ 8 and 10, respectively.
 On the right panel, blue circles and green diamonds represent the results for
 $l_{\text{max}} =$ 1 and 2, respectively.
 Insets show linear-linear (left panel) and log-log (right panel)
 plots.
 The vertical black dotted (gray dashed) lines indicate the cutoff values where
 deep trimers appear (are subtracted).
 Parameters from the LM2M2 potential are used as input.
\label{fig:F0_reg4_DA_log}
}
\end{figure*}
%%%%%%%%%%%%%%%%%%%%%%%%%%%%%%%%%%%%%%%%%%%%%%%%%%%%%%%%%%%%%%%%%%%%%%%%%%%%%%%%

%%%%%%%%%%%%%%%%%%%%%%%%%%%%%%%%%%%%%%%%%%%%%%%%%%%%%%%%%%%%%%%%%%%%%%%%%%%%%%%%
\begin{figure}[tb!]
\centering \includegraphics[width=0.95\linewidth]{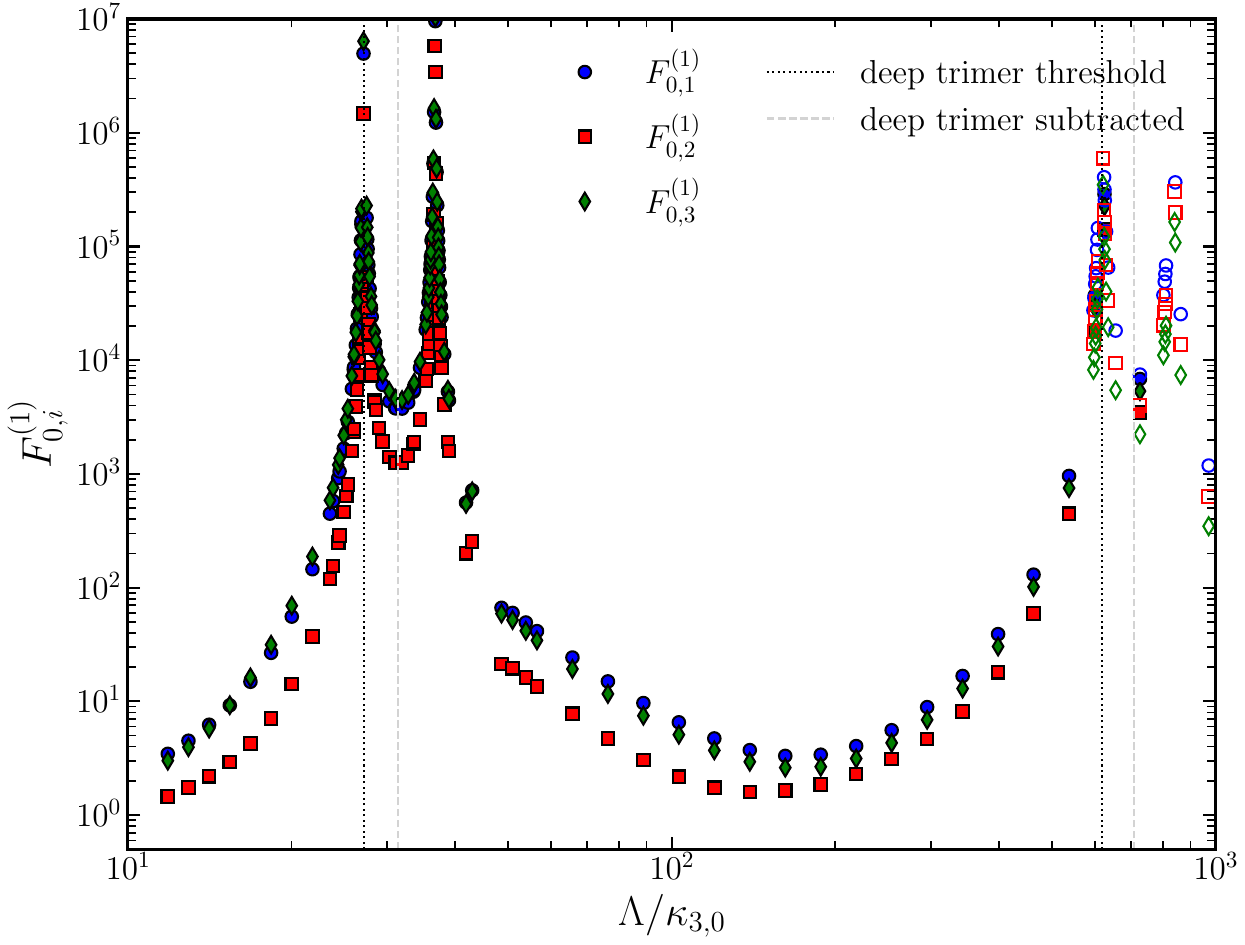}
\caption{Running of the dimensionless components $F_{0,1}^{(1)}$ (blue circles),
 $F_{0,2}^{(1)}$ (red squares), and $F_{0,3}^{(1)}$ (green diamonds) of the
 four-body LEC $F_0^{(1)}$, as functions of the momentum cutoff $\Lambda$ (in
 units of $\kappa_{3,0}$).
 The diagrammatic approach with a sharp cutoff is used here.
 Solid and hollow symbols are obtained with $l_{\text{max}}=$ 2 and 0, respectively.
 Vertical lines as in Fig.~\ref{fig:F0_reg4_DA_log}.
\label{fig:F0i}
}
\end{figure}
%%%%%%%%%%%%%%%%%%%%%%%%%%%%%%%%%%%%%%%%%%%%%%%%%%%%%%%%%%%%%%%%%%%%%%%%%%%%%%%%

The functional form of $H_{0,2}^{(1)}(\Lambda/\Lambda_\star)$ for a sharp cutoff
regulator in the STM equation can be derived analytically by evaluating the
logarithmic and linear divergences proportional to $r_2$~\cite{Ji:2011qg}.
With LO at unitarity, the logarithmically divergent term, which is proportional
to the LO two-body binding momentum, vanishes, and \footnote{Up to a factor of
 $-2$, $H_{0,2}^{(1)}(\Lambda/\Lambda_\star)$ is the function $h_{10}$ defined
 in Ref.~\cite{Ji:2011qg}. The parameter $\bar{\Lambda}$ used in
 Ref.~\cite{Ji:2011qg} is related to $\Lambda_\star$ through the relation
$\bar{\Lambda} = \Lambda_\star \exp(\pi/2 s_0)$.}
\begin{spliteq}
 &\quad H_{0,2}^{(1)}(\Lambda/\Lambda_\star) \\
 &= \frac{3\pi}{16}
 \frac{1+s_0^2}{1 - \cos(2s_0 \ln(\Lambda/\Lambda_\star) + 2\arctan(1/s_0))} \\
 &\null \quad \times \left[
  1+ \frac{\sin(2 s_0 \ln(\Lambda/\Lambda_\star)
  + \arctan(1/2s_0))}{ \sqrt{1+4 s_0^2}}
 \right] \,. \\
\label{H02_1}
\end{spliteq}
The log-periodicity is evident.
As shown on the right panel of Fig.~\ref{fig:H0_reg4}, this expression is in
excellent agreement with the numerical values of
$H_{0,2}^{(1)}(\Lambda/\Lambda_\star)$.
The derivation of $H_{0,1}^{(1)}(\Lambda/\Lambda_\star)$ involves finite
contributions that are analytically challenging to evaluate; therefore, only
numerical results are presented for this quantity.

Once the two-body effective range is included at NLO, a four-body force is
needed for renormalization in the four-body system. Similarly to $H_0$, we can
define a dimensionless coupling
\begin{equation}
 F_0^{(1)} \equiv \frac{\Lambda^4 E_0^{(1)}}{24 m^2C_0^{(0)3}}
 = {-}\frac{\Lambda^4\Delta^{(0)} g^{(1)}_4}{8\pi m \Omega^{(0)2}} \,.
\label{F0_1prime}
\end{equation}
The running of $F_0^{(1)}$ with $a_2$, $r_2$, and $B_{4,0}$ from the LM2M2
potential is plotted in Fig.~\ref{fig:F0_reg4_DA_log}, for both quartic
super-Gaussian in the FY formalism and the sharp cutoff in the diagrammatic
approach.
For the former, we include partial waves up to $l_{\text{max}}=10$, while for the
latter, convergence is reached already at $l_{\text{max}}=2$.
In the diagrammatic approach with a sharp cutoff, the four-body LEC is always
positive, whereas for the quartic super-Gaussian in the FY formalism it can take
negative values, as illustrated in an inset.
Apart from this small difference, the four-body LECs obtained from these
two approaches exhibit qualitatively similar behavior: near certain
special cutoffs, the four-body LEC develops a pole.
In the diagrammatic approach, one pole coincides with a pole in the three-body
LEC, \ie, deep-trimer threshold.
In this approach, the deep trimer is only subtracted in the four-body
calculation at the higher value of the cutoff where it enters.
In contrast, with the smoother regulators in the FY equations, poles in the
four-body LEC do not coincide with poles in the three-body LEC, where the deep
trimer is removed.
In both approaches the four-body LEC has additional poles at ``exceptional
cutoffs'' where the relevant matrix element vanishes~\cite{Gasparyan:2022isg}.
We will come back to this issue when discussing observables.

Similarly to the three-body LEC correction, the four-body LEC can be written in
terms of universal functions of $\Lambda/\Lambda_\star$,
\begin{spliteq}
 \frac{\kappa_{4,0}^{(0)}}{\Lambda} F_0^{(1)}(\Lambda/\Lambda_\star)
 &= \left(a_2 \kappa_{4,0}^{(0)}\right)^{{-}1} \,
 F_{0,1}^{(1)}(\Lambda/\Lambda_\star) \\
 &\quad\null + \frac{r_2 \kappa_{4,0}^{(0)}}{2} \,
 F_{0,2}^{(1)}(\Lambda/\Lambda_\star) \\
 &\quad\null + \left(\frac{\kappa_{4,0}}{\kappa_{4,0}^{(0)}}-1\right)
 F_{0,3}^{(1)}(\Lambda/\Lambda_\star) \,.
\label{F0_1}
\end{spliteq}
In the presence of the $a_2^{{-}1}$ correction alone, four-body binding energies
converge as the cutoff increases~\cite{Konig:2016utl} and there is no need for a
four-body force.
When one adds a four-body force with the purpose of fitting $B_{4,0}$, a
component ($F_{0,1}^{(1)}$) is proportional to $a_2^{{-}1}$ even if $r_2=0$.
In addition, when a nonzero effective-range correction is included, a four-body
force is necessary~\cite{Bazak:2018qnu}.

Runnings of the various four-body LEC components as functions of $\Lambda$ are
shown in Fig.~\ref{fig:F0i} for a sharp cutoff in the diagrammatic approach.
Due to computational constraints, points at larger cutoffs
($\Lambda/\kappa_{3,0} \gtrsim 580$), extending into another branch, are
obtained with $l_{\text{max}} =0$ and are not fully converged, thus differing slightly
from those with $l_{\text{max}}=2$.
An approximate log-periodicity is apparent.
These three functions have the same poles as $F_0^{(1)}$ since they come from
zeros of $\delta\lambda^\text{4BF}_0$ in
Eq.~\eqref{eigenvalue_variation_from_4BF}, which only depend on the LO dynamics.
These components are universal in the sense that, although they depend on the
choice of regulator, they are independent of the values of $a_2^{-1}$, $r_2$,
and $\kappa_{4,0}$.

\section{Results}
\label{sec:Results}

In the previous section, we have shown the running of the LECs, which are not
observables and depend on the regulator $\Lambda$ such that observables do not
--- apart from some residual cutoff dependence that vanishes in the large cutoff
limit.
This section presents our main results for observables.

The expansion around the two-body unitarity limit is an expansion in powers of
the interaction range $R$ and the inverse two-body scattering length
$a_2^{{-}1}$.
To form dimensionless parameters, they should be accompanied by some
characteristic momentum associated with the $A$-body bound state.
If we assume that each particle contributes an equal energy $Q_A^2/2m$ to the
binding, an estimate for the typical one-particle momentum is $Q_A \sim
\sqrt{2/A}\,\kappa_A$~\cite{vanKolck:2017jon}.
With this estimate, we expect the two-body NLO corrections to be suppressed by
the small combinations $\epsilon_{3,0}^a = \sqrt{3/2}(a_2 \kappa_{3,0})^{{-}1}$
and $\epsilon_{3,0}^r = r_2 \kappa_{3,0}/\sqrt{6}$ in the three-body system, and
$\epsilon_{4,i}^a = \sqrt{2}(a_2 \kappa_{4,i}^{(0)})^{{-}1}$ and
$\epsilon_{4,i}^r = r_2\kappa_{4,i}^{(0)}/\sqrt{8}$ in the four-body system.
Moreover, there are four- and more-body forces at subleading orders.
Using as renormalization condition for the NLO four-body force that
$\kappa_{4,0}$ is reproduced, an estimate of its effects in the four-body system
is given by $\epsilon_{4}^\kappa=\abs{\kappa_{4,0} -
\kappa_{4,0}^{(0)}}/\kappa_{4,0}^{(0)}$.

As a benchmark, we compare our results for the expansion around the unitarity
limit with those obtained for \isotope[4]{He} clusters directly from the LM2M2
potential without consideration of the universality linked to the two-body
unitarity.
We use the potential model as proxy for experimental data: were more data on
small clusters available, we could substitute them for the input and outcome of
the potential model.
When appropriate, we also compare with the ``standard'' SREFT
expansion~\cite{Braaten:2004rn, Hammer:2019poc}, where $a_2$ is reproduced
exactly at LO through the renormalization of $C_0^{(0)}$ or, equivalently,
$\Delta^{(0)}$.
In this case, results for binding energies of up to six-particle clusters are
known to NLO~\cite{Bazak:2016wxm, Bazak:2018qnu, Contessi:2023yoz}.

The proximity to the unitarity limit is revealed not only by the large
\isotope[4]{He}-\isotope[4]{He} scattering length, but also by the pattern of
light clusters.
The single dimer state is shallow on the scale of the trimer ground-state
energy, and the tetramer ground and first excited states are in agreement with
unitarity expectations: $\xi_0 \simeq 2.103$ and $\xi_1 \simeq 1.004$ are very
close to the LO values, 2.15 and 1.00114, respectively, in
Eq.~\eqref{eq:4intermsof3}.
Indeed, with values from the LM2M2 potential, the NLO corrections are governed
by $\epsilon_{3,0}^a \simeq 0.12$, $\epsilon_{3,0}^r \simeq 0.31$ for the
three-body ground state; $\epsilon_{4,0}^a \simeq 0.064$,
$\epsilon_{4,0}^r\simeq 0.57$ for the four-body ground state; and
$\epsilon_{4,1}^a \simeq 0.14$, $\epsilon_{4,1}^r\simeq 0.26$ for the four-body
excited state.
The additional expansion parameter, which is the same for the four-body ground-
and excited states, is $\epsilon_{4}^\kappa\simeq 0.021$.

The unitarity limit also predicts geometric towers of excited states associated
to these states, with a ratio of consecutive energies $\exp(2\pi/s_0)\simeq
515$~\cite{Efimov:1970zz}.
However, for the LM2M2 potential the ratio $B_{3,0}/B_{3,1}\simeq 56$ is almost
an order of magnitude smaller.
The reason is that finite scattering-length effects are likely not perturbative for the
excited state, as $\epsilon_{3,1}^a=\sqrt{3/2}(a_2 \kappa_{3,1})^{{-}1}\simeq
0.9$ is of order unity.
Still, the trimer excited state, just as the virtual state observed in
neutron-deuteron scattering~\cite{Adhikari:1982zzb, Rupak:2018gnc}, is the
remnant of the first Efimov excited state.
Both states are well described in ``standard'' SREFT~\cite{Bazak:2016wxm,Bazak:2018qnu,
Rupak:2018gnc,Qin:2020hsc}, but are likely too 
shallow to be amenable to the unitarity 
expansion. In the ``standard" expansion where $a_2$ is included at LO, the excited-state energy can be used instead of the ground-state energy to fix the renormalization condition for the three-body force~\cite{Bazak:2016wxm,Bazak:2018qnu}. Because the physics of the trimer excited state, unlike that of the trimer ground state and its associated four-body bound state, is not dominated solely by $\Lambda_\star$, we do not consider this choice in this work.

In the following, we examine the three-body ground state and the four-body
ground and excited states with a combination of the FY formalism and the
diagrammatic approach, which have complementary strengths and limitations.
Calculations in an EFT are carried out up to a finite order and
are therefore inherently truncated.
When there is only one expansion parameter $\epsilon$, the EFT expansion for an
observable $O$ can be written as
\begin{equation}
 O = O_0 \left( 1+ \sum_{i=1}^{\infty} c_i \epsilon^i \right) \equiv O^{[\nu]}
 + \delta O^{[\nu]} \,,
\label{O_EFTexp}
\end{equation}
in which $O^{[\nu]}$ includes all the contributions up to order
$\nu$, and $\delta O^{[\nu]} = O_0 \sum_{i=\nu+1}^{\infty}
c_i \epsilon^i$ is the EFT truncation error at order $\nu$.
As a rough estimate, assuming natural-sized coefficients $c_{i} \sim \OO(1)$,
the truncation error is estimated as
\begin{equation}
 \delta O^{[\nu]} \sim \text{max}(O^{[0]}, \ldots, O^{[\nu]})
 \times \epsilon^{\nu+1} \,.
\end{equation}
When multiple expansion parameters are present, we adopt the
largest one for $\epsilon$ to ensure a conservative estimate: $\epsilon_{3,0}^r$
for the three-body ground state; $\epsilon_{4,0}^r$ for the four-body ground
state; and $\epsilon_{4,1}^r$ for the four-body excited state.

With a finite regulator $\Lambda$, $O^{[\nu]}$ depends explicitly on $\Lambda$.
We show that $O^{[\nu]}(\Lambda)$, does indeed converge to a finite value, $O^{[\nu]}$, as
the cutoff $\Lambda$ increases, by considering extrapolations of the type
\begin{equation}
 O^{[\nu]}(\Lambda) = O^{[\nu]} \Big[
  1 + \frac{\kappa_{3,0}}{\Lambda}\, \Big(
   c_{0\nu}+ c_{1\nu} \sin(2 s_0 \ln (\Lambda/\Lambda_{1\nu}))
  \Big)
 \Big] \,,
\label{fit_formula}
\end{equation}
where $c_{0\nu}$, $c_{1\nu}$, and $\Lambda_{1\nu}$ are fit
parameters.
As discussed in Appendix~\ref{app:reg_dep}, this form incorporates the
oscillatory convergence expected from DSI and observed in some quantities.
It captures the leading inverse power of $\Lambda$ and the modulation by the
first harmonics.
As a check, we confirm that $c_{0\nu}$ and $c_{1\nu}$ are
normally no larger than ${\cal O}(1)$.

To estimate the extrapolation error of this simplified form, we
perform fits starting at different cutoff values.
The resulting set of fits defines an envelope that forms a band around the
calculated data.
We take the midpoint of this band in the large cutoff limit as the central
value, and the width of the band as the extrapolation uncertainty.
The extrapolation uncertainty determined in this manner depends on the chosen
fit range and fitting ansatz, and may vary if these are modified.
However, we achieve such high cutoff values that this variation is typically
much smaller than the EFT truncation error.

Our final results for the \isotope[4]{He} trimer and tetramer binding energies
and radii are summarized in Tables~\ref{table:Bs} and~\ref{table:rs} and
described in more detail in the rest of this section.

%%%%%%%%%%%%%%%%%%%%%%%%%%%%%%%%%%%%%%%%%%%%%%%%%%%%%%%%%%%%%%%%%%%%%%%%%%%%%%%%
\begin{table}[!tb]
\setlength{\extrarowheight}{1.2pt}
\caption{Values of the \isotope[4]{He} trimer ground-state ($B_{3,0}$), tetramer
 ground-state ($B_{4,0}$) and tetramer first-excited-state ($B_{4,1}$) energies
 (in mK) at LO and NLO, compared to values obtained directly from the LM2M2
 potential~\cite{Hiyama:2011ge}.
 An asterisk denotes input to the EFT calculation.
 U and S denote the expansion around unitarity and the ``standard'' expansion
 with finite $a_2$ at LO, respectively.
 Numbers in parentheses indicate extrapolation uncertainties, while the
 uncertainties following the $\pm$ sign correspond to EFT truncation errors.
\label{table:Bs}
}
\begin{center}
\begin{tabular}{c c c c}
\hline
\hline
 & $B_{3,0}$ (mK) & $B_{4,0}$ (mK) & $B_{4,1}$ (mK) \\
 \hline
 LO (U) &  126.40 (*)  &
 582(1)$\pm$332  & 126.68(2)$\pm$33  \\
 NLO (U) & 126.40 (*) & 558.98 (*)  &  127.49(2)$\pm$9 \\
 \hline
 LO (S) &  126.40 (*)  &
 518(1)$\pm$319  &  126.41(1)$\pm$33 \\
 NLO (S) & 126.40 (*) & 558.98 (*)  &  126.68(4)$\pm$9 \\
 \hline
 LM2M2 & 126.40  & 558.98  & 127.33 \\
  \hline
  \hline
\end{tabular}
\end{center}
\end{table}
%%%%%%%%%%%%%%%%%%%%%%%%%%%%%%%%%%%%%%%%%%%%%%%%%%%%%%%%%%%%%%%%%%%%%%%%%%%%%%%%

%%%%%%%%%%%%%%%%%%%%%%%%%%%%%%%%%%%%%%%%%%%%%%%%%%%%%%%%%%%%%%%%%%%%%%%%%%%%%%%%
\begin{table}[!tb]
\setlength{\extrarowheight}{1.2pt}
\caption{Values of the \isotope[4]{He} trimer ($r_{3,0}$) and tetramer
 ($r_{4,0}$) ground-state radii (in \AA) at LO and NLO, compared to values
 obtained directly from the LM2M2 potential~\cite{Hiyama:2011ge}.
 U and S are the same as in Table~\ref{table:Bs}.
\label{table:rs}
}
\begin{center}
\begin{tabular}{c c c}
\hline
\hline
 & $r_{3,0}$ (\AA) & $r_{4,0}$ (\AA) \\
 \hline
 LO (U) &  4.605(4)$\pm$1.8  & 3.24(6)$\pm$2.7 \\
 NLO (U)&  5.7(1)$\pm$0.6 &  4.7(1)$\pm$1.5\\
 \hline
 LO (S) &  4.806(3)$\pm$1.8  &  3.47(6)$\pm$2.7 \\
 NLO (S)&  5.7(1)$\pm$0.6 &  4.71(5)$\pm$1.5\\
 \hline
 LM2M2 & 6.326 & 5.16 \\
  \hline
  \hline
\end{tabular}
\end{center}
\end{table}
%%%%%%%%%%%%%%%%%%%%%%%%%%%%%%%%%%%%%%%%%%%%%%%%%%%%%%%%%%%%%%%%%%%%%%%%%%%%%%%%

\subsection{Three-body ground state}
\label{sec:3bgs}

Since the three-body ground-state binding energy is used to fix the three-body
LEC, here we focus on the radius, $r_{3,0}$, defined as the square root of the
average squared distance of a particle from the center of mass, denoted by
$\braket{ r_{3,0}^2}$.
Since we deal with an identical-particle system, $r_{3,0}$ is the same for any
one of the particles.
A convenient choice is the third particle since
\begin{equation}
 \vec{x}_3 - \vec{R}_3 = \frac{2}{3} \vec{r}_2 \,,
\label{x3CM}
\end{equation}
where $\vec{x}_i$ denotes the position of the $i$th particle, $\vec{R}_3 =
\sum_{i=1}^3{\vec{x}_i}/3$ is the center of mass of the three-body system, and
\begin{equation}
 \vec{r}_2 = \vec{x}_3 - \frac{1}{2} (\vec{x}_1+\vec{x}_2)
\end{equation}
is the Jacobi coordinate associated with the Jacobi momentum $\vec{u}_2$.
From Eq.~\eqref{x3CM}, we find
\begin{equation}
 \braket{ r_{3,0}^2 } = \frac{4}{9} \braket{ \vec{r}_2^2 } \,.
\label{r30_r2}
\end{equation}

The \isotope[4]{He} atom is neutral, but we can pretend that it has a unit
charge that does not induce extra interactions between \isotope[4]{He} atoms.
The radius is then obtained from the pseudo-charge form factor
\begin{equation}
 F_{C,3}(q^2) = \braket{ \Psi_3 | \hat{\rho}(\vec{q}) | \Psi_3 } \,,
 \label{Fc3}
\end{equation}
where $\vec{q}$ is the momentum transfer and $\hat{\rho}$ is the pseudo-charge operator for explicit coupling only to the third particle,
\ie, the momentum $\vecq$ is transferred explicitly only
to the subsystem described by the Jacobi momentum $\vec{u}_2$, and we have
\begin{equation}
 \braket{ \vec{u}_1 \vec{u}_2 |\hat{\rho}(\vec{q}) | \vec{u}_1'\vec{u}_2'}
 = \delta^{(3)}(\vec{u}_1-\vec{u}_1')
 \, \delta^{(3)}(\vec{u}_2-\vec{u}_2'- 2\vec{q}/3) \,.
\end{equation}
The representation of this operator in the three-body partial-wave basis
is discussed in Refs.~\cite{Konig:2019xxk,Lyu:2025yhz,Andis:2025fsg}.
In particular, the expressions relevant for this work can be obtained from
those in the latter two references by setting all spin and isospin quantum
numbers to zero.
The mean square radius is given by the usual relation,
\begin{equation}
 \braket{ r_{3,0}^2 } = {-}\frac{1}{6} \frac{\dd}{\dd q^2} F_{C,3}(q^2) \,.
\label{r_Fc3}
\end{equation}

The form factor should also be calculated in a perturbative expansion,
\begin{subalign}
 F_{C,3}^{(0)}(q^2)
 &= \braket{ \Psi_{3,0}^{(0)} | \hat{\rho}(\vec{q}) | \Psi_{3,0}^{(0)} } \,,
\label{FC3_0} \\
 F_{C,3}^{(1)}(q^2)
 &= 2\braket{ \Psi_{3,0}^{(0)} | \hat{\rho}(\vec{q}) | \Psi_{3,0}^{(1)} } \,.
\label{FC3_1}
\end{subalign}
Denoting
\begin{equation}
 \braket{ r_{3,0}^{2} }^{(\nu)}
 = {-}\frac{1}{6} \frac{\dd}{\dd q^2} F_{C,3}^{(\nu)}(q^2) \,,
\label{r30_nu}
\end{equation}
the LO radius is
\begin{equation}
 r_{3,0}^{(0)} = \sqrt{\braket{ r_{3,0}^{2} }^{(0)} } \,,
\label{r30_0}
\end{equation}
while the NLO correction is
\begin{equation}
 r_{3,0}^{(1)} = \frac{ \braket{ r_{3,0}^{2} }^{(1)} }{2r_{3,0}^{(0)}} \,.
\label{r30_1}
\end{equation}

\begin{figure}[tb!]
 \centering
 \includegraphics[width=0.95\linewidth]{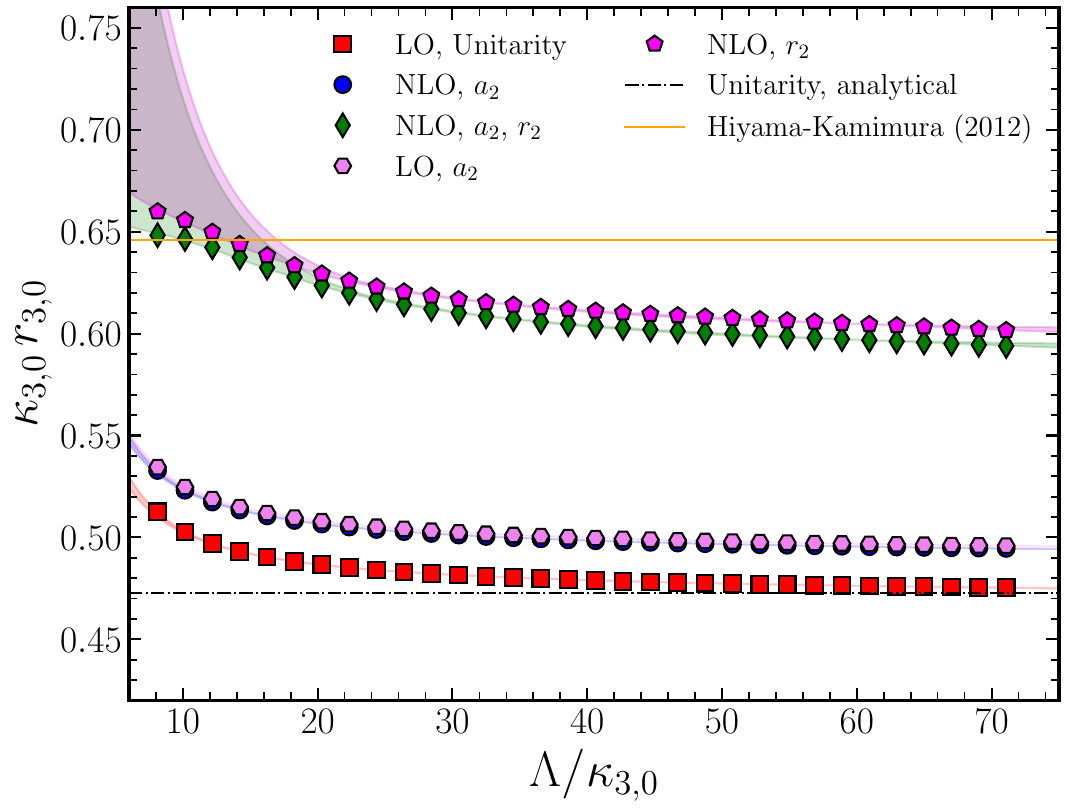}
 \caption{Three-body ground-state radius $r_{3,0}$ (in units of
  $\kappa_{3,0}^{-1}$) as a function of the momentum cutoff $\Lambda$ (in units
  of $\kappa_{3,0}$) for the quartic super-Gaussian regulator in the Faddeev
  equation with $l_{\text{max}}=12$.
  Red squares, blue circles (barely visible under the violet hexagons), and
  green diamonds represent, respectively, the LO results at unitarity, the
  (incomplete) NLO results when only the scattering length $a_2$ is included,
  and the full NLO results when both $a_2$ and the effective range $r_2$ are
  considered.
  Violet hexagons and magenta pentagons denote, respectively, the LO results
  with $a_2$ treated nonperturbatively and the corresponding NLO results with
  $r_2$ included perturbatively.
  The bands are the envelopes of fits to points with $\Lambda / \kappa_{3,0}
  \geq$ 8, 10, 15, 20, and 25. Input from the LM2M2 potential is used.
  The horizontal black dash-dotted line and the orange solid line represent,
  respectively, the radius at unitarity from Eq.~\eqref{mB3r2} and the direct
  calculation~\cite{Hiyama:2011ge} with the LM2M2 potential.
 \label{fig:3b_radii}
 }
\end{figure}

Figure~\ref{fig:3b_radii} shows the three-body ground-state radius as a function
of the momentum cutoff for the quartic super-Gaussian regulator in the F
equation at LO and NLO.
In the unitarity limit, the natural length scale is set by
$\kappa_{3,0}^{{-}1}$.
The bands are the envelopes of fits to numerical results with $\Lambda /
\kappa_{3,0} \geq$ 8, 10, 15, 20, and 25 using Eq.~\eqref{fit_formula}.
At NLO, these fits indicate larger uncertainties in the regime of
small $\Lambda$, while at LO they are good even at low values of
$\Lambda/\kappa_{3,0}$.
Convergence at high cutoff values indicates that no new LECs appear at these
orders.

%%%%%%%%%%%%%%%%%%%%%%%%%%%%%%%%%%%%%%%%%%%%%%%%%%%%%%%%%%%%%%%%%%%%%%%%%%%%%%%%
%%%%%%%%%%%%%%%%%%%%%%%%%%%%%%%%%%%%%%%%%%%%%%%%%%%%%%%%%%%%%%%%%%%%%%%%%%%%%%%%

At exact unitarity, the trimer radius and binding energy are linked by the
universal relation~\cite{Braaten:2004rn, Vanasse:2015fph}
\begin{equation}
 \left. m B_{3,0} \braket{ r_{3,0}^2 }
 \right|_{\rm unitarity}= \frac{1+s_0^2}{9} \,.
\label{mB3r2}
\end{equation}
Our LO numerical results are in excellent agreement with this relation as
$\Lambda$ increases.
We obtain $\kappa_{3,0}r_{3,0}^{(0)} = 0.470(1)$, only 0.6\% off the value from
Eq.~\eqref{mB3r2}.
The corresponding value (in units of \AA) for the \isotope[4]{He} trimer is
given in Table~\ref{table:rs}.

The incomplete NLO, where only the scattering length $a_2$ is included, gives an
extrapolated value 4.793(3)~\AA.
It amounts to a $\lesssim 5\%$ correction, somewhat smaller than the expected
magnitude of the expansion parameter $\epsilon_{3,0}^a$.
When the effective range $r_2$ is included for a complete NLO
calculation, further corrections arise, and the extrapolated NLO radius is
5.7(1)~\AA.
These corrections are $\sim 24\%$ of the LO value --- a similar
magnitude to $\epsilon_{3,0}^r$, but a bit smaller.
This happens because the NLO three-body force, which fixes the ground-state
energy at the right position, partially counteracts the effects of the
scattering length and effective range.

The smallness of the expansion parameters suggests that a nonperturbative
treatment of $a_2$ will not significantly improve on the perturbative treatment.
Indeed, our results for the ``standard'' SREFT expansion, also shown in
Fig.~\ref{fig:3b_radii}, are very close to those of the expansion around unitarity. In fact, the extrapolated value at NLO is the same, within
errors, as the NLO value of the unitarity expansion.
The small difference between these two expansions reflects the proximity of the
three-\isotope[4]{He} system to the unitarity limit.

It is the two-body effective-range correction, then, that is responsible for the
largest change at NLO, as evident in Fig.~\ref{fig:3b_radii} and
Table~\ref{table:rs}.
As expected, at NLO the predicted radius shifts toward the value calculated
directly from the LM2M2 potential~\cite{Hiyama:2011ge}.
The convergence pattern is dictated by the conventional EFT expansion in
$Q_3 R \sim \epsilon_{3,0}^r$.
At NLO, the difference to the direct result is $\sim 10\%$, which is consistent
with $(\epsilon_{3,0}^r)^2$. Hence, we expect it to diminish at higher orders.

The extrapolated values given in Table~\ref{table:rs} are in reasonable
agreement with the LO and NLO values of $r_{3,0}$, 4.82(19)~\AA and 5.94(1)~\AA,
obtained from the average bond length calculated in Ref.~\cite{Qin:2020hsc} (using
the conversion for identical-particle systems derived in the
appendix of Ref.~\cite{Wu:2023mhg}).
Differences stem from different inputs in fixing the two- and three-body forces:
in Ref.~\cite{Qin:2020hsc}, the three-body LEC is fixed by the atom–dimer
scattering length and the trimer ground-state binding energy shifts by $\sim
8\%$ at NLO, in contrast to our fixed ground-state binding energy.
Although different inputs are used, the results at
the same order still agree within the EFT truncation uncertainty.

\subsection{Four-body ground state}
\label{sec:4bgs}

The tetramer ground-state energy is a prediction at LO, but used to fix
the four-body LEC when the range correction is included at NLO.
To test the expansion at NLO, we consider here also the radius, which remains a
prediction at both LO and NLO.

As the cutoff increases and we approach the zero-range limit, the LO three-body
LEC transitions from one branch of the limit cycle to another, accompanied by
the emergence of a deep trimer, which evolves into an Efimov trimer in the large
cutoff limit.
The convergence behavior of four-body observables at large cutoffs can be
studied after the deep trimers are removed.
While in the diagrammatic approach deep trimers are addressed with the
principal-value prescription (Appendix~\ref{app:diagrammatic}), in the FY
formalism the projection potential $V_{P,3}$ defined in Eq.~\eqref{pseudopotential} contains an
unphysical parameter $\eta$, which physical observables should remain
independent of.
It is verified in Appendix~\ref{app:eta_dep} that this holds with good
accuracy in our calculation.
In the following, we present results for $\eta=1000$, which are essentially
indistinguishable from the limit $\eta\to \infty$.

\begin{figure*}[tb!]
\centering
\includegraphics[width=0.45\linewidth]{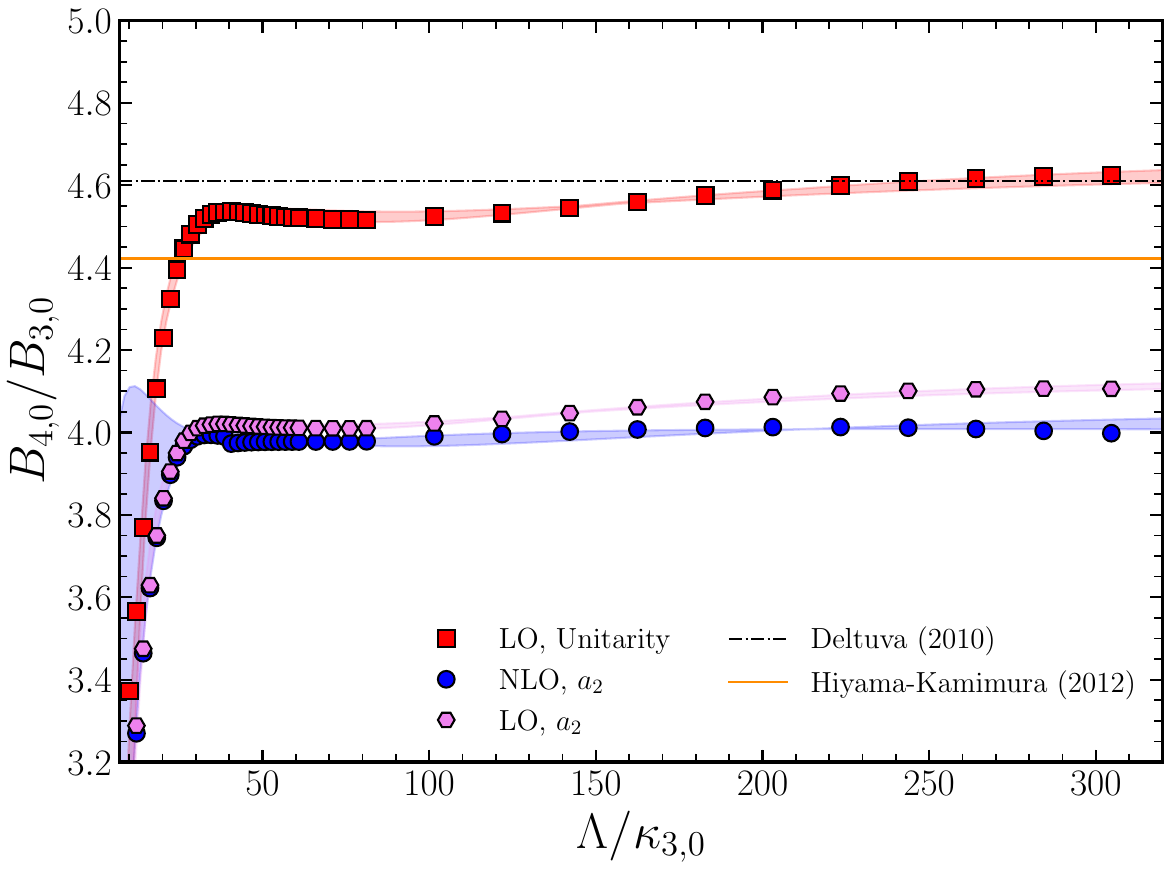}\hfil
\includegraphics[width=0.45\linewidth]{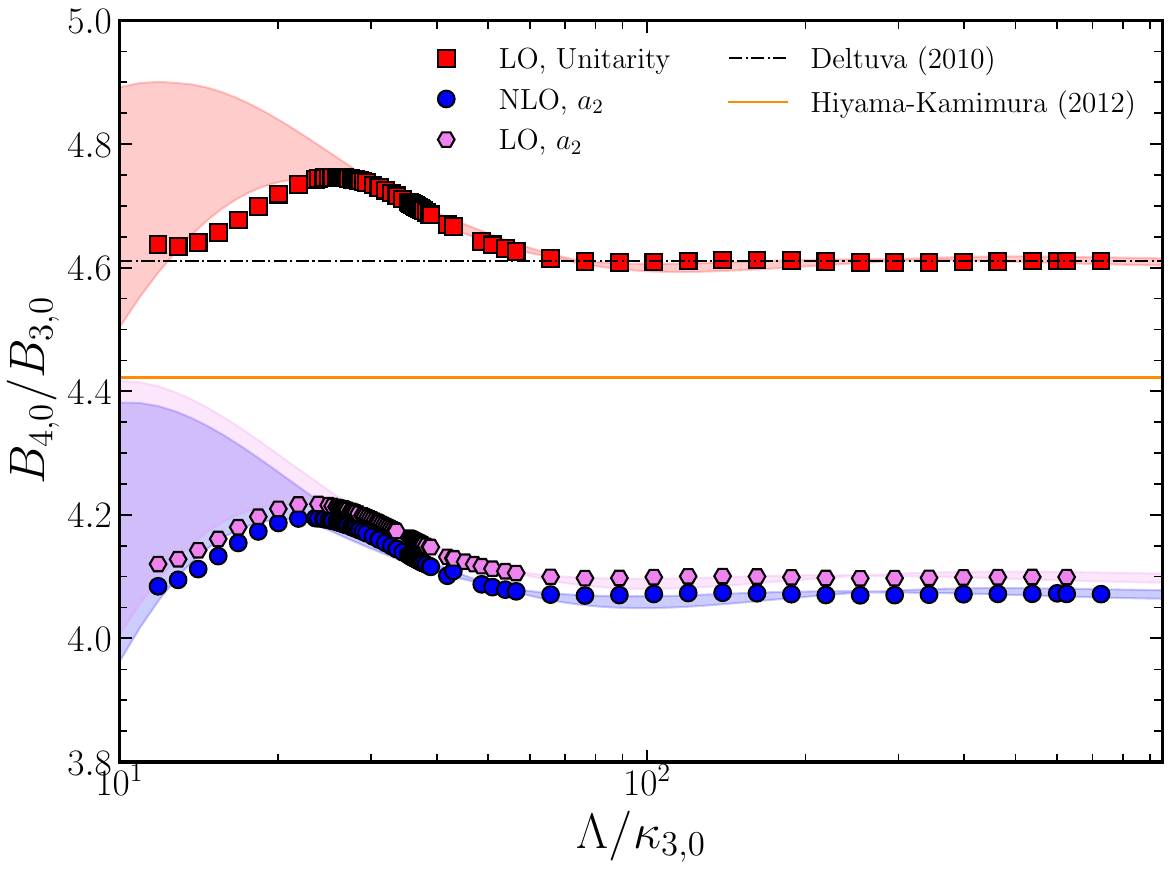}
\caption{Four-body ground-state binding energy $B_{4,0}$ (in units of $B_{3,0}$)
 to NLO as a function of the momentum cutoff $\Lambda$ (in units of
 $\kappa_{3,0}$) obtained by solving the FY equations with $l_{\text{max}}=4$ (left
 panel, linear scale) and from the diagrammatic approach with $l_{\text{max}}=2$ (right
 panel, semilogarithmic scale).
 Symbols and horizontal orange line are as in Fig.~\ref{fig:3b_radii}.
 The bands are the envelopes of fits to points with $\Lambda/\kappa_{3,0} \geq$
 10, 20, 30, and 40.
 The horizontal black dash-dotted line is the result at unitarity from
 Ref.~\cite{Deltuva:2010xd}.
\label{fig:B40_NLO}
}
\end{figure*}

In Fig.~\ref{fig:B40_NLO}, the four-body ground-state binding energy (in units
of $B_{3,0}$) is plotted as a function of the cutoff $\Lambda$ for the solution
of the FY equations with the quartic super-Gaussian regulator and the
diagrammatic approach with a sharp cutoff.
For the FY equations, cutoff values up to $\Lambda \sim 300\,\kappa_{3,0}$,
which reach beyond where the first deep trimer would appear, can be achieved.
The diagrammatic approach allows us to extend the analysis to even larger
sharp-cutoff values, $\Lambda \sim 800 \, \kappa_{3,0}$, beyond the threshold
for the second deep trimer.
Convergence with respect to the maximum angular momentum $l_{\text{max}}$ is discussed
in Appendix~\ref{app_sec:convergence}.
The FY results with $l_{\text{max}} = 4$ are almost identical to those with $l_{\text{max}} =
2$, and convergence is even faster in the diagrammatic approach.

%%%%%%%%%%%%%%%%%%%%%%%%%%%%%%%%%%%%%%%%%%%%%%%%%%%%%%%%%%%%%%%%%%%%%%%%%%%%%%%%
%%%%%%%%%%%%%%%%%%%%%%%%%%%%%%%%%%%%%%%%%%%%%%%%%%%%%%%%%%%%%%%%%%%%%%%%%%%%%%%%

The two methods are in broad agreement for large cutoff values.
Results at low cutoffs differ, which is expected due to the use of
different regulator functions.
As discussed in Appendix~\ref{app:reg_dep}, noticeable differences exist among
various regulators even within the FY approach.
They are most visible before the appearance of the first deep trimer, which
makes removing deep trimers all the more important.
The bands in Fig.~\ref{fig:B40_NLO} are the envelopes of fits with
Eq.~\eqref{fit_formula} to points with $\Lambda/\kappa_{3,0} \geq$ 10, 20, 30,
and 40.
Like for the three-body radius, the LO fits to FY results are good even
at low cutoff values.
However, without higher harmonics, they can only capture the corresponding NLO
values and the diagrammatic-approach points that are included in the fit.

The LO results from the diagrammatic approach converge to the universal
ratio~\cite{Lin:2023zqw}
\begin{equation}
 \left. \frac{B_{4,0}}{B_{3,0}}\right|_{\text{unitarity}} = 4.606(6) \,,
\label{eq:B40(0)}
\end{equation}
which is consistent with the result in Ref.~\cite{Deltuva:2010xd}.
In the FY case, we obtain a slightly larger value, 4.66(2).
The small difference could be an artifact of the extrapolation
using relatively small cutoff values and/or from neglecting the imaginary part
of the energy.
According to Refs.~\cite{Deltuva:2010xd, Lin:2023zqw}, the ratio between the
half-width (imaginary part) and $B_{4,0}$ is $\approx$ 0.3\%.
In our calculations, the deep trimers are removed and the four-body state
becomes stable.
This can potentially lead to a small energy shift, as observed in
Ref.~\cite{Lin:2023zqw}.
Moreover, small oscillations are found in the running of the energy as a
function of the cutoff, also confirming the observation in
Ref.~\cite{Lin:2023zqw}.
The extrapolated value for the \isotope[4]{He} tetramer based on the calculation
of the diagrammatic approach is listed in Table~\ref{table:Bs} with
a large LO truncation error.

At an incomplete NLO where only the two-body scattering length is included, the
four-body ground-state binding energy is still a prediction. The extrapolated values obtained from the FY formalism and the diagrammatic approach
are 510(4) mK and 514(1) mK, respectively.
The shift due to the scattering length is $\sim 12\%$ of the LO value,
consistent with an expansion in $2\epsilon_{4,0}^a$, where the factor 2 arises
because the energy scales with the square of the binding momentum.
We would thus expect this first-order scattering-length correction to differ
from treating $a_2$ exactly at LO, as in the standard expansion, by a few percent.
This expectation is confirmed by the results of an explicit calculation shown in
Fig.~\ref{fig:B40_NLO}.
Just like for the three-body ground-state radius, the expansion in
$1/Q_A a_2$ converges quickly.
Note that our central value for the exact treatment of $a_2$ given in Table \ref{table:Bs}
is smaller than the
universal value~\eqref{eq:B40(0)}, while the corresponding value in
Ref.~\cite{Bazak:2018qnu} is larger.
The difference likely lies in the input employed to fix the LECs: another
\isotope[4]{He}-\isotope[4]{He} potential and trimer excited-state energy
were used in Ref.~\cite{Bazak:2018qnu} compared to this work.
The expansion around unitarity cannot describe the three-body ground and excited
states simultaneously.

The shift induced in the four-body ground-state energy by a finite scattering
length goes in the direction of the result reported in Ref.~\cite{Hiyama:2011ge}
for the LM2M2 potential, which is also indicated in Fig.~\ref{fig:B40_NLO}.
However, it overshoots considerably, reflecting the omission of other NLO
effects, such as the two-body effective range. Results with only $a_2$ and $r_2$
included at NLO are shown in Appendix~\ref{app_sec:convergence}.
Changes are indeed dramatic, but the binding energy displays a strong cutoff
dependence.
It requires a four-body force for renormalization, confirming previous
findings~\cite{Bazak:2018qnu}.
The LEC associated with the four-body force is fixed by the four-body
ground-state binding energy, as shown in Fig.~\ref{fig:F0_reg4_DA_log}.
The four-body force effectively counteracts the two-body effective range, so that
overall the calculations converges at NLO.

To test the full expansion around the unitarity limit, we calculate the radius
of the four-body ground state and compare to the direct
calculation~\cite{Hiyama:2011ge} with the LM2M2 potential.
We present the result from solving the FY equations, as the formalism for
calculating the four-body radius using the diagrammatic approach has not yet
been developed.

The four-body ground-state radius $r_{4,0}$ can be defined and calculated in a
similar manner as the three-body radius. Now we single out the fourth particle
since
\begin{equation}
 \vec{x}_4 - \vec{R}_4 = \frac{3}{4} \vec{r}_3 \,,
 \label{x4CM}
\end{equation}
where $\vec{R}_4 = \sum_{i=1}^4{\vec{x}_i}/4$ is the center of mass of the
four-body system and
\begin{equation}
 \vec{r}_3 = \vec{x}_4 - \vec{R}_3
\end{equation}
is the Jacobi coordinate associated with the Jacobi momentum $\vec{u}_3$.
From Eq.~\eqref{x4CM},
\begin{equation}
 \braket{ r_{4,0}^2 } = \frac{9}{16}\braket{ \vec{r}_3^2 } \,.
\end{equation}
The pseudo-charge form factor $F_{C,4}(q^2)$ is calculated in an analogous
manner to Eqs.~\eqref{FC3_0} and~\eqref{FC3_1}, but with the four-body wave
function and the pseudo-charge operator,
\begin{multline}
 \quad \braket{ \vec{u}_1 \vec{u}_2 \vec{u}_3 |\hat{\rho}(\vec{q})
 | \vec{u}_1'\vec{u}_2'\vec{u}_3'} \\
 =\delta^{(3)}(\vec{u}_1-\vec{u}_1')
 \delta^{(3)}(\vec{u}_2-\vec{u}_2')
 \delta^{(3)}(\vec{u}_3-\vec{u}_3'- \frac{3}{4}\vec{q}) \,.
\end{multline}
As for three particles, the partial-wave matrix elements for this
operator are discussed in Refs.~\cite{Konig:2019xxk,Lyu:2025yhz,Andis:2025fsg},
and once the form factor is obtained, the corresponding four-body radius is
determined following the same procedure as in
Eqs.~\eqref{r30_nu},~\eqref{r30_0}, and~\eqref{r30_1}.

\begin{figure}[tb!]
\centering
\includegraphics[width=0.95\linewidth]{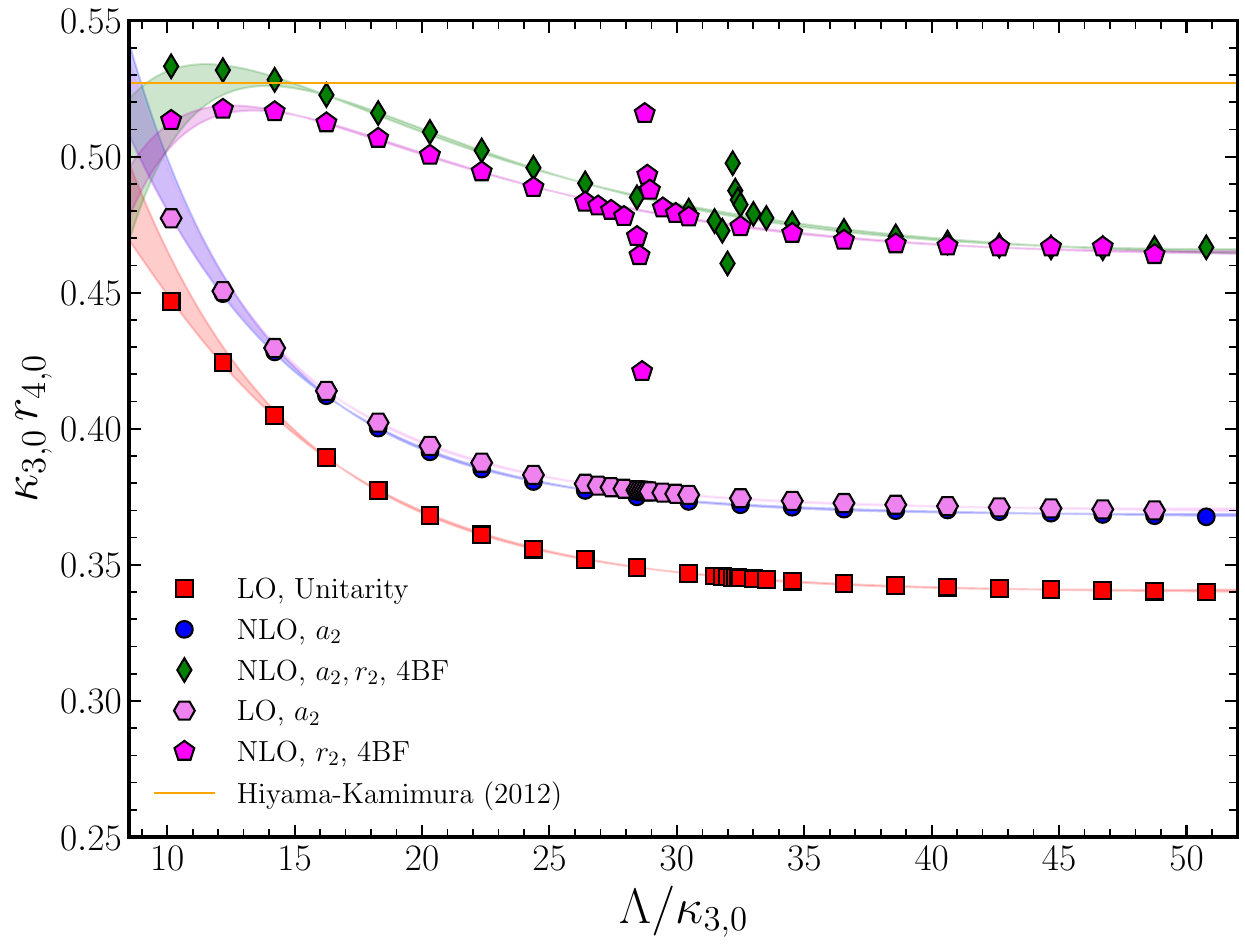}
\caption{Four-body ground-state radius $r_{4,0}$ (in units of
 $\kappa_{3,0}^{-1}$) as a function of the momentum cutoff $\Lambda$ (in units
 of $\kappa_{3,0}$) for the quartic super-Gaussian regulator in the FY equations
 with $l_{\text{max}}=10$.
 Symbols, input, and horizontal line are as in Fig.~\ref{fig:3b_radii}, except
 that the NLO results include also the four-body force.
 The bands are the envelopes of fits to points with $\Lambda/\kappa_{3,0} \geq$
 10 and 15, with small regions around NLO ``exceptional cutoffs'' excluded.
\label{fig:r40_U_NLO}
}
\end{figure}

Figure~\ref{fig:r40_U_NLO} shows the four-body ground-state radius as a function
of the momentum cutoff for the quartic super-Gaussian regulator.
We again use $\kappa_{3,0}^{-1}$ as the unit of length.
The bands are now the envelopes of fits to points with $\Lambda/\kappa_{3,0}
\geq$ 10 and 15.

%%%%%%%%%%%%%%%%%%%%%%%%%%%%%%%%%%%%%%%%%%%%%%%%%%%%%%%%%%%%%%%%%%%%%%%%%%%%%%%%
%%%%%%%%%%%%%%%%%%%%%%%%%%%%%%%%%%%%%%%%%%%%%%%%%%%%%%%%%%%%%%%%%%%%%%%%%%%%%%%%

The LO radius exhibits very good convergence with respect to both momentum and
angular momentum cutoffs, leading to an extrapolated value $\kappa_{3,0}
r_{4,0}^{(0)}= 0.33(1)$.
At exact unitarity,
\begin{equation}
 \left. m B_{4,0}
 \braket{ r_{4,0}^{2} } \right|_{\text{unitarity}}
 = 0.50(1)\,,
\label{m_B40_r2}
\end{equation}
which is universal for the deeper four-body states associated with each Efimov
three-body state.
This value is in good agreement with Ref.~\cite{Carlson:2017txq} and confirms
that at unitarity the four-body ground state is smaller than the three-body
ground state.
The corresponding value for the \isotope[4]{He} tetramer is given in
Table~\ref{table:rs}.

A similar convergence pattern is found for the NLO radius when only the
two-body scattering length is included.
It converges toward the LO value with $a_2$ treated non-perturbatively, the
extrapolated values being 3.44(7)~\AA and 3.47(6)~\AA, respectively.
The corresponding correction to the unitarity radius is $\sim$ 6\% of the LO
value, consistent with the expansion parameter $\epsilon_{4,0}^a$.
The inclusion of scattering-length effects leads to a modest improvement in the
calculated radius ($\approx$ 12\% of the gap between the LO result and the value
from Ref.~\cite{Hiyama:2011ge}), which indicates that additional corrections are
still required to remedy the remaining discrepancy.

When effective-range corrections are included, both large angular-momentum
($l_{\mathrm{max}}$) and high momentum ($\Lambda$) cutoffs are required to
achieve converged results.
Figure~\ref{fig:r40_U_NLO} presents converged results with $l_{\mathrm{max}}=$
10.
The irregular behavior observed near $\Lambda \simeq 32\,\kappa_{3,0}$ for the
expansion around unitarity corresponds to the pole of the four-body LEC shown in
Fig.~\ref{fig:F0_reg4_DA_log}.
A
similar divergence also occurs for the the standard expansion at $\Lambda
\approx 28.7\,\kappa_{3,0}$.
Such behavior is a common feature in the vicinity of ``exceptional cutoffs'',
first identified in the two-nucleon problem~\cite{Gasparyan:2022isg} and later
also in the three-nucleon system~\cite{Thim:2025vhe} when higher-order
interactions are treated in perturbation theory within Chiral EFT.
It should be emphasized that this is a regulator-dependent issue: whether
exceptional cutoffs appear, and where they occur, depends on the choice of the
regularization procedure.
One way to address this issue is to employ, or design, regulators that avoid
these exceptional cutoffs; examples are provided in Refs.~\cite{Yang:2024yqv,
PavonValderrama:2025zzk, PavonValderrama:2025azr}.

Alternatively, one can adjust the strategy for fixing the LECs to remain
consistent within EFT truncation errors~\cite{Peng:2024aiz}.
This freedom, which would not exist in a model where there is no systematic
expansion, is a reflection of the fact that ``exceptional cutoffs'' are indeed
exceptional: convergence is clear except in their immediate vicinity, and
various regulators will give the same extrapolated values within the truncation
uncertainty.
For simplicity, we stay away from these exceptional cutoffs, as our numerical
results show explicitly that a consistent trend persists for cutoff values
sufficiently far from them.
When fitting to Eq.~\eqref{fit_formula}, we exclude $29 \leq
\Lambda/\kappa_{3,0} \leq 34$ for NLO in the unitarity expansion and $26 \leq
\Lambda/\kappa_{3,0} \leq 30$ in the standard expansion.
Around $\Lambda \simeq 54\,\kappa_{3,0}$, one encounters the second exceptional
cutoff for the expansion around the unitarity, whereas in the standard expansion
it occurs at a smaller value.
We stop before reaching these cutoffs as larger momentum cutoffs would
require finer meshes and significantly more computational resources to achieve
convergence for high $l_{\mathrm{max}}$.

Within the considered cutoff range, the NLO radius exhibits good convergence,
providing clear evidence that a four-body force is enough to renormalize the
four-body system at NLO when effective-range corrections are taken into
account~\cite{Bazak:2018qnu}.
Results for NLO in the unitarity expansion are very close to those of the
standard expansion, which reflects the smallness of $a_2^{-1}$ effects,
particularly after the energy is fixed.
The extrapolated values for the \isotope[4]{He} tetramer are given in
Table~\ref{table:rs}.
The combined contributions of the effective range, the four-body force, and the
scattering length account for approximately 75\% of the discrepancy between the
LO radius at unitarity and the corresponding value obtained with the LM2M2
potential~\cite{Hiyama:2011ge}.
The residual difference is within the EFT truncation error and is
expected to be further reduced by higher-order corrections.
Relative to the LO value, the combined NLO contributions amount to about 45\%,
which is of similar size to $\epsilon_{4,0}^r$, the largest expansion parameter
among the three expansion parameters for the four-body ground state.

\subsection{Four-body excited state}
\label{sec:4bex}

The excited state of the \isotope[4]{He} tetramer is close to the threshold for
break-up into a trimer and an atom.
While this can be described well with Halo EFT~\cite{Wu:2023mhg},
here we perform a full \abinitio calculation in the framework of SREFT.
Since this state involves large distance scales, it becomes very difficult to
obtain converged results by solving the FY equations.
The diagrammatic approach provides a more efficient way to study this state.

\begin{figure}[tb!]
\centering
\includegraphics[width=0.95\linewidth]{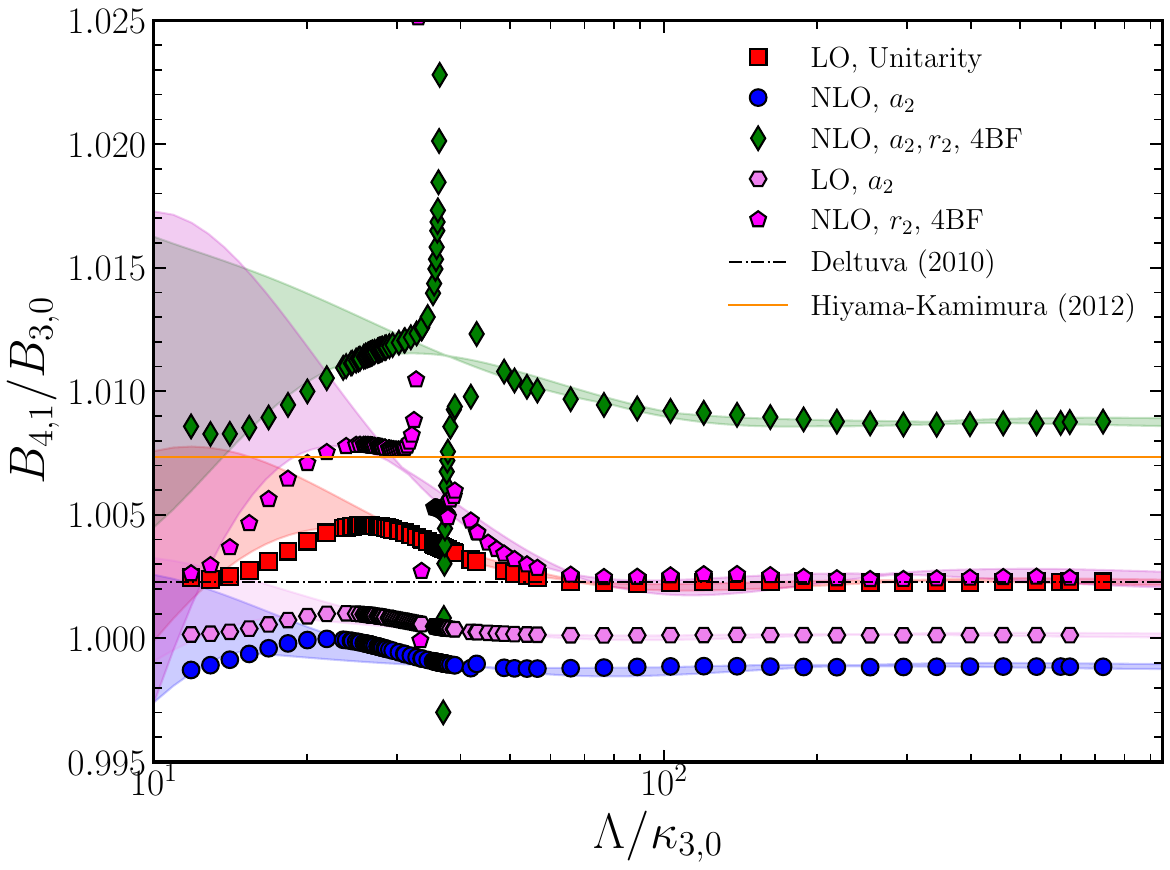}
\caption{Four-body excited-state binding energy $B_{4,1}$ (in units of
 $B_{3,0}$) to NLO as a function of the momentum cutoff $\Lambda$ (in units of
 $\kappa_{3,0}$) obtained in the diagrammatic approach with $l_{\text{max}}=2$.
 Symbols and horizontal lines are as in Figs.~\ref{fig:3b_radii}
 and~\ref{fig:B40_NLO}, and bands as in Fig.~\ref{fig:B40_NLO} with regions
 around NLO ``exceptional cutoffs'' excluded  ($28 \leq \Lambda/\kappa_{3,0}
 \leq 45$ for the green
 diamonds and $28 \leq \Lambda/\kappa_{3,0} \leq 40$  for magenta pentagons).
\label{fig:B41_NLO_DA} }
\end{figure}

The binding energy of the excited state to NLO is shown in
Fig.~\ref{fig:B41_NLO_DA}, in the same units as Fig.~\ref{fig:B40_NLO}.
The bands represent fits over the same cutoff values.
At exact unitarity, we obtain
\begin{equation}
 \left. \frac{B_{4,1}}{B_{3,0}}\right|_{\text{unitarity}} = 1.0022(1)
  \label{eq:B41(0)}
 \end{equation}
and reproduce the universal value~\cite{Deltuva:2010xd}.
For the \isotope[4]{He} tetramer, this translates into a $\sim 0.5$\% difference
to the LM2M2 value~\cite{Hiyama:2011ge}, as seen in Table~\ref{table:Bs}.

%%%%%%%%%%%%%%%%%%%%%%%%%%%%%%%%%%%%%%%%%%%%%%%%%%%%%%%%%%%%%%%%%%%%%%%%%%%%%%%%
%%%%%%%%%%%%%%%%%%%%%%%%%%%%%%%%%%%%%%%%%%%%%%%%%%%%%%%%%%%%%%%%%%%%%%%%%%%%%%%%

At an incomplete NLO where only $a_2^{{-}1}$ is included, the energy is shifted
in the opposite direction compared to the direct value of the LM2M2
potential~\cite{Hiyama:2011ge}.
The correction is only about 0.33\% of the LO value, much smaller than
$\epsilon_{4,1}^a$, suggesting the NLO three-body force provides a relatively
stronger counteracting effect for this state.
This may be related to the halo-like structure of the excited state: the
three-particle core, which provides the dominant contribution to its binding
energy, remains almost unchanged as the NLO three-body force is tuned to
preserve the LO trimer ground-state energy.
Although small in magnitude, this correction reduces the NLO binding energy of
the excited state below that of the trimer, rendering the state unstable at this
order.
Resumming $a_2^{{-}1}$ contributions, as it is done in the standard SREFT
expansion, produces slightly smaller corrections, which are nevertheless
sufficient for stability~\cite{Lin:2023zqw}, although the excited state is then
extremely shallow.
The corresponding value for the \isotope[4]{He} tetramer is given in
Table~\ref{table:Bs}.

As shown in Appendix~\ref{app_sec:convergence}, inclusion of $r_2$ without the
four-body force leads to a binding energy that decreases further, while
oscillating with $\Lambda$, just as for the ground-state energy.
When the four-body force, which is fixed by the four-body ground-state binding
energy, is included, we observe that the binding energy diverges around
exceptional cutoffs, which are located at $\Lambda \simeq 36.7\,\kappa_{3,0}$
for the expansion around the unitarity limit and $\Lambda \approx
33.0\,\kappa_{3,0}$ for the standard expansion.
The divergence in the four-body LEC at the pole of the three-body LEC, located
at $\Lambda \approx 26.8\,\kappa_{3,0}$, does not propagate to the observable.
This is known as a ``factorizable zero''~\cite{Yang:2024yqv,
PavonValderrama:2025zzk} and is not problematic.

Away from the neighborhood of the exceptional cutoff, the oscillation induced by
the inclusion of $r_2$ nearly disappears.
The full NLO corrections move results in the direction of the
LM2M2 potential.
For the expansion around the unitarity limit, the state is not only stable but
lies slightly
below that of the LM2M2 potential.
In the standard expansion the state remains stable at NLO but
its energy differs very little from the unitarity value.
Excluding $28 \leq \Lambda/\kappa_{3,0} \leq 45$ and $28 \leq
\Lambda/\kappa_{3,0} \leq 40$ for, respectively, the unitarity and standard
expansions, the extrapolated full NLO values are displayed in
Table~\ref{table:Bs}.
The NLO corrections are significantly smaller than the expected expansion
parameters $\epsilon_{4,1}^a$, $\epsilon_{4,1}^r$, and $\epsilon_4^{\kappa}$,
indicating that substantial cancellations occur among these contributions for
the four-body excited state.
The difference at NLO between the two expansions, while small
($\sim$ 0.6\%) in absolute scale, is significant relative to the atom-trimer
threshold, indicating that the properties of the halo structure of this state
are sensitive to LO dynamics.

A comparison between the expansion around the unitarity limit
and the standard expansion indicates that the former not only
emphasizes universality, but also provides a
better starting point for the four-body excited state.
Because the $1/Qa_2$ expansion converges, scattering-length
effects can be included at LO within its uncertainty.
However, not doing so generates a state that is more bound at LO, thereby simplifying the NLO calculation.
This provides an example of
``improved action''~\cite{Contessi:2023yoz, Contessi:2024vae,
Contessi:2025xue}, where the input is chosen within uncertainties to provide
a quicker convergence for central values at low orders.

\section{Summary and outlook}
\label{sec:Conclusion}

The unitarity limit serves as a non-trivial and universal reference point for a
wide class of quantum systems characterized by large size and short-range
interactions.
In this limit, multi-body spectra are determined by a single three-body
parameter and discrete scale invariance.
Around this limit, deviations from perfect universality can be systematically
incorporated in a perturbative expansion.
Effective field theory provides a natural language for describing and organizing
these effects in a model-independent way.

In this work, we studied the \isotope[4]{He} trimer and tetramer as benchmark
examples to demonstrate the idea of expansion around the unitarity limit.
Within the framework of Short-Range EFT, we employed both the Faddeev-Yakubovsky
formalism and a diagrammatic approach to tackle the few-body problem.
In the four-body calculations, we carefully removed contributions from deep
trimer states, which are outside the EFT's regime of validity and can complicate
the problem.
This removal is essential and enabled us to safely push the cutoff to large
values, allowing for a cleaner investigation of the convergence behavior and
renormalization structure of the theory.
The subtraction technique is general and can be applied to a broad class of
few-body problems where unwanted or spurious states interfere with the physical
spectrum.

Our analysis targeted two key types of observables: binding energies and
root-mean-square (rms) radii.
Our unitarity results are in agreement with analytical and
numerical calculations available in the literature.
Once the leading-order parameter is fixed to the trimer ground-state energy,
predictions for the tetramer energies and trimer and tetramer ground-state radii
capture the dominant universal features near unitarity.
Corrections from the large two-body scattering length are well
defined by themselves, and their perturbative and nonperturbative treatments
can be compared.
In all cases, their effects are found to be small, and the differences smaller
still.
This finding supports the assignment of the scattering length to higher orders.

Next-to-leading-order corrections incorporate not only the
two-body scattering length, but also the two-body effective range.
As seen in the ground-state trimer radius, the latter yields larger corrections
in the direction of potential-model results.
While the LO theory requires only a three-body force for consistent
renormalization, the inclusion of range corrections at NLO introduces additional
ultraviolet sensitivity in the four-body sector.
This necessitates the inclusion of a four-body force at the same order to absorb
the cutoff dependence and maintain renormalizability~\cite{Bazak:2018qnu}.
``Exceptional cutoffs", which can be traced to vanishing matrix
elements at isolated values of the cutoff, are observed in the four-body
sector, but away from these special points, a clear convergence pattern
persists.
The emergence of the four-body force at NLO and exceptional
cutoffs reflect the rich structure of few-body renormalization and highlights
the necessity of extending power-counting schemes to properly accommodate such
effects.
With the four-body force, a new scale enters the description of
four- and more-body systems.
After determining this scale from the tetramer ground-state energy, NLO
corrections significantly improve the description of tetramer observables.
Although subleading, they are quantitatively important for providing accurate
predictions in real physical systems.

Our results demonstrate that the EFT expansion around the unitarity limit is a
powerful and systematically improvable tool for studying few-body systems.
The methods developed here provide a strong foundation for future work.
Possible directions include extending the analysis to more complex systems, such
as nuclear or heteronuclear clusters, and generalizing the framework to
higher-body sectors.
Additionally, we anticipate that the expansion around unitarity will find
broader applications across a wide range of physical systems, from ultracold
atomic gases to halo nuclei and hadronic molecules.

\begin{acknowledgments}
We thank Martin Sch\"{a}fer and Johannes Kirscher for useful discussions and for 
sharing their results for comparison.
UvK acknowledges insightful discussions with Lucas Madeira and Francesco
Pederiva.
XL, UvK, and SK thank the Institute for Nuclear Theory (INT) at the University
of Washington for its kind hospitality and stimulating research environment.
This research was supported in part by the
INT's U.S.\ Department of Energy grant No.~DE-FG02-00ER41132.
FW and SK furthermore acknowledge the hospitality of the ECT*, where part of
this work was carried out.
The work of XL and SK was supported in part by the U.S.\ National Science
Foundation (Grant No.~PHY--2044632) and by the U.S.\ Department of Energy (Grant
No.~DE-SC0024622).
This material is based upon work supported by the U.S.\ Department of Energy,
Office of Science, Office of Nuclear Physics, under the FRIB Theory Alliance,
Award No.~DE-SC0013617, and under Award No.~DE-FG02-04ER41338.
Computational resources for parts of this work were provided by the
high-performance computing cluster operated by North Carolina State University
and by the Jülich Supercomputing Centre.
\end{acknowledgments}

\section*{Data Availability}

The numerical results shown in this work are openly available as data
files~\cite{Zenodo}.
The program codes used to compute these numerical results are not publicly
available, but they will be made available by the authors upon reasonable
request.

\appendix

\section{Projection method for solving an NLO bound-state equation}
\label{appen:projection_NLO}

For calculating the first-order correction to a bound-state wave function, we
need to solve an inhomogenous equation such as Eq.~\eqref{F_NLO} or
Eq.~\eqref{FY_NLO}.
After discretizing all integrals, it becomes a finite-dimensional inhomogeneous
linear equation, with the complication that the matrix on the left-hand side of
the equation, which involves the LO kernel, is singular.
However, it is singular in a very particular way: the LO state is a solution of
the homogeneous equation.
Since we assume no degeneracy, the relevant matrix is rank-deficient by exactly
1, and we know exactly what the nontrivial kernel is.
In this section, we discuss this problem from a general linear-algebra
perspective.

To this end, consider
\begin{equation}
 A\ket{x} = \ket{b} \,,
\label{eq:Axb}
\end{equation}
with $\ker A = \spn\{\ket{a_0}\}$.
To simplify the reasoning below, we assume that $\norm{a_0}=1$.
We furthermore assume that the problem~\eqref{eq:Axb} is consistent, \ie, that a
solution exists (otherwise the perturbative expansion does not exist).
Then, since $A$ is singular, there exist in fact infinitely many solutions.
Among these, we wish to find the solution $\ket{x_0}$ that is orthogonal to
$\ket{a_0}$, $\braket{x_0|a_0}=0$, as is standard practice in perturbation
theory.
For a complete argument we note that this solution is unique, for if there were
another vector $\ket{x_0'}$ that satisfies $A\ket{x_0'}=\ket{b}$ and
$\braket{x_0'|a_0} = 0$, then
\begin{equation}
 A\Big(\ket{x_0} - \ket{x_0'}\Big) = \ket{b} - \ket{b} = 0 \,,
\end{equation}
which implies $\ket{x_0} - \ket{x_0'} = \alpha\ket{a_0}$ with some $\alpha$.
However, we also know that $(\bra{x_0} - \bra{x_0'})\ket{a_0} = 0$ and therefore
it follows that $\alpha=0$, \ie, $\ket{x_0} - \ket{x_0'} = 0$.
In short, we are using the fact that $\ker A$ is one-dimensional and orthogonal
to the row space.

Numerically, the singular problem~\eqref{eq:Axb} is solved by finding a
least-squares solution with minimum norm, which can be written as
\begin{equation}
 \ket{x_0} \in \argmin_{\ket{x}} \Big[
 \norm{A\ket{x}-\ket{b}} \Big] \cap \argmin_{\ket{x}} \norm{{x}} \,.
\label{eq:x0-argmin}
\end{equation}
The claim is that there is exactly one such $\ket{x_0}$, and that $\ket{x_0}$ is
the unique solution of Eq.~\eqref{eq:Axb} orthogonal to $\ket{a_0}$.
This can be shown as follows:
\begin{enumerate}
 \item We already showed that a solution of Eq.~\eqref{eq:Axb} is uniquely
  determined by the requirement that it is orthogonal to $\ket{a_0}$.
 \item We also know, by the consistency assumption, that a solution exists.
  That implies the minimum value attained by the first term in
  Eq.~\eqref{eq:x0-argmin} is zero and, consequently, we only need to show that
  the particular solution $\ket{x_0}$ that is orthogonal to $\ket{a_0}$
  minimizes $\norm{{x}}$ among all possible solutions.
 \item To see this, let $\ket{x}$ be any solution of Eq.~\eqref{eq:Axb} and let
  $\lambda = \braket{x|a_0}$.
  If $\lambda = 0$, then $\ket{x} = \ket{x_0}$ by the established uniqueness of
  $\ket{x_0}$. Otherwise, if $\lambda \neq 0$, then $\ket{x} - \lambda\ket{a_0}
  = \ket{x_0}$.
  From the Cauchy-Schwarz inequality we get
  \begin{equation}
   \braket{x|x_0} \leq \norm{{x}}\cdot\norm{{x_0}} \,,
  \end{equation}
  but also it holds that
  \begin{equation}
   \braket{x|x_0} = \braket{x+\lambda a_0|x_0} = \braket{x_0|x_0}
   = \norm{{x_0}}^2 \,.
  \end{equation}
  Together it then follows that $\norm{{x_0}} \leq \norm{{x}}$ for any other
  solution $\ket{x}$.
\end{enumerate}

While the least-squares method is perfectly suited to solve for the NLO wave
function, we can unfortunately not easily use it in a situation where we do not
have the relevant operator $A$ represented as an explicit matrix.
That is, we are looking for methods where only the result of applying $A$ to
some vector $\ket{x}$ is required, \ie, iterative solvers.
Such solvers for linear systems are readily available, but in order to solve a
least-squares problem (as opposed to a linear system with non-singular $A$), one
generally needs not only $A$ applied to $\ket{x}$, but also its transpose,
$A^T$, applied to $\ket{x}$, and in our numerical approach we unfortunately
cannot easily evaluate $A^T\ket{x}$.

However, there is an alternative technique we can use. With the same definitions
as introduced above, let
\begin{equation}
 \tilde{A} = A + \beta \ket{a_0}\bra{a_0}
 \label{eq:Axb-tilde}
\end{equation}
and note that for any $\beta \neq 0$, $\tilde{A}$ is non-singular by
construction.
Therefore the modified problem
\begin{equation}
 \tilde{A}\ket{x} = \ket{b}
\end{equation}
has a unique solution.
If we consider our $\ket{x_0}$ as the unique solution of Eq.~\eqref{eq:Axb} that
is orthogonal to $\ket{a_0}$, then
\begin{equation}
 \tilde{A}\ket{x_0} = A\ket{x_0} + \beta \braket{a_0|x_0}\ket{a_0}
 = A\ket{x_0} = \ket{b} \,,
\end{equation}
and therefore $\ket{x_0}$ solves also Eq.~\eqref{eq:Axb-tilde}.
Since that equation has a \emph{unique} solution, we can numerically solve
Eq.~\eqref{eq:Axb-tilde} instead of Eq.~\eqref{eq:Axb}, using any iterative
solver that only needs the result of $\tilde{A}$ applied to a vector, and obtain
in this way the desired solution $\ket{x_0}$ of Eq.~\eqref{eq:Axb}.

\section{More on the diagrammatic approach}
\label{app:diagrammatic}

In this appendix we offer more details on the solution of the equations of the
diagrammatic approach~\cite{Lin:2023zqw}.

\subsection{Rewriting the integral equation and including the four-body force}

Defining
\begin{equation}
 \ket{\Gamma'_3} \equiv (1-K_{33})\ket{\Gamma_3} \,,
\end{equation}
Eq.~\eqref{4bodydiageqn-opform} can be written as
\begin{align}
 \label{4bodydiageqn-opform-app2}
 &K'
 \begin{pmatrix}
  \ket{\Gamma'_3} \\
  \ket{\Gamma_2}
 \end{pmatrix}
 =
 \begin{pmatrix}
  \ket{\Gamma'_3} \\
  \ket{\Gamma_2}
 \end{pmatrix}\,,
\end{align}
where
\begin{equation}
 K' =
 \begin{pmatrix}
  P_{34}K_{33}(1-K_{33})^{-1} & (1+P_{34})K_{32} \\
  (1+P_{d})K_{23}(1-K_{33})^{-1}&  (1+P_{d})K_{22}
 \end{pmatrix}\,.
\end{equation}
Now we define
\begin{equation}
 \begin{pmatrix}
  \ket{\widetilde{\Gamma}_3} \\
  \ket{\widetilde{\Gamma}_2}
 \end{pmatrix} \equiv
 \begin{pmatrix}
  1 & -K_{32}\\
  0&  1
 \end{pmatrix}
 \begin{pmatrix}
  \ket{\Gamma'_3} \\
  \ket{\Gamma_2}
 \end{pmatrix}
\end{equation}
and plug it into Eq.~\eqref{4bodydiageqn-opform-app2}, yielding
\begin{align}
 \widetilde{K}\begin{pmatrix}
  \ket{\widetilde{\Gamma}_3} \\
  \ket{\widetilde{\Gamma}_2}
  \end{pmatrix} = \begin{pmatrix}
  \ket{\widetilde{\Gamma}_3} \\
  \ket{\widetilde{\Gamma}_2}
 \end{pmatrix}\,,
\end{align}
where
\begin{align}
 \widetilde{{K}} =  {K}'\begin{pmatrix}
  1 & K_{32}\\
  0&  1
  \end{pmatrix} - \begin{pmatrix}
  0 & K_{32}\\
  0&  0
 \end{pmatrix}\,.
\end{align}
It is straightforward to show that this yields the expressions in
Eqs.~\eqref{4bodydiageqn2-opform} and~\eqref{4bodydiageqn2-kerneldef-1}
through~\eqref{4bodydiageqn2-kerneldef-4}.

At NLO there are contributions from the four-body force in
Fig.~\ref{fig:4bodydiagrams-w4BF} expressed in Eqs.~\eqref{eq:4BFinDA}
and~\eqref{eq:4BFinDAeffect}.
Adding them to Eq.~\eqref{4bodydiageqn-opform} and including permutations gives
\begin{align}
 \begin{pmatrix}
  1+P_{34} & 0 \\
  0&  1+P_{d}
 \end{pmatrix}
 \begin{pmatrix}
  K^{\text{w4BF}}_{33}  & K^{\text{w4BF}}_{32} \\
  K^{\text{w4BF}}_{23} &  K^{\text{w4BF}}_{22}
 \end{pmatrix}
 \begin{pmatrix}
  \ket{\Gamma_3} \\
  \ket{\Gamma_2}
 \end{pmatrix}
 =
 \begin{pmatrix}
  \ket{\Gamma_3} \\
  \ket{\Gamma_2}
 \end{pmatrix} \,.
\end{align}
Following the same procedure as before, we introduce
\begin{align}
 \begin{pmatrix}
  \ket{\widetilde{\Gamma}_3} \\
  \ket{\widetilde{\Gamma}_2}
 \end{pmatrix} \equiv
 \begin{pmatrix}
  1-K^{\text{w4BF}}_{33} & -K^{\text{w4BF}}_{32}\\
  0&  1
 \end{pmatrix}
 \begin{pmatrix}
  \ket{\Gamma_3} \\
  \ket{\Gamma_2}
 \end{pmatrix}
\end{align}
to obtain an equation analogous to Eq.~\eqref{4bodydiageqn2-opform} in the
presence of the four-body force,
\begin{align}
\label{4bodydiageqn2-opform-w4BF}
 \begin{pmatrix}
  P_{34} & 0\\
  0&  1+P_{d}
 \end{pmatrix}
 \begin{pmatrix}
  \widetilde{K}^{\text{w4BF}}_{33} & \widetilde{K}^{\text{w4BF}}_{32}\\
  \widetilde{K}^{\text{w4BF}}_{23}&  \widetilde{K}^{\text{w4BF}}_{22}
 \end{pmatrix}
 \begin{pmatrix}
  \ket{\widetilde{\Gamma}_3} \\
  \ket{\widetilde{\Gamma}_2}
 \end{pmatrix}
 =
 \begin{pmatrix}
  \ket{\widetilde{\Gamma}_3} \\
  \ket{\widetilde{\Gamma}_2}
 \end{pmatrix}\,,
\end{align}
with the kernels in Eqs.~\eqref{4bodydiageqn2-kerneldef-w4BF-1}
through~\eqref{4bodydiageqn2-kerneldef-w4BF-4}.
Equation~\eqref{4bodydiageqn2-opform-w4BF} can be cast into a similar form as
Eq.~\eqref{4bodydiageqn3-opform} by eliminating the $\ket{\widetilde{\Gamma}_3}$
component, resulting in Eqs.~\eqref{4bodydiageqn3-opform-w4BF-main}
and~\eqref{K_tilde_w4BF-main}.
Note that these expressions include the four-body force non-perturbatively.
The perturbative approach as discussed in Sec.~\ref{sec:diagrammatic-NLO}
requires expanding Eqs.~\eqref{4bodydiageqn2-opform-w4BF}
and~\eqref{K_tilde_w4BF-main}.

\subsection{Subtracting deep trimer poles}

Near its pole at $E = -B_3$, the boson-dimer amplitude in Fig.~\ref{fig:t3body}
can be written as (see, \eg, Refs.~\cite{Ji:2011qg,Lin:2023zqw})
\begin{spliteq}
\label{t-near-pole}
 t_3(E)
 &= \frac{R_3}{E+B_3 + \ii\epsilon} + \cdots \\
 &= R_3\left({-}\ii\pi \delta(E + B_3)
 + \text{PV} \frac{1}{E + B_3}\right) + \cdots \,,
%\raisetag{1.5em}
\end{spliteq}
where $E$ is the boson-dimer center-of-mass energy and $R_3$ is the residue of
$t_3(E)$ at the pole.
The dependence of the amplitude on the incoming and outgoing momenta between the
dimer and single boson is suppressed.
The residue, $R_3$, can be computed purely numerically or semi-analytically.
To subtract the pole in a four-body calculation, one can keep the principal
value (PV) in Eq.~\eqref{t-near-pole} but drop the term ${-}\ii\pi \delta(E +
B_3)$.
This process should be repeated for each pole of $t_3(E)$ if it contains
multiple poles.
Doing this subtraction, for example at some tetramer binding energy $B_4$,
amounts to removing from the spectrum the state made of the trimer (with a
binding energy $B_3$) plus a single boson spectator with a total energy
\begin{align}
 \label{trimer-pole-kinematic}
 {-}B_3 + \frac{2p_0^2}{3m} = {-}B_4 \,,
\end{align}
where $p_0$ is the relative momentum between the trimer and the single boson
spectator, and $2p_0^2/3m$ is the kinetic energy.
Using this method, one removes only this single state, into which the tetramer
could otherwise decay, which is sufficient to retain the tetramer as a bound
state.
This is in contradistinction to the pseudopotential projection method where
\emph{all} continuum states made of a trimer plus a spectator boson are pushed
to infinitely high energies as $\eta\to\infty$ in Eq.~\eqref{pseudopotential}.
Nevertheless, as a deep trimer is by definition well above the EFT cutoff, the
impact of subtracting either one or more states of this sort on the low-energy
physics should be equivalent, from an EFT point of view, up to higher-order
corrections.

At NLO the three-body force receives a correction and can be fitted to one of
the trimer poles of $t_3(E)$.
If a trimer binding energy, $B_3$, is used to fit the three-body force at both
LO and NLO, then this $B_3$ receives no NLO correction and for this pole the
subtraction used in Eq.~\eqref{t-near-pole} is still sufficient at NLO.
However, if $t_3(E)$ contains multiple trimer poles, the binding energies of
those poles that are not used to fit the NLO three-body force typically receive
non-zero NLO corrections.
These corrections contain double poles when treated perturbatively in $t_3(E)$.
Mathematically this means
\begin{spliteq}
\label{t-near-pole-NLO}
 t_3(E)
 &= \frac{R_3}{E+B_3 + \delta B_3 + \ii\epsilon} + \cdots \\
 &= \frac{R_3}{E+B_3+ \ii\epsilon} - \frac{R_3}{E+B_3+ \ii\epsilon}
 \frac{\delta B_3}{E+B_3+ \ii\epsilon} + \cdots \,,
\end{spliteq}
where $\delta B_3$ is a perturbative correction to $B_3$.
In practice it is $p_0$ in Eq.~\eqref{trimer-pole-kinematic} that receives an
NLO correction.
In an NLO four-body calculation, such poles in the NLO correction to the
boson-dimer amplitude, $t_3(E)$, must be addressed to avoid numerical
instability.
This can be done conveniently by directly including the perturbation $\delta
B_3$ when computing the principal value of the corresponding simple poles at LO,
and the double poles at NLO are taken care of by the automated chain rule.
For example,
\begin{multline}
 \int_a^{b} \dd E\, \frac{1}{E+B_3 + \delta B_3 + \ii\epsilon}
 = \ln\left(
  \frac{b + B_3 + \delta B_3+ \ii\epsilon}{a+B_3 + \delta B_3+ \ii\epsilon}
 \right) \\
 =\ln\left(\frac{b + B_3+\ii\epsilon}{a+B_3+ \ii\epsilon} \right)
 + \delta B_3 \frac{a-b}{(b+B_3)(a+B_3)} + \cdots \,.
\end{multline}
The first term on the last line implicitly contains an imaginary piece,
${-}\ii\pi$, for $a + B_3 <0$ and $b+B_3>0$.
This term corresponds to the singular piece in Eq.~\eqref{t-near-pole} and is
dropped by taking the absolute value of the argument in the logarithm.
The second term on the last line corresponds to the double pole in
Eq.~\eqref{t-near-pole-NLO} and is always finite in this calculation.

%%%%%%%%%%%%%%%%%%%%%%%%%%%%%%%%%%%%%%%%%%%%%%%%%%%%%%%%%%%%%%%%%%%%%%%%%%%%%%%%
\begin{figure*}[tb!]
\centering
\includegraphics[width=0.45\linewidth]{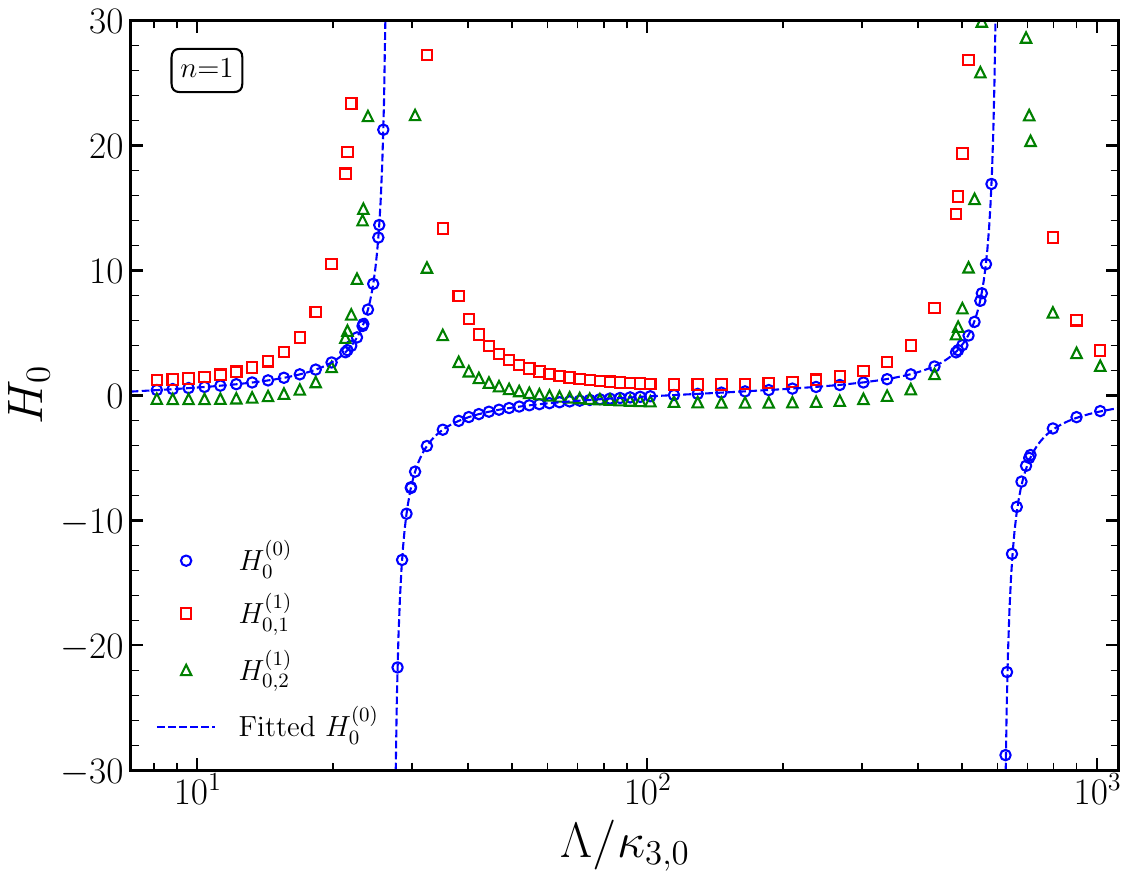}\hfil
\includegraphics[width=0.45\linewidth]{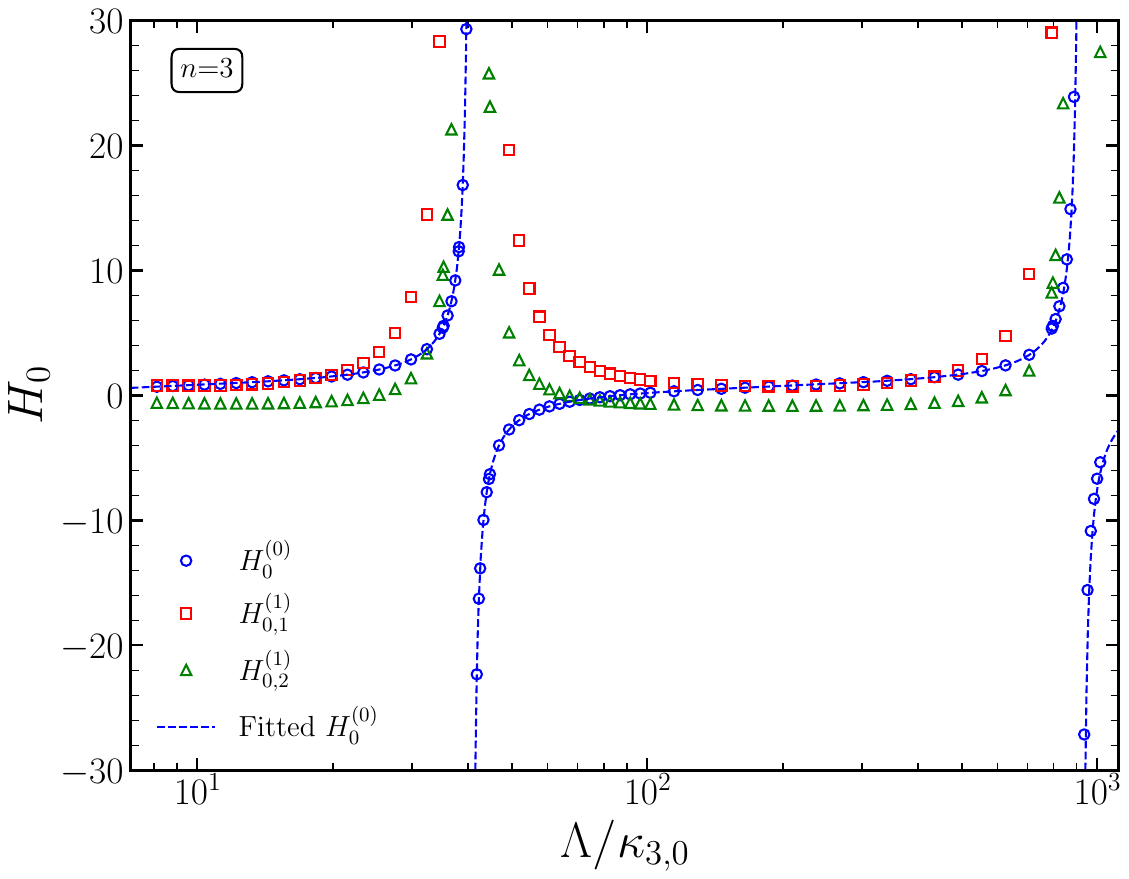}
\caption{Same as the left panel of Fig.~\ref{fig:H0_reg4} but for the standard
  Gaussian ($n=1$, left panel) and the sextic super-Gaussian ($n=3$, right
 panel).
 \label{fig:H0_reg2_6}
}
\end{figure*}
%%%%%%%%%%%%%%%%%%%%%%%%%%%%%%%%%%%%%%%%%%%%%%%%%%%%%%%%%%%%%%%%%%%%%%%%%%%%%%%%

\section{Regulator dependence}
\label{app:reg_dep}

The exact form of the running of the LECs depends on the regulator function, and
so do results for observables at small cutoff values.
In the main text, results of the FY approach were given using
the quartic super-Gaussian regulator.
Here we illustrate regulator dependence with the standard Gaussian and sextic
super-Gaussian regulators.

The dependence of the two-body LECs with the regulator is manifest in the
regulator factors $\theta_i$ introduced in Sec.~\ref{sec:MethodsFY}.
Figure~\ref{fig:H0_reg2_6} plots the running of the three-body LEC $H_0$ at LO
and NLO as a function of the cutoff for the standard Gaussian and sextic
super-Gaussian regulators.
Comparison with Fig.~\ref{fig:H0_reg4} for the quartic super-Gaussian shows the
same qualitative behavior.
The LO results are fitted with Eq.~\eqref{H0_0_expr}, where the parameters
$\{b_0, \delta_0, h_0\} \simeq $ $\{2.3965, 0.7094, 0.7976\}$ and $\{2.5985,
0.3766, 1.0037\}$ for the standard Gaussian and sextic super-Gaussian,
respectively.
The fit is again excellent, confirming the validity of Eq.~\eqref{H0_0_expr} for
separable regulators~\cite{Chen:2025rti}.
At NLO, differences in the two- and three-body LECs lead to corresponding
differences in the running of the four-body LEC $F_{0}^{(1)}$, as
shown in Fig.~\ref{fig:F0_reg4_DA_log} in the main text.

\begin{figure}[tb!]
\centering
\includegraphics[width=0.95\linewidth]{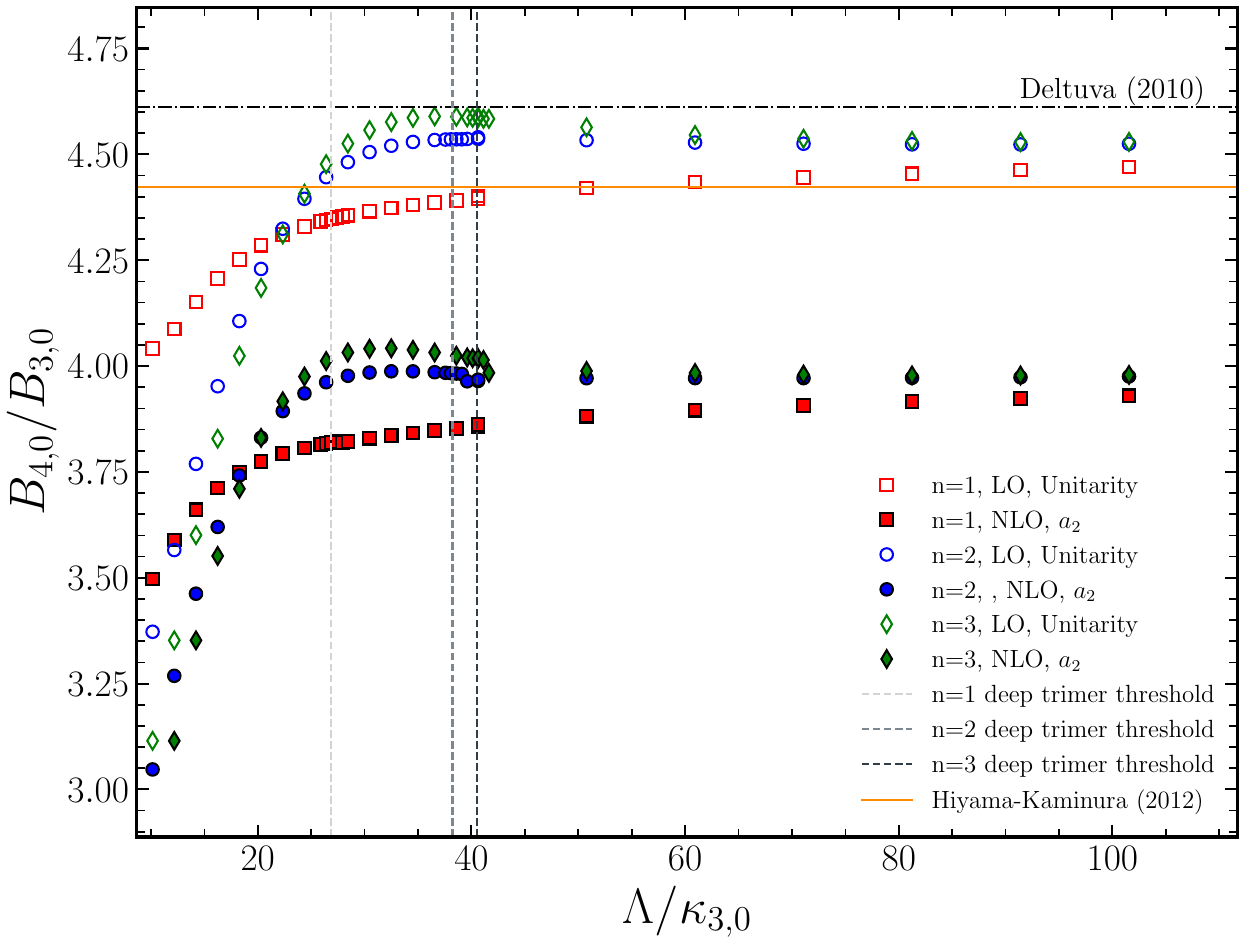}
\caption{Four-body ground-state binding energy $B_{4,0}$ (in units of $B_{3,0}$)
 as a function of the momentum cutoff $\Lambda$ (in units of $\kappa_{3,0}$)
 obtained by solving the FY equations with $l_{\text{max}}=2$ using different
 regulators.
 Red squares, blue circles, and green diamonds are the LO results at unitarity
 with the standard ($n=1$), quartic ($n=2$), and sextic ($n=3$) (super-)Gaussian
 regulators, respectively.
 Full points with the same color and shape are the corresponding (incomplete)
 NLO results when only the scattering length is included.
 The vertical lines mark the deep-trimer thresholds for different regulators,
 which appear at larger $\Lambda$ as $n$ increases.
 The horizontal black dashed and orange solid lines are the results at unitarity
 from Ref.~\cite{Deltuva:2010xd} and for the LM2M2 potential from
 Ref.~\cite{Hiyama:2011ge}, respectively.
 \label{fig:B40_NLO_regs}
}
\end{figure}

In observables, the arbitrariness in the choice of regulator decreases as the
cutoff $\Lambda$ is increased.
However, significant differences are seen at low cutoff values.
As an illustration, Fig.~\ref{fig:B40_NLO_regs} shows the four-body ground-state
binding energy at LO and incomplete NLO for $\Lambda \lesssim 100
\,\kappa_{3,0}$ obtained by solving the FY equations with different regulator
functions and $l_{\text{max}}=2$.
Results from different regulator functions converge to the same values, as they
should.
However, this convergence is evident only beyond the region of cutoffs where the
first deep trimer appears.
Depending on the regulator function, an extrapolation on the basis of points
below the first deep-trimer threshold could be deceiving.
This highlights the benefits of the deep-trimer removal we perform in
order to reach the regime of relatively large cutoff values.

%%%%%%%%%%%%%%%%%%%%%%%%%%%%%%%%%%%%%%%%%%%%%%%%%%%%%%%%%%%%%%%%%%%%%%%%%%%%%%%%
%%%%%%%%%%%%%%%%%%%%%%%%%%%%%%%%%%%%%%%%%%%%%%%%%%%%%%%%%%%%%%%%%%%%%%%%%%%%%%%%

Renormalization ensures that any observable $O^{[\nu]}(\Lambda)$
at order $\nu$ converges to a finite value $O^{[\nu]}$ as the cutoff increases.
The $\Lambda$ dependence must have the form
\begin{spliteq}
 O^{[\nu]}(\Lambda) = O^{[\nu]}\Bigg[ &
  1 + \frac{\kappa_{3,0}}{\Lambda}\, F_1^{(\nu)}(\Lambda/\Lambda_\star) \\
  &+ \left(\frac{\kappa_{3,0}}{\Lambda}\right)^2
  F_2^{(\nu)}(\Lambda/\Lambda_\star) + \cdots
 \Bigg] \,,
\label{fit_formula_full}
\end{spliteq}
where $F_i^{(\nu)}(x)$ are bounded functions, except perhaps for isolated points
(``exceptional cutoffs'').

In the two-body unitarity limit, DSI requires that the functions $F_i^{(0)}(x)$
be invariant when $x\to \exp(\pi/s_0)x$ or, alternatively, when $2s_0 \ln x$
changes by multiples of $2\pi$.
We can thus expand in a Fourier series,
\begin{equation}
 F_i^{(\nu)}(x) = \gamma_{i0}^{(\nu)}
 + \sum_{l=1}^\infty \gamma_{il}^{(\nu)}
 \sin\left(2 l s_0 \ln x +\phi_{il}^{(\nu)} \right) \,,
\label{F}
\end{equation}
where $\gamma_{i0}^{(\nu)}$, $\gamma_{il}^{(\nu)}$, and $\phi_{il}^{(\nu)}$
are real numbers.

Unless $\gamma_{il}^{(\nu)}\ll \gamma_{i0}^{(\nu)}$, we should
see oscillations as $\Lambda$ increases, as indeed is the case for the
four-body ground-state energy in Fig.~\ref{fig:B40_NLO}.
We do not have an \apriori argument to neglect higher harmonics, but fits of
our results with both the $l= 1$ and 2 harmonics
indicate that $\gamma_{i2}^{(\nu)}$ is normally smaller than
$\gamma_{i1}^{(\nu)}$.
Since the number of parameters in Eq.~\eqref{F} increases rapidly, we settle for
fits of the form of Eq.~\eqref{fit_formula} of the main text.
We see they reproduce the cutoff dependence well over a large range of cutoff
values.
The values of the parameters for selected fits and
observables are listed in Table~\ref{table:fit_paras}.
As expected, $c_{0\nu}$ and $c_{1\nu}$ are normally no larger than $\OO(1)$.

%%%%%%%%%%%%%%%%%%%%%%%%%%%%%%%%%%%%%%%%%%%%%%%%%%%%%%%%%%%%%%%%%%%%%%%%%%%%%%%%
\begin{table}[!tb]
\setlength{\extrarowheight}{1.2pt}
\caption{Values of $O^{[\nu]}$, $c_{0\nu}$, $c_{1\nu}$, and $\Lambda_{1\nu}$
 in Eq.~\eqref{fit_formula} for various observables, determined by fitting to
 the full set of calculated data for the expansion around the unitarity limit.
 $\nu =1$ denotes the full NLO result.
 ``FY'' and ``DA''  refer to the Faddeev-Yakubovsky equations and the
 diagrammatic approach, respectively.
 \label{table:fit_paras}
}
\begin{center}
\begin{tabular}{c c c c c c}
\hline
\hline
 & $\nu$ & $O^{[\nu]}$ & $c_{0\nu}$ & $c_{1\nu}$
 & $\Lambda_{1\nu}/\kappa_{3,0}$ \\
 \hline
 $\kappa_{3,0}r_{3,0}$& 0  & 0.47 & 0.76  & $-$0.056 & 6.5  \\
 & 1 & 0.59 &0.83  & 0.32 &8.5 \\
 $B_{4,0}/B_{3,0}$ & 0 {\scriptsize (FY)} &4.64  & $-$2.0 &1.03 & 18.3 \\
 & 0 {\scriptsize (DA)} &4.61  &0.16  &0.61 &14.1 \\
 $\kappa_{3,0}r_{4,0}$ & 0 & 0.34 &1.96 & 1.41 & 4.2 \\
 & 1 & 0.47 & 0.54 & 1.34 & 7.3 \\
 $B_{4,1}/B_{3,0}$ & 0 & 1.0023 & 0.007 & 0.051 & 13.8 \\
 & 1 & 1.0086 & 0.048 & 0.063 &22.0 \\
 \hline
 \hline
\end{tabular}
\end{center}
\end{table}
%%%%%%%%%%%%%%%%%%%%%%%%%%%%%%%%%%%%%%%%%%%%%%%%%%%%%%%%%%%%%%%%%%%%%%%%%%%%%%%%

\section{Dependence of observables on $\eta$}
\label{app:eta_dep}

The pseudopotential introduced in Eq.~\eqref{pseudopotential} in order to
remove deep trimer states from the spectrum in the FY approach involves an
unphysical parameter $\eta$, on which physical observables should not depend.
With this pseudopotential, the full three-body Green's function at
LO becomes
\begin{spliteq}
 \tilde{\mathcal{G}}_3^{(0)}(E) = \mathcal{G}_3^{(0)}(E)
 &- \sum_{i=1}^{N_3} \frac{ \ket{ \Psi_{3,{-}i}^{(0)} }
 \bra{ \Psi_{3,-i}^{(0)}} }{ (E+B_{3,{-}i}^{(0)})^{2} } \\
 &\times \left(
  \frac{1}{E+B_{3,-i}^{(0)}} - \frac{1}{\eta B_{3,{-}i}^{(0)}}
 \right)^{\!{-}1} \,,
\label{Green3}
\end{spliteq}
where $\mathcal{G}_3^{(0)}(E)$ is the three-body Green's function with the
original LO three-body potential, \ie, without the $V_{P,3}$ term.
As $\eta$ goes to infinity, the deep trimers are removed by moving them
to infinite positive energy.
In practice, it is not necessary to take this exact limit as long as
$\eta$ is large enough to move the deep trimer states far away from the physical
region of interest.
In this scenario, it is then expected to find a residual dependence on
$\eta$ that behaves like $1/\eta$.
In the following, we verify this expectation through explicit numerical
calculations, for specifically the removal of the
first deep trimer.

With the original LO three-body potential, the first deep trimer
appears as the cutoff passes a threshold, and its binding energy approaches
$B_{3,{-}1}^{(0)} \simeq 515 \, B_{3,0}$ as the cutoff increases further.
Since $B_{4,0}^{(0)}$ is only a few times
larger than $B_{3,0}$,
the deep trimer, although pushed up by the
pseudopotential, remains below the four-body ground state in the region $\eta
\lesssim$ 0.99.
For $\eta>$ 1, the deep trimer is shifted above the four-particle threshold,
rendering the four-body ground state stable and enabling the application of
bound-state formalism to this state.
We illustrate $\eta>1$ effects with the quartic super-Gaussian regulator in the
FY equations for a cutoff above the first deep-trimer threshold, $\Lambda =
50.77\kappa_{3,0}$.

%%%%%%%%%%%%%%%%%%%%%%%%%%%%%%%%%%%%%%%%%%%%%%%%%%%%%%%%%%%%%%%%%%%%%%%%%%%%%%%%
\begin{figure}[tb!]
\centering
\includegraphics[width=0.95\linewidth]{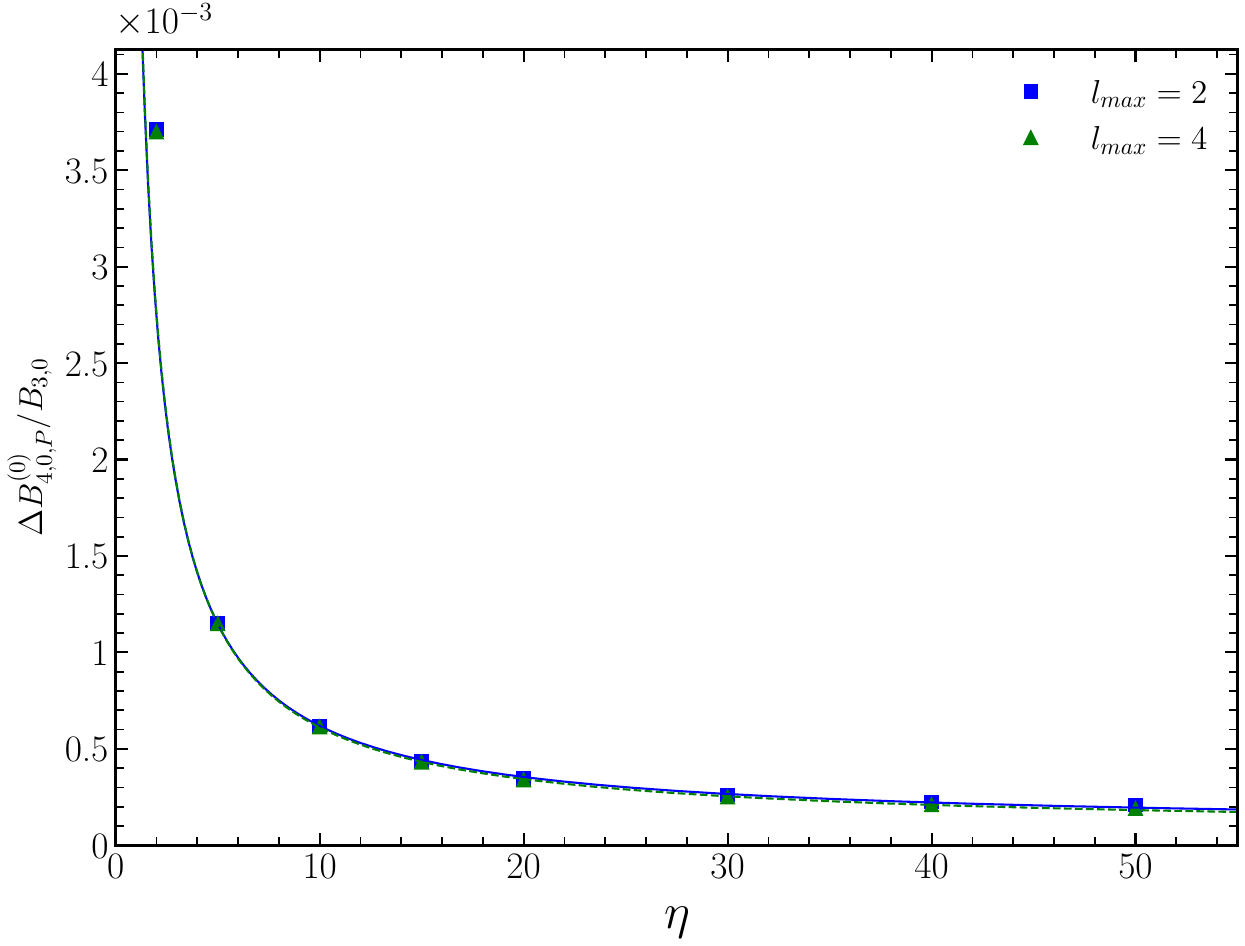}
\caption{
 Shift $\Delta B_{4,0,P}^{(0)}$ in the four-body ground-state energy due to the
 pseudopotential~\eqref{pseudopotential} (in units of $B_{3,0}$) as a function
 of the dimensionless projection factor $\eta$, for a quartic super-Gaussian
 regulator in the FY equations with $\Lambda = 50.77\kappa_{3,0}$.
 Blue squares and green triangles are the results for $l_{\text{max}}=$ 2 and 4,
 respectively.
 The blue solid and green dashed lines are determined by fitting the data points
 of $l_{\text{max}}=$ 2 and 4, respectively, to Eq.~\eqref{B40P_fit}.
 \label{fig:VP3_ME_eta}
}
\end{figure}
%%%%%%%%%%%%%%%%%%%%%%%%%%%%%%%%%%%%%%%%%%%%%%%%%%%%%%%%%%%%%%%%%%%%%%%%%%%%%%%%

%%%%%%%%%%%%%%%%%%%%%%%%%%%%%%%%%%%%%%%%%%%%%%%%%%%%%%%%%%%%%%%%%%%%%%%%%%%%%%%%
\begin{figure}[tb!]
\centering
\includegraphics[width=0.95\linewidth]{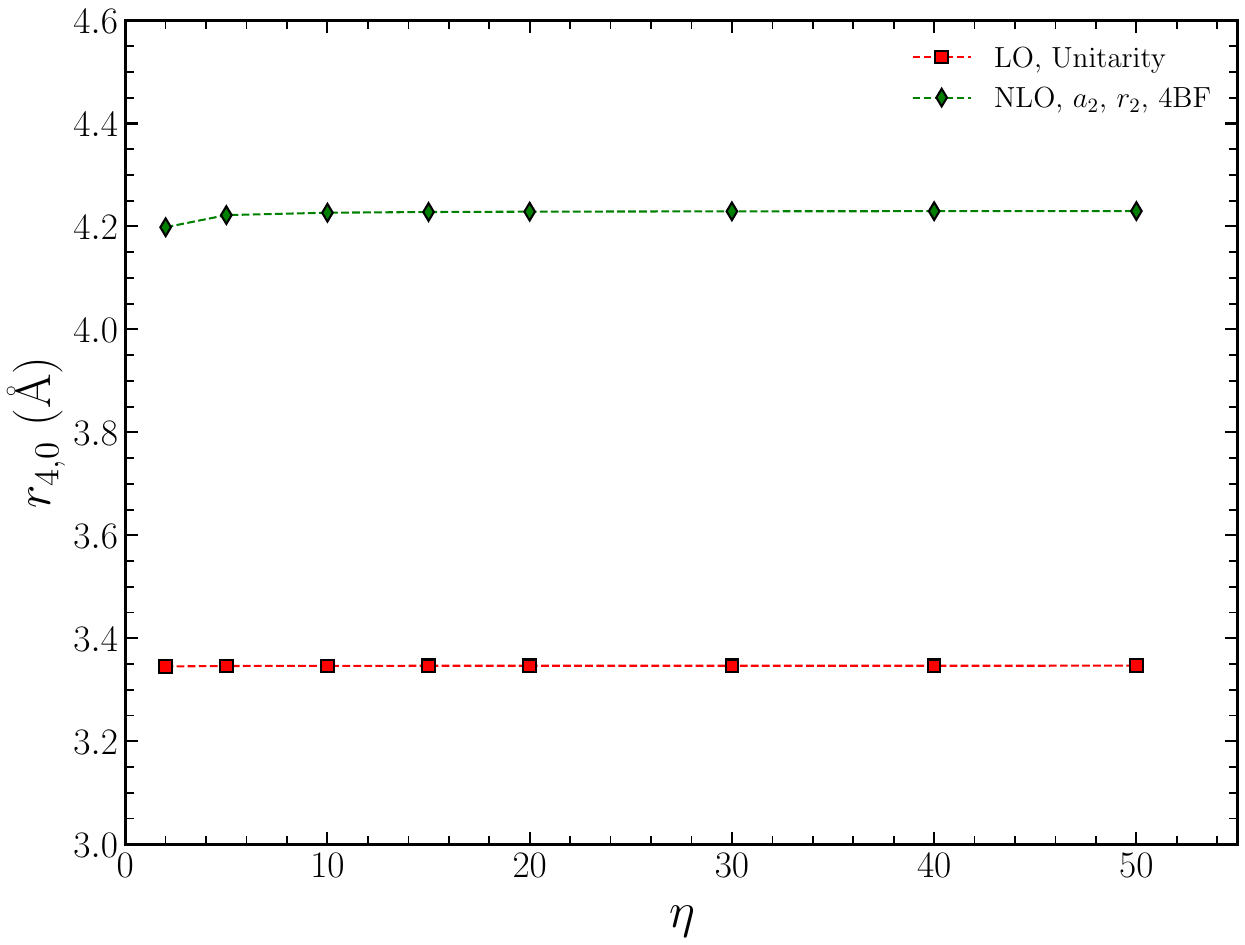}
\caption{Ground-state radius $r_{4,0}$ (in \AA) of the \isotope[4]{He} tetramer as a
 function of the dimensionless projection factor $\eta$, for a quartic
 super-Gaussian regulator in the FY equations with $\Lambda =
 50.77\kappa_{3,0}$.
 Red squares and green diamonds correspond to LO and full NLO, respectively, for
 $l_{\text{max}}=2$.
 \label{fig:r40_eta_reg4}
}
\end{figure}
%%%%%%%%%%%%%%%%%%%%%%%%%%%%%%%%%%%%%%%%%%%%%%%%%%%%%%%%%%%%%%%%%%%%%%%%%%%%%%%%

The contribution of the pseudopotential to the LO four-body binding energy,
\begin{equation}
 \Delta B_{4,0,P}^{(0)}(\eta)
 = 4\braket{{\Psi}_{4,0}^{(0)} | V_{P,3} | {\Psi}_{4,0}^{(0)} } \,,
\end{equation}
where ${\Psi}_{4,0}^{(0)}$ is the LO four-body ground-state wave function
obtained with the pseudopotential~\eqref{pseudopotential}, must be finite
as $\eta$ increases.
In Fig.~\ref{fig:VP3_ME_eta}, the shift $\Delta B_{4,0,P}^{(0)}(\eta)$ is
plotted as a function of $\eta$.
The decrease with $\eta$ is clearly seen and can be quantified by a fit
\begin{equation}
 \frac{\Delta B_{4,0,P}^{(0)} (\eta)}{B_{3,0}} = d_1 + \frac{d_2}{\eta} \,.
\label{B40P_fit}
\end{equation}
We find $\{d_1, d_2\}$ = \{0.0087, 0.514\} and \{0.0073, 0.520\} for
$l_{\text{max}} =$ 2 and 4, respectively.
Since $d_1$ is so small, the energy shift induced by the pseudopotential can be safely neglected as $\eta$ goes to infinity.
In practical calculations, we use $\eta$ sufficiently large so that the residual
contribution becomes negligible compared to the truncation error.

The $\eta$ dependence of other observables are also expected to be mild.
This is confirmed in Fig.~\ref{fig:r40_eta_reg4}, where we present the
ground-state radius of the \isotope[4]{He} tetramer at LO and NLO for
$l_{\text{max}} = 2$.
The radius in this region becomes nearly independent of $\eta$, as desired.

\section{Convergence with maximal angular momentum and special cutoffs}
\label{app_sec:convergence}

%%%%%%%%%%%%%%%%%%%%%%%%%%%%%%%%%%%%%%%%%%%%%%%%%%%%%%%%%%%%%%%%%%%%%%%%%%%%%%%%
\begin{figure}[tb!]
\centering
\includegraphics[width=0.95\linewidth]{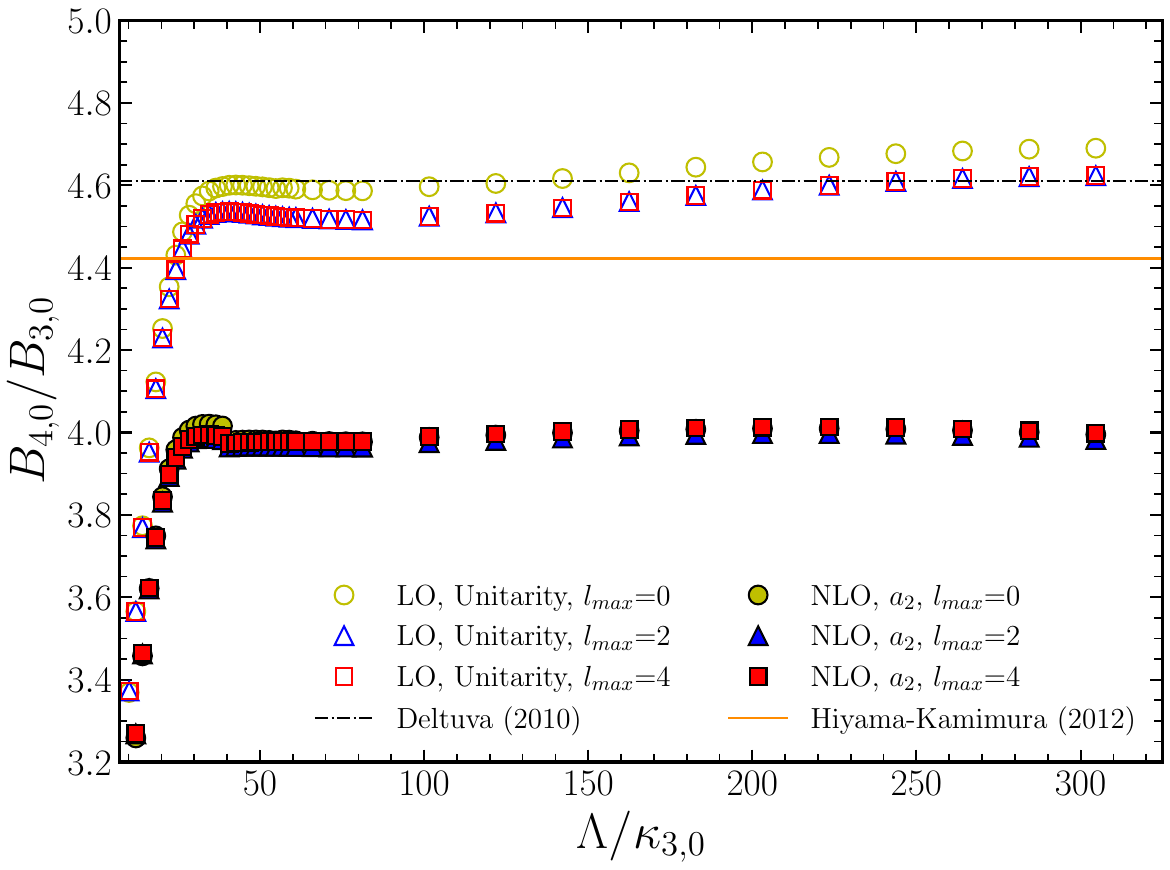}
\caption{
 Tetramer ground state binding energy as a function of the regulator $\Lambda$
 (in units of $\kappa_{3,0}$) obtained by solving the FY equations with the
 quartic super-Gaussian regulator.
 Yellow circles, blue triangles, and red squares denote LO results at unitarity
 for $l_{\text{max}}=$ 0, 2 and 4, respectively.
 The full points are the corresponding NLO values with $a_2$ only.
 The horizontal black dash-dotted line and the orange solid line are the results
 at unitarity from Ref.~\cite{Deltuva:2010xd} and for the LM2M2 potential from
 Ref.~\cite{Hiyama:2011ge}, respectively.
 \label{fig:B40_lmax_FY}
}
\end{figure}
%%%%%%%%%%%%%%%%%%%%%%%%%%%%%%%%%%%%%%%%%%%%%%%%%%%%%%%%%%%%%%%%%%%%%%%%%%%%%%%%

Four-body calculations are carried out up to a maximum angular momentum cutoff
$l_{\text{max}}$.
In this appendix we show explicitly the convergence with respect to
$l_{\text{max}}$ in observables for both FY and diagrammatic approaches.

%%%%%%%%%%%%%%%%%%%%%%%%%%%%%%%%%%%%%%%%%%%%%%%%%%%%%%%%%%%%%%%%%%%%%%%%%%%%%%%%
\begin{figure}[tb!]
\centering
\includegraphics[width=0.95\linewidth]{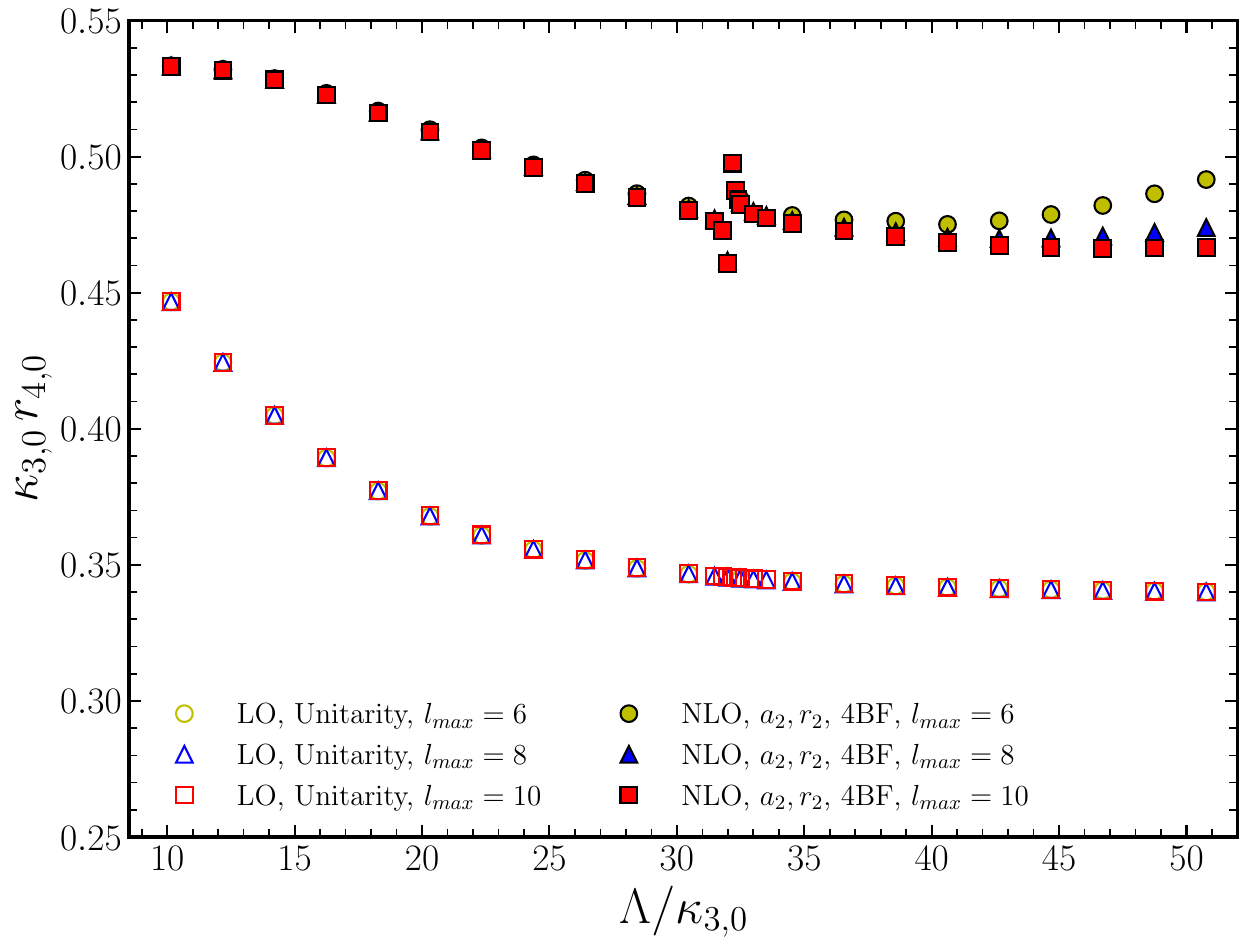}
\caption{
 Tetramer ground-state radius $r_{4,0}$ (in units of $\kappa_{3,0}^{{-}1}$)
 as a function of the momentum cutoff $\Lambda$ (in units of $\kappa_{3,0}$)
 for the quartic super-Gaussian regulator in the FY equations.
 Yellow circles, blue triangles, and red squares denote LO results at unitarity
 for $l_{\text{max}}=$ 6, 8 and 10, respectively.
 The full points are the corresponding full NLO values with $a_2$, $r_2$, and
 the four-body force included.
 \label{fig:r40_lmax_FY}
}
\end{figure}
%%%%%%%%%%%%%%%%%%%%%%%%%%%%%%%%%%%%%%%%%%%%%%%%%%%%%%%%%%%%%%%%%%%%%%%%%%%%%%%%

%%%%%%%%%%%%%%%%%%%%%%%%%%%%%%%%%%%%%%%%%%%%%%%%%%%%%%%%%%%%%%%%%%%%%%%%%%%%%%%%
\begin{figure*}[tb!]
\centering
\includegraphics[width=0.45\linewidth]{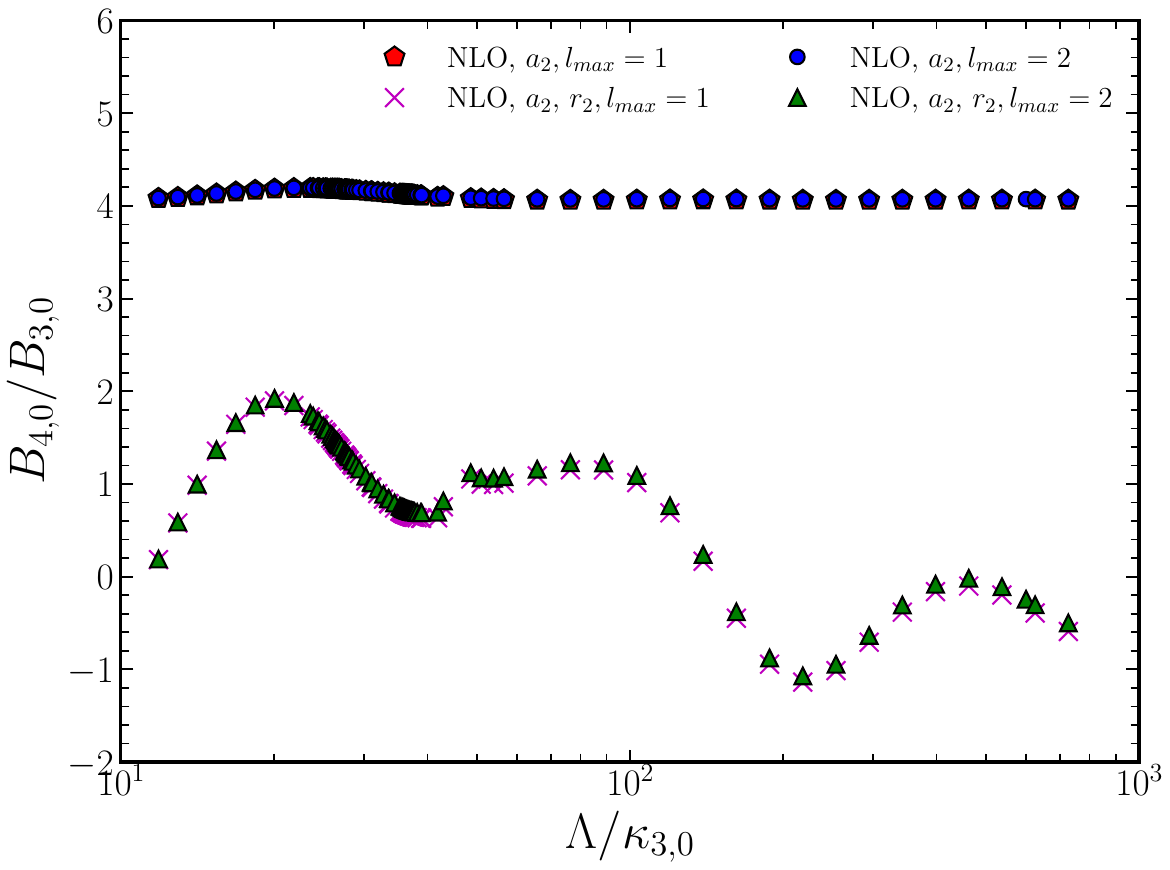}
\includegraphics[width=0.45\linewidth]{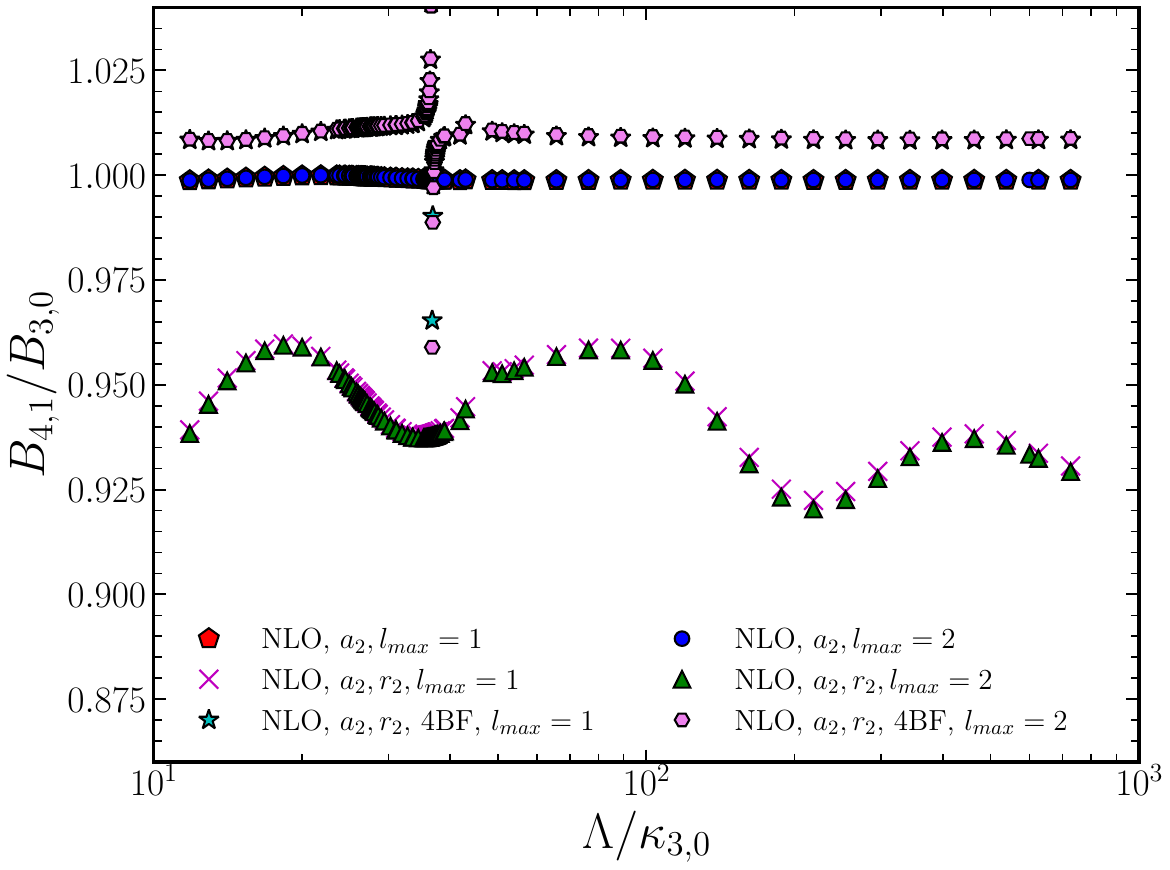}
\caption{
 Tetramer ground- (left panel) and excited- (right panel) binding energies at
 NLO as a function of the sharp cutoff $\Lambda$ (in units of $\kappa_{3,0}$)
 using the diagrammatic approach.
 Red pentagons and blue circles denote results for $l_{\text{max}}=$ 1 and 2,
 respectively, with the scattering length $a_2$ only.
 The corresponding results including also the effective range $r_2$ are
 represented by magenta crosses and and green triangles, while those for the
 full NLO ($a_2$, $r_2$ and four-body force), by cyan stars and violet hexagons
 (right panel only).
 \label{fig:B4i_NLO_DA_lmax}
}
\end{figure*}
%%%%%%%%%%%%%%%%%%%%%%%%%%%%%%%%%%%%%%%%%%%%%%%%%%%%%%%%%%%%%%%%%%%%%%%%%%%%%%%%

%%%%%%%%%%%%%%%%%%%%%%%%%%%%%%%%%%%%%%%%%%%%%%%%%%%%%%%%%%%%%%%%%%%%%%%%%%%%%%%%
\begin{figure}[tb!]
\centering
\includegraphics[width=0.95\linewidth]{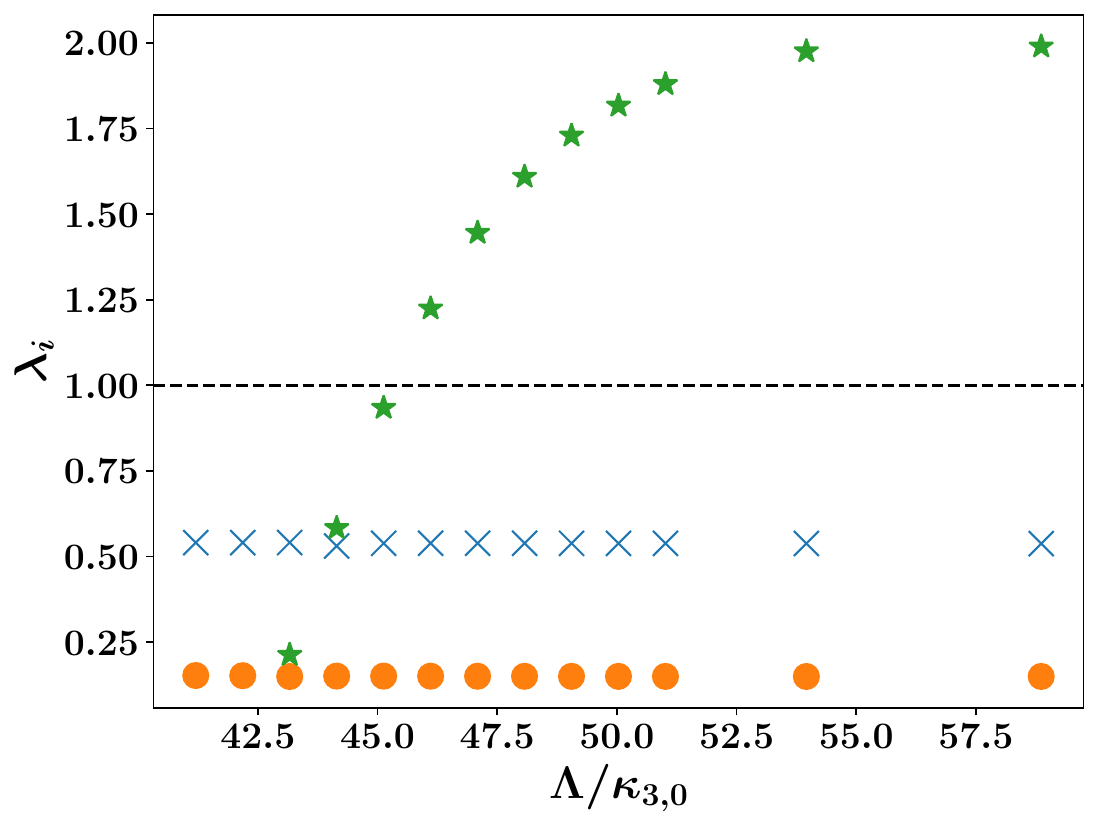}
\caption{
 Largest three (dimensionless) eigenvalues $\lambda_i$ of $\widetilde{K}_{33}$
 with $l_{\text{max}} = 1$ as a function of the cutoff $\Lambda$ (in units of
 $\kappa_{3,0}$).
 \label{fig:eig_PK33}
}
\end{figure}
%%%%%%%%%%%%%%%%%%%%%%%%%%%%%%%%%%%%%%%%%%%%%%%%%%%%%%%%%%%%%%%%%%%%%%%%%%%%%%%%

In Fig.~\ref{fig:B40_lmax_FY}, the four-body ground-state binding energy (in
units of $B_{3,0}$) to NLO for expansion around the unitarity limit obtained by
solving the FY equations with the quartic super-Gaussian regulator is plotted up
to $\Lambda \approx 300\kappa_{3,0}$ for $l_{\text{max}} \leq 4$.
If only the $s$ wave is included ($l_{\text{max}}=0$), the LO four-body
ground-state binding energy quickly approaches the universal
ratio~\cite{Deltuva:2010xd, Lin:2023zqw}.
However, as the cutoff increases, it gradually deviates from this value.
The inclusion of higher partial waves reduces the deviation at large cutoffs and
leads to convergence toward the universal value.
In the accessible cutoff range, results up to NLO for $l_{\text{max}}=2, 4$ are
nearly the same, indicating that convergence for the ground-state binding energy
is achieved for $l_{\text{max}}=4$ in the FY calculation.

The NLO radius, however, requires a larger $l_{\text{max}}$ to
achieve convergence when the range correction and four-body force are included.
As shown in Fig.~\ref{fig:r40_lmax_FY}, the LO results for $l_{\text{max}}$ = 6,
8, and 10 are nearly identical, indicating convergence already at
$l_{\text{max}} = 6$.
In contrast, the full NLO radius, with $a_2$, $r_2$, and the four-body force
included, only converges at $l_{\text{max}} = 10$, leading to a substantial
increase in computational cost.

Figure~\ref{fig:B4i_NLO_DA_lmax} shows the tetramer ground- and excited-state
binding energies obtained using the diagrammatic approach with maximum angular
momentum $l_{\text{max}} = 1, 2$.
Results with different $l_{\text{max}}$ are indistinguishable on the scale used
in the figure.
The truncation error at $l_{\text{max}} = 2$ is estimated to be $<0.1\%$, around
the same order as the numerical uncertainty.
In particular, a much faster convergence with respect to
$l_{\text{max}}$ is observed compared to the results obtained using the FY
equation.
This is likely because the diagrammatic approach uses single-particle
coordinates, where the symmetry between the two single bosons (other than the
dimer) is exact at any $l_{\text{max}}$, whereas the FY equations in this work
use Jacobi coordinates, where this symmetry is only recovered as
$l_{\text{max}}\to \infty$.

The results of incomplete NLO with $a_2$ only in Fig.~\ref{fig:B4i_NLO_DA_lmax}
vary little with the cutoff.
In contrast, when $r_2$ is included, there is wild cutoff variation, as
documented previously~\cite{Bazak:2018qnu}.
The inclusion of a four-body force fitted to $B_{4,0}$ ensures convergence also
of $B_{4,1}$.
There is, however, a divergence near $\Lambda \simeq 37\kappa_{3,0}$,
associated with an ``exceptional cutoff''.
In addition, there is also a bump at $\Lambda\simeq 45\kappa_{3,0}$.
This corresponds to a numerical artifact caused by $1-\widetilde{K}_{33}$
becoming singular near this cutoff, where Eq.~\eqref{4bodydiageqn3-opform}
technically becomes ill-defined.
To demonstrate this issue, the three largest eigenvalues of $\widetilde{K}_{33}$
are shown in Fig.~\ref{fig:eig_PK33}.
Two of the three eigenvalues exhibit little cutoff dependence and are always
smaller than one.
However, the last one reaches 1 near $\Lambda/\kappa_{3,0} \approx 45$.
While this eigenvalue of $\widetilde{K}_{33}$ appears in the spectrum as the
deep trimer emerges at sufficiently large cutoffs, it does not correspond to any
physical tetramer state since $\widetilde{K}_{33}$ is only one component of the
four-body kernel in the four-body integral
equation~\eqref{4bodydiageqn2-opform}.
The full four-body equation~\eqref{4bodydiageqn2-opform} is free of such
numerical artifacts, and the RG invariance of the tetramer binding energies is
not affected.

\bibliography{refs}

\end{document}

%% file: defs.tex
\newcommand{\simge}{\hspace*{0.2em}\raisebox{0.5ex}{$>$}
	\hspace{-0.8em}\raisebox{-0.3em}{$\sim$}\hspace*{0.2em}}

\newcommand{\ie}{\textit{i.e.}\xspace}
\newcommand{\eg}{\textit{e.g.}\xspace}

\newcommand{\apriori}{\textit{a priori}\xspace}
\WithSuffix\newcommand\apriori*{\textit{a-priori}\xspace}

\newcommand{\etc}{\textit{etc.}\xspace}

\newcommand{\abinitio}{\textit{ab initio}\xspace}

\renewcommand{\vec}[1]{\mathbf{#1}}

\newcommand{\vecq}{\mathbf{q}}

\newcommand{\ii}{\mathrm{i}}

\newcommand{\OO}{\mathcal{O}}

\let\angstr\AA
\renewcommand{\AA}{\angstr\xspace}

\newcommand{\spn}{\text{span}}
\newcommand{\argmin}{\text{arg min}}

\newcommand{\CG}[6]{\ensuremath{C^{#5#6}_{#1#2,#3#4}}}
\WithSuffix\newcommand\CG*[3]{\ensuremath{C^{#3}_{#1,#2}}}

\stackMath
\newcommand\reallywidehat[1]{%
\savestack{\tmpbox}{\stretchto{%
  \scaleto{%
    \scalerel*[\widthof{\ensuremath{#1}}]{\kern-.6pt\bigwedge\kern-.6pt}%
    {\rule[-\textheight/2]{1ex}{\textheight}}%WIDTH-LIMITED BIG WEDGE
  }{\textheight}%
}{0.5ex}}%
\stackon[1pt]{#1}{\tmpbox}%
}

\newcommand*\rvec[1]%
{\ensuremath{\overset{\smash{\raisebox{-1.5pt}{\tiny$\rightarrow$}}}{#1}}}
\newcommand*\lvec[1]%
{\ensuremath{\overset{\smash{\raisebox{-1.5pt}{\tiny$\leftarrow$}}}{#1}}}

\NewEnviron{subalign}[1][]{%
\begin{subequations}\begin{align}
  \BODY
\end{align}\label{#1}\end{subequations}
}

\NewEnviron{spliteq}{%
\begin{equation}\begin{split}
  \BODY
\end{split}\end{equation}
}

\NewEnviron{mleq}{%
\begin{equation}
  \BODY
\end{equation}
}

\definecolor{green}{rgb}{0.0, 0.5, 0.0}